\documentclass[a4paper]{article}
\usepackage[a4paper,top=3cm,bottom=2cm,left=2.5cm,right=2.5cm,marginparwidth=1.75cm]{geometry}
\usepackage{amsfonts}
\usepackage{bm}
\usepackage{extarrows}
\usepackage{amssymb}
\usepackage{color}
\usepackage[all]{xy}
\usepackage{graphicx}
\usepackage{braket}
\usepackage{amsmath}
\usepackage{appendix}
\usepackage{graphicx}
\usepackage[colorinlistoftodos]{todonotes}
\usepackage{amsthm}
\usepackage{mathrsfs}
\usepackage{amssymb}
\usepackage{accents}
\usepackage{extarrows}
\usepackage{makecell}
\usepackage{authblk}
\usepackage{url}
\usepackage{bbm}
\usepackage{hyperref}
\usepackage{cite}
\usepackage{eufrak}
\hypersetup{colorlinks,
            citecolor=black,
            linkcolor=black,
            urlcolor=green,
            pdftex}

\usepackage[english]{babel}
\usepackage[utf8x]{inputenc}
\usepackage[T1]{fontenc}

\newcommand{\vev}[1]{\left\langle #1 \right\rangle}

\usepackage[normalem]{ulem}

\title{Twisted geometry coherent states in all dimensional loop quantum gravity: II. Ehrenfest Property}
\author[1,2]{Gaoping Long \footnote{201731140005@mail.bnu.edu.cn}\thanks{corresponding author}}
\affil[1]{Department of Physics, South China University of Technology, Guangzhou 510641, China}
\affil[2]{Department of Physics, Beijing Normal University, Beijing 100875, China}

\date{}

\begin{document}

\maketitle

\begin{abstract}
In the preceding paper of this series of articles we constructed the twisted geometry coherent states in all dimensional loop quantum gravity and established their peakedness properties. In this paper we establish the ``Ehrenfest property'' of these coherent states which are labelled by the twisted geometry parameters. By this we mean that the expectation values of the
polynomials of the elementary operators as well as the operators which are not polynomial functions of the
elementary operators, reproduce, to zeroth order in $\hbar$, the values of the corresponding classical functions at the twisted geometry space point where the coherent state is peaked.

\end{abstract}

\section{Introduction}
In our companion paper \cite{Long:2021lmd}, we constructed a new family of coherent state in all dimensional loop quantum gravity (LQG) and studied its basic properties. This new family of coherent state, whose analogue in (1+3)-dimensional $SU(2)$ LQG is proposed and studied in \cite{Rovelli_2006,Bianchi_2009,Bianchi_2010,Calcinari_2020,Freidel:2010aq}, is called twisted geometry coherent state since they are labeled by the twisted geometry variables which parametrize the $SO(D+1)$ holonomy-flux phase space of all dimensional LQG. As we explained in \cite{Long:2021lmd}, we consider the twisted geometry coherent state in all dimensional LQG instead of the heat-kernel coherent state which is frequently used in (1+3)-dimensional $SU(2)$ LQG \cite{1994The,ThiemannComplexifierCoherentStates,Thomas2001Gauge,2001Gauge,2000Gauge,Han_2020, Han_2020semiclassical,Zhang:2021qul,Long:2021izw}, because the specific studies of heat-kernel coherent state in all dimensional LQG are confronted with some technical problems. Nevertheless, the twisted geometry coherent state in all dimensional LQG take much simpler formulation than the heat-kernel one \cite{Long:2021xjm}, which ensures that its related calculations only involves the familiar Gaussian summation and $SO(D+1)$ coherent intertwiner which has been fully studied in \cite{Long:2020euh} \cite{Livine:2007Nsfv}. Thus, if one can verify that the twisted geometry coherent state in all dimensional LQG possesses a well-behaved peakedness property and ``Ehrenfest property'' in the $SO(D+1)$ holonomy-flux phase space, then the twisted geometry coherent state can be used to study many issues involving the semi-classicality in all dimensional LQG, e.g. the effective dynamics based on coherent state. In our companion paper \cite{Long:2021lmd}, we have shown that the twisted geometry coherent states in all dimensional LQG provide an over-completeness basis of the kinematic Hilbert space in which edge-simplicity constraint is solved, and the expectation values of holonomy and flux operators with respect to the twisted geometry coherent states coincide with the corresponding classical values given by the labels of the coherent states, up to some gauge degrees of freedom. Besides, the peakedness of the wave functions of the twisted geometry coherent state in holonomy, momentum and phase space representations is studied and its well controlled by a semi-classical parameter which is proportional to $\hbar$.

The main result of the present article is that the ``Ehrenfest property'', to zeroth order, indeed holds for the twisted geometry coherent states in all dimensional LQG. In other words, the expectation values of polynomials of the elementary operators as well as the operators which are not polynomial functions of the elementary operators, reproduce, to zeroth order in $\hbar$, the values of the corresponding classical functions at the twisted geometry space point where the coherent state is peaked. To achieve this goal, we consider the monomials of the holonomy and flux operators. By using the completeness relation of the twisted geometry coherent state, the calculation of the monomials of the holonomy and flux operators is converted into the calculation of the matrix elements of the holonomy and flux operators in the twisted geometry coherent state basis. Then, the matrix elements of the holonomy and flux operators in the twisted geometry coherent state basis can be calculated by using the techniques developed in the calculation of overlap function of twisted geometry coherent state \cite{Long:2021lmd}.  To complete this calculation, the properties of the Clebsch-Gordan coefficients related to the Perelomov type coherent state of $SO(D+1)$, as well as the derivative of the overlap functions of Perelomov type coherent state of $SO(D+1)$ are studied as the key points. Besides, similar to that of the heat-kernel coherent state in $SU(2)$ LQG \cite{2000Gauge},  the expectation values of non-polynomial operators with respect to twisted geometry coherent state in all dimensional LQG will be studied by reformulating it as the Hamburger moment problem.

This paper is organized as follows. In section \ref{sec2}, we will review the Kinematic structures of all dimensional LQG and some basic properties of twisted geometry coherent state which are necessary for the studies in this article. Beginning with the structure of classical phase spaces, including the connection phase space and the holonomy-flux phase space of all dimensional LQG, the twisted geometry parametrization and the analysis of gauge degrees of freedom with respect to simplicity constraint will be reviewed. Then, the quantum Hilbert space with its coherent-intertwiner-spin-network basis and the elementary operators in all dimensional LQG will be pointed out. Besides, as key ingredients to construct the ``Ehrenfest property'' of the twisted geometry coherent state, the completeness relation of twisted geometry coherent states and the overlap functions of both Perelomov type coherent state and twisted geometry coherent state in all dimensional LQG will be introduced explicitly. Based on this foundation, we will construct the ``Ehrenfest property'' of the twisted geometry coherent state explicitly in section \ref{sec3}. We will first construct the ``Ehrenfest property'' for operator monomials by proving the matrix elements of the elementary operators with respect to twisted geometry coherent state are well evaluated by their corresponding expectation values, and then construct it for non-polynomial operators by using the Hamburger theorem.  Finally, a short conclusion will be given in section \ref{sec4}.

\section{Kinematic structure of all dimensional loop quantum gravity}\label{sec2}
\subsection{Classical phase space of all dimensional loop quantum gravity}\label{sec2.1}
The (1+D)-dimensional Lorentzian LQG is constructed by canonically quantizing GR based on the
Yang-Mills phase space coordinatized by the conjugate pair $(A_{aIJ}, \pi^{bKL})$ with the non-vanishing Poisson bracket \cite{Bodendorfer:Ha,Bodendorfer:La,Bodendorfer:Qu}
\begin{equation}\label{Poisson1}
\{{A}_{aIJ}(x), \pi^{bKL}(y)\}=2\kappa\beta\delta_a^b\delta_{[I}^K\delta_{J]}^{L}\delta^{(D)}(x-y),
\end{equation}
 where the connection $A_{aIJ}$ and its canonical conjugate momentum $\pi^{bKL}$ are $so(D+1)$ valued fields on a D-dimensional spatial manifold $\Sigma$.  $\kappa$ and $\beta$ represent the gravitational constant and Babero-Immirze parameter respectively. Here we use $I,J,K,...$ for the internal vector
index in the definition representation space of $SO(D+1)$ and $a,b,c,...$ for the spatial index.
The dynamics of this Hamiltonian system is governed by the Gaussian, Simplicity, vector and scalar constraints, which read
\begin{equation}
\mathcal{G}^{IJ}\equiv \partial_a\pi^{aIJ}+2{A}_{aK}^{[I}\pi^{a|K|J]}\approx0,
\end{equation}
\begin{equation}\label{Sdef}
\quad S^{ab[IJKL]}\equiv\pi^{a[IJ}\pi^{|b|KL]}\approx0,
\end{equation}
\begin{equation}
\quad  C_a\approx0, \ \text{and}\    C\approx0
\end{equation}
respectively.
Based on the Poisson structure \eqref{Poisson1} of the connection phase space, one can check that these constraints obey a first class constraint algebra.
Furthermore, one can also check that the Gauss constraint generates the $SO(D+1)$ gauge transformation of this Yang-Mills gauge theory, while the simplicity constraint restricts the degrees of freedom of $\pi^{aIJ}$ to that of a D-frame $E^{aI}$ which describes the spatial geometry and generates some other gauge transformations. 
In other words, one can solve the simplicity constraint and get the solution $\pi^{aIJ}=2n^{[I}E^{|a|J]}$ with $n^{I}E^{a}_{I}=0, n^In_I=1$ and  $E^{aI}$ being the densitized D-frame which gives the spatial metric $q_{ab}$ by $qq^{ab}=E^{aI}E^b_I$, where $q$ is the determinant of $q_{ab}$ \cite{Bodendorfer:Ha}. Besides, one can reconstruct
the densitized extrinsic curvature $\tilde{{K}}_{ab}=q_{bc}\tilde{{K}}_a^{\ c}$ of the spatial manifold $\sigma$ by
\begin{equation}
\tilde{{K}}_a^{\ b}\approx{ K}_{aIJ}\pi^{bIJ}\equiv \frac{1}{\beta}({A}_{aIJ}-\Gamma_{aIJ})\pi^{bIJ}
\end{equation}
 on the Gaussian and simplicity constraint surface, where $\Gamma_{aIJ}$ is purely constructed from $\pi^{aIJ}$ and it is the spin connection of $E^{aI}$ exactly on the simplicity constraint surface \cite{Bodendorfer:Ha}. To clarify the gauge degrees of freedom corresponding to simplicity simplicity, let us decompose ${K}_{aIJ}
:=\frac{1}{\beta}({A}_{aIJ}-\Gamma_{aIJ})$ as
 \begin{equation}
 {K}_{aIJ}=2n^{[I}{K}_a^{J]}+\bar{{K}}_{a}^{IJ},
 \end{equation}
 where $\bar{{K}}_{a}^{IJ}:=\bar{\eta}^I_K\bar{\eta}^J_L{K}_{a}^{KL}$ with $\bar{\eta}_I^J:=\delta_I^J-n_In^J$ and $\bar{{K}}_{a}^{IJ}n_I=0$. Based on Eqs.\eqref{Poisson1} and \eqref{Sdef}, one can check that the component $2n^{[I}{K}_a^{J]}$ and $\pi^{aIJ}$ Poisson commutes with the simplicity constraint while $\bar{{K}}_{a}^{IJ}$ does not. Hence, the simplicity constraint fixes both $\tilde{{K}}_a^{\ b}$ and $q_{ab}$ and it exactly introduce extra gauge degrees of freedom represented by $\bar{{K}}_{a}^{IJ}$. The details of these discussion can be found in the Ref.\cite{Bodendorfer:Ha} and it is shown that, the standard symplectic reduction procedures with respect to Gaussian and simplicity constraint in the $SO(D+1)$ connection phase space leads to the ADM phase space of $(1+D)$-dimensional GR, with the coordinates $\tilde{{K}}_a^{\ b}$ and $q_{ab}$ of the ADM phase space are Dirac observables with respect to Gaussian and simplicity constraints. It should be emphasized that $\bar{{K}}_{a}^{IJ}$ are pure gauge components with respect to the simplicity constraint, which only contributes gauge degrees of freedom in this $SO(D+1)$ Yang-Mills theory. As we will show, the counterpart of $\bar{{K}}_{a}^{IJ}$ in the discrete phase of all dimensional LQG is a critical point of the results of this paper.

Apart from the different gauge group which however is compact and the additional simplicity constraint,
the $SO(D+1)$ connection formulation of (1+D)-dimensional GR is precisely the same as $SU(2)$ connection formulation of (1+3)-dimensional GR, and the quantisation of the $SO(D+1)$ connection formulation is therefore in complete analogy with (1+3)-dimensional $SU(2)$ LQG \cite{Ashtekar2012Background,thiemann2007modern,rovelli2007quantum,RovelliBook2,Han2005FUNDAMENTAL}. By following any standard text on LQG such as \cite{thiemann2007modern,rovelli2007quantum}, the loop quantization of the $SO(D+1)$ connection formulation of (1+D)-dimensional GR leads to a kinematical Hilbert space $\mathcal{H}$ \cite{Bodendorfer:Qu}, which can be regarded as a union of the Hilbert spaces $\mathcal{H}_\gamma=L^2((SO(D+1))^{|E(\gamma)|},d\mu_{\text{Haar}}^{|E(\gamma)|})$ on all possible finite graphs $\gamma$ embedded in $\Sigma$,  where $E(\gamma)$ denotes the set composed by the independent edges of $\gamma$ and $d\mu_{\text{Haar}}^{|E(\gamma)|}$ denotes the product of the Haar measure on $SO(D+1)$. In this sense, on each given $\gamma$ there is a discrete phase space $(T^\ast SO(D+1))^{|E(\gamma)|}$, which is coordinatized by the elementary discrete variables---holonomies and fluxes. The holonomy of $A_{aIJ}$ along an edge $e\in\gamma$ is defined by
 \begin{equation}
h_e[A]:=\mathcal{P}\exp(\int_eA)=1+\sum_{n=1}^{\infty}\int_{0}^1dt_n\int_0^{t_n}dt_{n-1}...\int_0^{t_2} dt_1A(t_1)...A(t_n),
 \end{equation}
 where $A(t):=\frac{1}{2}\dot{e}^aA_{aIJ}\tau^{IJ}$, $\dot{e}^a$ is the tangent vector field of $e$, $\tau^{IJ}$ is a basis of $so(D+1)$ given by $(\tau^{IJ})^{\text{def.}}_{KL}=2\delta^{[I}_{K}\delta^{J]}_{L}$ in definition representation space of $SO(D+1)$, and $\mathcal{P}$ denoting the path-ordered product.
The flux $F^{IJ}_e$ of $\pi^{aIJ}$ through the (D-1)-dimensional face dual to edge $e$ is defined by
\begin{equation}\label{F111}
 F^{IJ}_e:=-\frac{1}{4}\text{tr}\left(\tau^{IJ}\int_{e^\star}\epsilon_{aa_1...a_{D-1}}h(\rho^s_e(\sigma)) \pi^{aKL}(\sigma)\tau_{KL}h(\rho^s_e(\sigma)^{-1})\right),
 \end{equation}
 where $e^\star$ is the (D-1)-face traversed by $e$ in the dual lattice of $\gamma$, $\rho^s(\sigma): [0,1]\rightarrow \Sigma$ is a path connecting the source point $s_e\in e$ to $\sigma\in S_e$ such that $\rho_e^s(\sigma): [0,\frac{1}{2}]\rightarrow e$ and $\rho_e^s(\sigma): [\frac{1}{2}, 1]\rightarrow S_e$. Similarly, we can define the dimensionless flux $X^{IJ}_e$ as
 \begin{equation}
 X^{IJ}_e=-\frac{1}{4\beta a^{D-1}}\text{tr}\left(\tau^{IJ}\int_{e^\star}\epsilon_{aa_1...a_{D-1}}h(\rho^s_e(\sigma)) \pi^{aKL}(\sigma)\tau_{KL}h(\rho^s_e(\sigma)^{-1})\right),
 \end{equation}
 where $a$ is an arbitrary but fixed constant with the dimension of length. Since $SO(D+1)\times so(D+1)\cong T^\ast SO(D+1)$, this new discrete phase space $\times_{e\in \gamma}(SO(D+1)\times so(D+1))_e$, called the phase space of $SO(D+1)$ loop quantum gravity on the fixed graph $\gamma$, is a direct product of $SO(D+1)$ cotangent bundles. Finally, the complete phase space of the theory is given by taking the union over the phase spaces of all possible graphs.
In the discrete phase space associated to $\gamma$, the constraints are expressed by the smeared variables. The discretized Gauss constraints is given by
 \begin{equation}
 G_v:=\sum_{b(e)=v}X_e-\sum_{t(e')=v}h_{e'}^{-1}X_{e'}h_{e'}\approx0.
 \end{equation}
The discretized simplicity constraints are separated as two sets. The first one is the edge-simplicity constraint $S^{IJKL}_e\approx0$ which takes the form \cite{Bodendorfer:Qu}\cite{Bodendorfer:SgI}
\begin{equation}
\label{simpconstr}
S_e^{IJKL}\equiv X^{[IJ}_e X^{KL]}_e\approx0, \ \forall e\in \gamma
\end{equation}
and the second one is the vertex-simplicity constraint $S^{IJKL}_{v,e,e'}\approx0$ which is given by \cite{Bodendorfer:Qu}\cite{Bodendorfer:SgI}
\begin{equation}\label{simpconstr2}
\quad S_{v,e,e'}^{IJKL}\equiv X^{[IJ}_e X^{KL]}_{e'}\approx0,\ \forall e,e'\in \gamma, s(e)=s(e')=v.
\end{equation}
The symplectic structure of the discrete phase space can be expressed by the Poisson algebra between the elementary variables $(h_e, X^{IJ}_e)$, which reads
 \begin{eqnarray}
 &&\{h_e, h_{e'}\}=0,\quad\{h_e, X^{IJ}_{e'}\}=\delta_{e,e'}\frac{\kappa}{a^{D-1}} \frac{d}{dt}(e^{t\tau^{IJ}}h_e)|_{t=0}, \\\nonumber
 && \{X^{IJ}_e, X^{KL}_{e'}\}=\delta_{e,e'}\frac{\kappa}{a^{D-1}}(\delta^{IK}X_e^{JL}+\delta^{JL }X^{IK}_e-\delta^{IL}X_e^{JK}-\delta^{JK}X_e^{ IL}).
 \end{eqnarray}
Then, by using this Poisson algebra, it is easy to verify that $G_v\approx0$ and $S_e\approx0$ form a first class constraint system as
\begin{eqnarray}
\label{firstclassalgb}
\{S_e, S_e\}\propto S_e\,,\,\, \{S_e, S_v\}\propto S_e,\,\,\{G_v, G_v\}\propto G_v,\,\,\{G_v, S_e\}\propto S_e,\,\,\{G_v, S_v\}\propto S_v, \quad b(e)=v,
\end{eqnarray}
where the algebras among $G_v\approx0$ are isomorphic to the $so(D+1)$ algebra, and the ones involving $S_e\approx0$ weakly vanish. Especially, the algebras among the vertex-simplicity constraint are the problematic ones, with the open anomalous brackets \cite{Bodendorfer:2011onthe}
\begin{eqnarray}
\label{anomalousalgb}
\{S_{v,e,e'},S_{v,e,e''}\}\propto \emph{anomaly terms}
\end{eqnarray}
where the $ ``\emph{anomaly terms}''$ are not proportional to any of the existing constraints in the phase space.

In fact, a similar simplicity constraint is widely studied in the 4-dimensional spin-foam theory \cite{Dupuis:2010RSC,Kaminski:2009SFA,Livine:2007Nsfv,Engle:2007va,Engle:2007LQGv,Freidel:2007NSF}. In all dimensional LQG, the treatment of this anomalous vertex simplicity constraint in both quantum theory and classical discrete theory is a critical problem. It has been shown that the strong imposition of vertex simplicity constraint eliminates the physical degrees of freedom erroneously \cite{FreidelBFDescriptionOf,Barrett:1997Rsn}. Thus, one need to construct a new treatment of the anomalous simplicity constraint and explain its geometric meaning to ensure the correctness.
The generalized twisted geometric parametrization of the discrete phase space of all dimensional LQG is such a scheme, which leads us to solve the anomalous simplicity constraint and ensure the physical degrees of freedom and gauge degrees of freedom are separated correctly.  As shown in Ref. \cite{PhysRevD.103.086016}, with the Gaussian constraint, simplicity constraint and the $(D-1)$-faces' shape matching condition being imposed appropriately, the generalized twisted geometric parameters reproduce the Regge geometries on the D-dimensional spatial manifold $\Sigma$ correctly. Let us review the generalized twisted geometric parametrization of the discrete phase space in all dimensional LQG briefly as follows. Recall that the discrete phase space associated to a given graph $\gamma$ is denoted by $\times_{e\in \gamma}T^\ast SO(D+1)_e$. In this phase space, one can first solve the  edge-simplicity constraint equation and it leads to the constraint surface defined by
\begin{equation}
\times_{e\in \gamma}T_{\text{s}}^\ast SO(D+1)_e:=\{(...,(h_e,X_e),...)\in \times_{e\in \gamma}T^\ast SO(D+1)_e|X_{e}^{[IJ}X_{e}^{KL]}=0\}.
\end{equation}
To simplify the statements,  one can first consider the edge-simplicity constraint surface $T_{\text{s}}^\ast SO(D+1)$ for one copy of the edge only.
Then, the generalized twisted geometry variables $(V,\tilde{V},\xi^o, \eta,\bar{\xi}^\mu)$ can be introduced to reparametrize the edge-simplicity constraint surface $T_{\text{s}}^\ast SO(D+1)$. These generalized twisted geometry variables $(V,\tilde{V},\xi^o, \eta,\bar{\xi}^\mu)$ and their space
 \begin{equation}
 P:=Q_{D-1}\times Q_{D-1}\times T^*S^1\times SO(D-1)
 \end{equation}
 are constructed as follows.  The bi-vector $V$ or $\tilde{V}$ with fixed norm constitutes the space $Q_{D-1}:=SO(D+1)/(SO(2)\times SO(D-1))$, where $SO(2)\times SO(D-1)$ is the maximum subgroup of $SO(D+1)$ which preserves the bi-vector $\tau_o:=2\delta_1^{[I}\delta_2^{J]}$. The real number $\eta$ combining with $\xi^o\in [-\pi,\pi)$ constitute the space $T^*S^1$.  Besides, $e^{\bar{\xi}^\mu\bar{\tau}_\mu}=:\bar{u}$ is an element of $SO(D-1)$ which preserves the bi-vector $\tau_o$, with $\bar{\tau}_\mu$ being a basis of $so(D-1)$ and $\mu\in\{1,...,\frac{(D-1)(D-2)}{2}\}$. In order to capture the intrinsic curvature by these parameters, it is necessary to specify one pair of the $SO(D+1)$ valued Hopf sections $u(V)$ and $\tilde{u}( \tilde{V})$, which satisfy $V=u(V)\tau_ou(V)^{-1}$ and $\tilde{V}=-\tilde{u}(\tilde{V})\tau_o\tilde{u}(\tilde{V})^{-1}$. Then, the generalized twisted geometry parametrization for one copy of the edge can be established by the map
\begin{eqnarray}\label{para}
P\ni(V,\tilde{V},\xi^o,\eta,\bar{\xi}^\mu)\mapsto(h, X)\in T_{\text{s}}^\ast SO(D+1):&& X=\frac{1}{2}\eta V=\frac{1}{2}\eta u(V)\tau_ou(V)^{-1}\\\nonumber
&&h=u(V)\,e^{\bar{\xi}^\mu\bar{\tau}_\mu}e^{\xi^o\tau_o}\,\tilde{u}(\tilde{V})^{-1}.
\end{eqnarray}
 It is easy to check that, the two points $(V,\tilde{V},\xi^o, \eta,\bar{\xi}^\mu)$ and $(-V,-\tilde{V},-\xi^o,-\eta,\dot{\xi}^\mu)$ in $P$ are mapped to the same point $(h, X)\in T_{s}^\ast \!SO(D+1)$ by the map \eqref{para}, where $e^{\dot{\xi}^\mu\bar{\tau}_{\mu}}=e^{-2\pi\tau_{13}}e^{\bar{\xi}^\mu\bar{\tau}_{\mu}}e^{2\pi\tau_{13}}$ and $\tau_{13}=\delta_1^{[I}\delta_3^{J]}$.  Thus, the map \eqref{para} gives a two-to-one double covering of the image. A more detailed study shows that \cite{PhysRevD.103.086016}, a bijection map can be constructed in the region $|X|\neq0$ by selecting either branch among the two-to-one double covering \eqref{para}. Moreover, the new parameters also simplify the Poisson structures of the discrete phase space. For instance, the non-vanishing Poisson bracket between $\xi^o$ and $\eta$ can be given by
 \begin{equation}\label{xiN}
 \{\xi^o, \eta\}=\frac{2\kappa}{a^{D-1}},
 \end{equation}
 with $\xi^o$ and $\eta$ representing a portion of the degrees of freedom of extrinsic and intrinsic geometry respectively. Now we can get back to the discrete phase space of all dimensional LQG on the whole graph $\gamma$, which is just the Cartesian product of the discrete phase space on each single edge of $\gamma$. Then, the twisted geometry parametrization of the discrete phase space on one copy of the edge can be generalized to that of the whole graph $\gamma$ directly. Futhermore, the twisted geometry parameters $(V,\tilde{V},\xi^o, \eta)$ take the interpretation of the discrete geometry describing the dual lattice of $\gamma$, which can be explained explicitly as follows. We first note that $\frac{1}{2}\eta _e V_e$ and $\frac{1}{2}\eta _e \tilde{V}_e$ represent the area-weighted outward normal bi-vectors of the $(D-1)$-face dual to $e$ in the perspective of source and target points of $e$ respectively, with $\frac{1}{2}\eta _e$ being the dimensionless area of the $(D-1)$-face dual to $e$. Then, the holonomy $h_e=u_e(V_e)\,e^{\bar{\xi}_e^\mu\bar{\tau}_\mu}e^{\xi_e^o\tau_o}\,\tilde{u}^{-1}_e(\tilde{V}_e)$ takes the interpretation that it rotates the inward normal $-\frac{1}{2}\eta _e\tilde{V}_e$ of the (D-1)-face  dual to $e$ in the perspective of the the target point of $e$, into the outward normal $\frac{1}{2}\eta _e{V}_e$ of the (D-1)-face  dual to $e$ in the perspective of the source point of $e$, wherein $u_e(V_e)$ and $\tilde{u}_e(\tilde{V}_e)$ capture the contribution of intrinsic curvature, and $e^{\xi_e^o\tau_o}$ captures the contribution of extrinsic curvature to this rotation. Moreover, $e^{\bar{\xi}_e^\mu\bar{\tau}_\mu}$ represents some redundant degrees of freedom for reconstructing the discrete geometry. Finally, with the Gaussian and vertex simplicity constraint being imposed at the vertices of $\gamma$, one can get the closed twisted geometry which describes the D-polytopes and their gluing method in the dual lattice of $\gamma$ \cite{PhysRevD.103.086016,Freidel:2010aq,Bianchi:2010Polyhedra}.

The discrete geometric interpretation of the twisted geometry parametrization points out a proper treatment of the anomalous vertex simplicity constraint in the discrete phase space of all dimensional LQG. It has been shown that, by considering some kinds of the continuum limit, one can establish the relation between the constraint surface defined by both of the edge simplicity and anomalous vertex simplicity constraints (CSEVSC) in the discrete phase space and the constraint surface defined by the non-anomalous simplicity constraint (CSNASC) in the connection phase space. Especially, on these two constraint surfaces, the gauge transformation induced by the edge simplicity constraint corresponds to that induced by the non-anomalous simplicity constraint  exactly in the continuum limit, which can be illustrated as
  \begin{equation}
\bar{\xi}^\mu_e|_{\text{CSEVSC}}\xrightarrow{\text{continuum limit}}
\bar{K}_{aIJ}|_{\text{CSNASC}}
  \end{equation}
  where $\bar{\xi}^\mu_e$ and $\bar{K}_{aIJ}$ capture the pure gauge degrees of freedom with respect to simplicity constraint in holonomy $h_e$ and connection $A_{aIJ}$ respectively.
  In summary, the implementation of the Gaussian and anomalous simplicity constraints in discrete phase space contains two steps: (i)  Execute the symplectic reduction with respect to edge simplicity constraint and Gaussian constraint; (ii) Solve the vertex simplicity constraint equation to get the constraint surface. As we mentioned before,  the resulting space is parametrized by the so-called constrained twisted geometry space, which covers the degrees of freedom of internal and external Regge geometry on the $D$-dimensional spatial manifold $\Sigma$, with the twisted geometry parameters being endowed with certain geometric interpretations in Regge geometry \cite{PhysRevD.103.086016}.
\subsection{Spin network basis of the kinematic Hilbert space in all dimensional loop quantum gravity}

The Hilbert space $\mathcal{H}$ of all dimensional LQG is given by the completion of the space of cylindrical functions on the quantum configuration space, which can be decomposed into the sectors --- the Hilbert spaces constructed on graphs. For a given graph $\gamma$ with $|E(\gamma)|$ edges, the related Hilbert space is given by $\mathcal{H}_\gamma=L^2((SO(D+1))^{|E(\gamma)|}, d\mu_{\text{Haar}}^{|E(\gamma)|})$. This Hilbert space associates to the classical phase space $\times_{e\in\gamma}T^\ast SO(D+1)_e$ aforementioned. A basis of this space is given by the spin-network functions which are labelled by (1) an $SO(D+1)$ representation $\Lambda$ assigned to each edge; and (2) an intertwiner $i_v$ assigned to each vertex $v$. Each basis state $\Psi_{\gamma,{\vec{\Lambda}}, \vec{i}}(\vec{h}_e)$, as a wave function on $\times_{e\in\gamma}SO(D+1)_e$, is then given by
\begin{eqnarray}
\Psi_{\gamma,{\vec{\Lambda}}, \vec{i}}(\vec{h}(A))\equiv \bigotimes_{v\in\gamma}{i_v}\,\, \rhd\,\, \bigotimes_{e\in\gamma} \pi_{\Lambda_e}(h_{e}(A)),
\end{eqnarray}
where $\vec{h}(A):=(...,h_e(A),...), \vec{\Lambda}:=(...,\Lambda_e,...), e\in\gamma$, $\vec{i}:=(...,i_v,...), v\in\gamma$ , $\pi_{\Lambda_e}(h_{e})$ denotes the matrix of holonomy $h_e$ associated to edge $e$ in the representation labelled by $\Lambda_e$, and $\rhd$ denotes the contraction of  the representation matrixes of holonomies with the intertwiners. Hence, the wave function $\Psi_{\gamma,{\vec{\Lambda}}, \vec{i}}(\vec{h}(A))$ is simply the product of the functions given by specified components of the holonomy matrices, selected by the intertwiners at the vertices.
The action of the elementary operators---holonomy operator and flux operator---on the spin-network functions can be given as
\begin{eqnarray}
 \hat{ h}_{e}(A)\circ \Psi_{\gamma,{\vec{\Lambda}}, \vec{i}}(\vec{h}(A)) &=& { h}_e(A) \Psi_{\gamma,{\vec{\Lambda}}, \vec{i}}(\vec{h}(A)) \\\nonumber
  \hat{F}_e^{IJ}\circ\Psi_{\gamma,{\vec{\Lambda}}, \vec{i}}(\vec{h}(A)) &=&-\mathbf{i}\hbar\kappa\beta R_e^{IJ}\Psi_{\gamma,{\vec{\Lambda}}, \vec{i}}(\vec{h}(A))
\end{eqnarray}
with $R_{e}^{IJ}:=\text{tr}((\tau^{IJ}h_e)^{\text{T}}\frac{\partial}{\partial h_e})$  being the right
invariant vector fields on $SO(D+1)$ associated to the edge $e$, and $\text{T}$ denoting the transposition of the matrix. Then, the other operators in all dimensional LQG, such as spatial geometric operators and Hamiltonian operator, can be constructed based on these elementary operators \cite{long2020operators,Long:2020agv,Zhang:2015bxa}.

In order to obtain the kinematic physical Hilbert space, one needs to solve the kinematic constraints, including Gaussian constraint, edge-simplicity constraint and vertex simplicity constraint in $\mathcal{H}$. Following the results given in Sec.\ref{sec2.1}, the Gaussian constraint and edge-simplicity constraint are imposed strongly. The resulting space is spanned by the edge-simple and gauge invariant spin-network states, whose edges are labelled by the simple representations of $SO(D+1)$ and vertices are labelled by the gauge invariant intertwiners. Besides, the anomalous vertex simplicity constraints are imposed weakly and the corresponding weak solutions are given by the spin-network states whose vertices are labelled by the simple coherent intertwiners \cite{long2019coherent}. A typical spin-network state whose vertices are labelled by the gauge invariant simple coherent intertwiners can be given as
\begin{equation}
\Psi_{\gamma,\vec{N},\vec{\mathcal{I}}_{\text{s.c.}}}(\vec{h}(A))=\text{tr}(\otimes_{e\in\gamma} \pi_{N_e}(h_e(A))\otimes_{v\in\gamma}\mathcal{I}_v^{\text{s.c.}})
\end{equation}
where $\pi_{N_e}(h_e(A))$ denotes the representation matrix of $h_e(A)$ with $N_e$ is an non-negative integer labeling a simple representation of $SO(D+1)$, and $\vec{\mathcal{I}}_{\text{s.c.}}$ is defined by $\vec{\mathcal{I}}_{\text{s.c.}}:=(...,\mathcal{I}_v^{\text{s.c.}},...)$ with $\mathcal{I}_v^{\text{s.c.}}$ being the so-called gauge invariant simple coherent intertwiner labeling the vertex $v\in\gamma$.
\subsection{Perelomov type coherent state of $SO(D+1)$ and coherent intertwiner}
 In order to give the details of the construction of simple coherent intertwiners,  we must first introduce some concepts of the simple representation of $SO(D+1)$ and the homogeneous harmonic functions on the $D$-sphere.
 The homogeneous harmonic functions with degree $N$ on the $D$-sphere ($S^D$) compose a space $\mathfrak{H}_{D+1}^{N}$ with dimensionality
  \begin{equation}
  \dim(\mathfrak{H}_{D+1}^{N})=\dim(\pi_N)=\frac{(D+N-2)!(2N+D-1)}{(D-1)!N!},
  \end{equation}
  and $\mathfrak{H}_{D+1}^{N}$ is a realization of the simple representation space of $SO(D+1)$ labelled by $N$. Introduce a subgroup series $SO(D+1)\supset SO(D)\supset SO(D-1)\supset ... \supset SO(2)_{\delta_1^{[I}\delta_2^{J]}}$ with $SO(2)_{\delta_1^{[I}\delta_2^{J]}}$ being the one-parameter subgroup of SO$(D+1)$ generated by $\tau_o:=2\delta_1^{[I}\delta_2^{J]}$. Then, an orthogonal and normalized basis of the space $\mathfrak{H}_{D+1}^{N}$ can be given as
  \begin{equation}
  \{\Xi_{D+1}^{N,\mathbf{M}}(\bm{x})|\mathbf{M}:=M_1,M_2,...,M_{D-1}, N\geq M_1 \geq M_2\geq...\geq|M_{D-1}|, \bm{x}\in S^D\}
  \end{equation}
  or equivalently,  in Dirac bracket notation can be denoted by $\ket{N,\mathbf{M}}$, where $\mathbf{M}:=M_1,M_2,...,M_{D-1}$ with $N\geq M_1 \geq M_2\geq...\geq|M_{D-1}|$, and $N,M_1,... M_{D-2}\in\mathbb{N}$, $M_{D-1}\in \mathbb{Z}$. The labels $N, \mathbf{M}$ of the function $\Xi_{D+1}^{N,\mathbf{M}}(\bm{x})$ is interpreted as that $\Xi_{D+1}^{N,\mathbf{M}}(\bm{x})$ belongs to the series of subspaces $\mathfrak{H}_{2}^{M_{D-1}}\subset \mathfrak{H}_{3}^{M_{D-2}}\subset...\subset\mathfrak{H}_{D}^{M_{1}}\subset\mathfrak{H}_{D+1}^{N}$ which are the irreducible simple representation spaces labeled by $M_{D-1},...,M_2,M_1, N$ of the series of subgroups $SO(2)_{\delta_1^{[I}\delta_2^{J]}} \subset SO(3)\subset ... \subset SO(D)\subset SO(D+1)$ respectively \cite{vilenkin2013representation}. With this convention, the orthogonal and normalized property of this basis can be expressed as
\begin{equation}
\langle N,\mathbf{M}|N,\mathbf{M}'\rangle:=\int_{S^D} d\bm{x} \, \overline{\Xi_{D+1}^{N,\mathbf{M}}(\bm{x})}\Xi_{D+1}^{N,\mathbf{M}'}(\bm{x})=\delta_{\mathbf{M},\mathbf{M}'}
\end{equation}
with $\delta_{\mathbf{M},\mathbf{M}'}=1$ if $\mathbf{M}=\mathbf{M}'$ and zero otherwise,
where $d\bm{x}$ is the normalized invariant measure on $S^D$. An element $g\in$ SO$(D+1)$ acts on a spherical harmonic function $f(\bm{x})$ on D-sphere as
  \begin{equation}\label{gfx}
  g\circ f(\bm{x})=f(g^{-1}\circ\bm{x}),
  \end{equation}
  where $g\circ \bm{x}$ denotes the action of $g\in SO(D+1)$ on the point $\bm{x}\in S^{D}$ by its definition.
 Correspondingly, the basis $\{\tau_{IJ}\}$ of $so(D+1)$, defined by $(\tau_{IJ})^{\text{def.}}:=2\delta_I^{[K}\delta_J^{L]}$ in the definition representation space of $SO(D+1)$, are operators in $\mathfrak{H}_{D+1}^{N}$ and they act on the spherical harmonic function as
 \begin{equation}
 \tau_{IJ}\circ f(\bm{x}):=\frac{d}{dt}f(e^{-t\tau_{IJ}}\circ\bm{x})|_{t=0}.
 \end{equation}
This action also gives a representation of the Lie algebra
 \begin{equation}\label{[JJ]}
[\tau_{IJ},\tau_{KL}]=\delta_{IL}\tau_{JK}+\delta_{JK}\tau_{ IL}-\delta_{IK}\tau_{JL}-\delta_{JL }\tau_{IK}.
\end{equation}

The general scheme of the construction of Perelomov type coherent state for compact Lie algebra is introduced in Ref.\cite{GeneralizedCoherentStates}. For the case of $SO(D+1)$ involved in this article, let us consider  the state $|N,\mathbf{N}\rangle\in\mathfrak{H}_{D+1}^{N}$ which corresponds to the highest weight vector with $\mathbf{N}=\mathbf{M}|_{M_1=...=M_{D-1}=N}$. Then, the Perelomov type coherent states in $\mathfrak{H}_{D+1}^{N}$ can be defined by \cite{Long:2020euh}
\begin{equation}
|N,V\rangle:=u(V)|N,\mathbf{N}\rangle.
\end{equation}
 Equivalently, the Perelomov type coherent state $|N,V\rangle$ can also be defined by
  \begin{equation}
  |N,V\rangle:=u(-V)|N,\bar{\mathbf{N}}\rangle,
  \end{equation}
  wherein the state $|N,\bar{\mathbf{N}}\rangle\in \mathfrak{H}_{D+1}^{N}$ corresponds to the lowest weight vector with $\mathbf{N}=\mathbf{M}|_{M_1=...=M_{D-2}=N, M_{D-1}=-N}$. It has been proved that the Perelomov type coherent state $|N,V\rangle$ of $SO(D+1)$ processes well peakedness properties for the operators $\tau_{IJ}$ \cite{Long:2020euh}, i.e. minimize the uncertainty of the expectation value $\langle N,V|\tau_{IJ}|N,V\rangle=\mathbf{i}N V_{IJ}$ and the Heisenberg uncertainty relation of the operators $\tau_{IJ}$: the inequality
      \begin{equation}
      \left(\bigtriangleup\!\vev{\tau_{IJ}}\right)^2 \left(\bigtriangleup\!\vev{\tau_{KL}}\right)^2\ \geq\ \frac{1}{4}\left|\vev{[\tau_{IJ},\tau_{KL}]}\right|^2
      \end{equation}
      is saturated for the Perelomov type coherent state $|N, V\rangle$, where we used the shorthand $\vev{\hat{o}}\equiv\langle N,V|\hat{o}|N,V\rangle$ and $\bigtriangleup\!\vev{\hat{o}}\equiv\sqrt{\vev{\hat{o}^2}-\vev{\hat{o}}^2}$. The family of Perelomov type coherent states $\{|N, V\rangle\}$ also composes an over-complete basis of $\mathfrak{H}_{D+1}^N$, which reads
\begin{equation}\label{iden1}
  \dim\left(\mathfrak{H}_{D+1}^N\right)\int_{Q_{D-1}}dV|N,V\rangle\langle N,V|=\mathbb{I}_{\mathfrak{H}_{D+1}^N},
\end{equation}
where $\int_{Q_{D-1}}dV=1$ with $dV$ being the invariant measure on $Q_{D-1}$ induced by the Haar measure on $SO(D+1)$. One can also check the nonorthogonal property of this type of coherent state, that is, the coherent states $|N, V\rangle$ and $|N, V'\rangle$ with $V\neq V'$ are not mutually orthogonal unless $[V^{IJ}\tau_{IJ},V'^{KL}\tau_{KL}]=0$. This means that we have
\begin{equation}
0\leq|\langle N,V|N, V'\rangle|\leq1
\end{equation}
with $\langle N,V|N, V'\rangle=1$ if $V=V'$, $\langle N,V|N, V'\rangle=0$ if $[V^{IJ}\tau_{IJ},V'^{KL}\tau_{KL}]=0 \ \text{and}\ V\neq V'$. Moreover, it is worth to introduce is the matrix element $\langle N,V|\tau_{IJ}|N, V'\rangle$ of the operator $\tau_{IJ}$ in the Perelomov type coherent state basis. Let us take $V'^{KL}=2\delta_1^{[K}\delta_2^{L]}$ without loss of generality, then we have
\begin{equation}
\langle N,V|\tau_{12}|N, V'\rangle=\mathbf{i}N\langle N,V|N, V'\rangle,
\end{equation}
\begin{equation}
\langle N,V|\tau_{IJ}|N, V'\rangle=0,\ \ \text{for}\ I,J\neq 1,2,
\end{equation}
\begin{equation}
\langle N,V|\tau_{IJ}|N, V'\rangle=N\langle 1,V|\tau_{IJ}|1, V'\rangle\langle N-1,V|N-1, V'\rangle,\ \ \text{for}\ I\in\{1,2\}\ \text{and}\ J\neq 1,2,
\end{equation}
where $\langle 1,V|\tau_{IJ}|1, V'\rangle=0$ if $V=V'$ with $I\in\{1,2\}$ and $J\neq 1,2$. We are interested in the derivatives of $\langle 1,V|\tau_{IJ}|1, V'\rangle$ with $I\in\{1,2\}$ and $J\neq 1,2$, which can be evaluated by
\begin{equation}\label{deri}
0\leq|\langle 1,V|\tau_{KL}\tau_{IJ}|1, V'\rangle|\leq 1,\ \ \text{for}\ K,L\in\{1,...,D+1\}, I\in\{1,2\}\ \text{and}\ J\neq 1,2.
\end{equation}
Thus, we can conclude that $\langle 1,V|\tau_{IJ}|1, V'\rangle$ with $I\in\{1,2\}$ and $J\neq 1,2$ as functions of $V$ on $Q_{D-1}$ vanish at $V=V'$ and the growth of their modules are restricted by their derivatives evaluated by Eq.\eqref{deri} as $V$ being transformed by $e^{t\tau_{KL}}\in SO(D+1)$.

Now let us introduce the details of the coherent intertwiner constructed by the Perelomov type coherent state of $SO(D+1)$ at a vertex $v\in\gamma$. Without loss of generality, we re-orient the edges linked to $v$ to be outgoing at $v$ in $\gamma$. With this setting, the gauge fixed coherent intertwiners, as elements of the tensor product space $\mathcal{H}^{\vec{N}_e}_v:=\otimes_{ b(e)=v}\overline{\mathfrak{H}}^{N_{e},D+1}$, are defined as  $\check{\mathcal{I}}_v^{\text{c.}}(\vec{N},\vec{V}):=\otimes_{e: b(e)=v}\langle N_e,V_e|$, where $\bar{\mathfrak{H}}^{N_{e},D+1}$ is the dual space of homogeneous harmonic functions with degree $N_e$ on the D-sphere and $|N_{e},V_{e}\rangle:=u(V_{e})|N_{e},\mathbf{N}_{e}\rangle$ with $u(V_{e})$ being specific $SO(D+1)$ valued function of $V_e$ satisfying $V_e=u(V_{e})\tau_ou(V_{e})^{-1}$. Then, the gauge invariant coherent intertwiners $\mathcal{I}_v^{\text{c.}}$ can be defined as the group averaging of $\check{\mathcal{I}}_v^{\text{c.}}$, which means $\mathcal{I}_v^{\text{c.}}(\vec{N},\vec{V}):=\int_{SO(D+1)}dg\otimes_{e: b(e)=v}\langle N_e,V_e|g$. Specifically, the so-called simple coherent intertwiners $\check{\mathcal{I}}_v^{\text{s.c.}}$ (or $\mathcal{I}_v^{\text{s.c.}}$ in gauge invariant case) is defined by requiring $V_e^{[IJ}V_{e'}^{KL]}=0$ with $b(e)=b(e')=v$ in their definitions. It has been proved that the expectation value of vertex simplicity constraint operator vanishes with respect to the simple coherent intertwiners, hence they weakly solve the vertex simplicity constraint \cite{long2019coherent}. Besides, it has been shown that the gauge invariant simple coherent intertwiners can be regarded as quantum D-polytopes in all dimensional LQG \cite{Long:2020agv},  hence it is reasonable to weakly solve the anomalous vertex simplicity constraint.

\subsection{Generalized twisted geometry coherent states in all dimensional loop quantum gravity}
The generalized twisted geometry coherent states in all dimensional LQG is introduced in our companion paper \cite{Long:2021lmd} firstly, which is given by
\begin{eqnarray}\label{TGCS}
\breve{\Psi}_{\gamma,\vec{\mathbb{H}}^o_e}(\vec{h}_e)&:=&\prod_e\breve{\Psi}_{\mathbb{H}^o_e}(h_e)
\end{eqnarray}
with
\begin{eqnarray}
\breve{\Psi}_{\mathbb{H}^o_e}(h_e)&:=&\sum_{N_e}(\dim(\pi_{N_e}))^{3/2} e^{-tN_e(N_e+D-1)}(e^{(\eta_e -\mathbf{i}\xi^o_e)(N_e+\frac{D-1}{2})}\langle N_e,\mathbf{N}|u_e^{-1}h_e\tilde{u}_e|N_e,\mathbf{N}\rangle\\\nonumber
&&+e^{(-\eta_e +\mathbf{i}\xi^o_e)(N_e+\frac{D-1}{2})}\langle N_e,\bar{\mathbf{N}}|u_e^{-1}h_e\tilde{u}_e|N_e,\bar{\mathbf{N}}\rangle).
\end{eqnarray}
This coherent state associated to edge $e$ can also be reformulated as
\begin{eqnarray}\label{twcs}
\nonumber\breve{\Psi}_{\mathbb{H}^o_e}(h_e)&:=&\sum_{N_e}(\dim(\pi_{N_e}))^{3/2} e^{\frac{(\eta_e)^2+t^2(D-1)^2}{4t}}\left(\exp(-t(\frac{\eta_e}{2t}-d_{N_e})^2)e^{-\mathbf{i}\xi^o_ed_{N_e}}\langle N_e,\mathbf{N}|u_e^{-1}h_e\tilde{u}_e|N_e,\mathbf{N}\rangle\right.\\
&&+\left.\exp(-t(\frac{\eta_e}{2t}+d_{N_e})^2)e^{\mathbf{i}\xi^o_ed_{N_e}}\langle N_e,\bar{\mathbf{N}}|u_e^{-1}h_e\tilde{u}_e|N_e,\bar{\mathbf{N}}\rangle\right)
\end{eqnarray}
 where $d_{N_e}\equiv (N_e+\frac{D-1}{2})$, $\mathbb{H}^o_e:=(V_e,\tilde{V}_e,\xi_e^o, \eta_e)$ are the twisted geometry parameters, $\eta_e$ represents the module of the dimensionless flux $X_e$, and $t\equiv\frac{\kappa\hbar}{a^{D-1}}$.
\subsubsection{Resolution of the identity}
The system of twisted geometry coherent state spans an over-complete basis of the solution space of edge simplicity constraint.
Denoted by $\mathcal{H}^{\text{s.}}_\gamma$ the space composed by the spin-network functions constructed on $\gamma$ and labelled by simple representations on their edges. Then, the system of generalized twisted geometry coherent state provides a resolution of the identity in $\mathcal{H}^{\text{s.}}_\gamma$, which reads
\begin{equation}\label{resoid}
\mathbbm{1}_{\mathcal{H}^{\text{s.}}_\gamma}=
\int_{\times_{e\in\gamma}\check{P}_e}d\vec{\mathbb{H}}_e^o|\breve{\Psi}_{\gamma,\vec{\mathbb{H}}^o_e}\rangle\langle \breve{\Psi}_{\gamma,\vec{\mathbb{H}}^o_e}|, 
\end{equation}
wherein $\check{P}_e:=(\mathbb{R}_+\times S^1\times Q_{D-1}\times Q_{D-1})_e$ is space of the twisted geometry parameters $\mathbb{H}^o_e$, and the measure $d\vec{\mathbb{H}}^o$ is defined by
\begin{equation}
d\vec{\mathbb{H}}_e^o:=\prod_{e\in\gamma}\frac{d\eta_e}{\sqrt{2\pi t}}e^{-\frac{\eta_e^2+t^2(D-1)^2}{2t}} \prod_{e\in\gamma}du_ed\tilde{u}_e\prod_{e\in\gamma}\frac{d\xi^o_{e}}{2\pi},
\end{equation}
where $d\eta_e$ is the Lebesgue measure on $\mathbb{R}$, $d\xi^o_{e}$ is the measure on $S^1$, and $du_e$ or $d\tilde{u}_e$ is the measure on $Q_{D-1}$.

  We are also interested in the Hilbert space $\mathcal{H}^{\text{s.}}_e$ spanned by the spin-network functions  constructed on a single edge $e$ and labelled by simple representations. The twisted geometry coherent state associated to edge $e$ also provides an over-complete basis of $\mathcal{H}^{\text{s.}}_e$, which reads
  \begin{equation}\label{resoide}
\mathbbm{1}_{\mathcal{H}^{\text{s.}}_e}=
\int_{\check{P}_e}d{\mathbb{H}}^o_e|\breve{\Psi}_{{\mathbb{H}}^o_e}\rangle\langle \breve{\Psi}_{{\mathbb{H}}^o_e}|, 
\end{equation}
wherein the measure $d{\mathbb{H}}^o_e$ is defined by
\begin{equation}
d{\mathbb{H}}^o_e:=\frac{d\eta_e}{\sqrt{2\pi t}}e^{-\frac{\eta_e^2+t^2(D-1)^2}{2t}} du_ed\tilde{u}_e\frac{d\xi^o_{e}}{2\pi}.
\end{equation}

Though the terms corresponding to lowest weights in the twisted geometry coherent states \eqref{TGCS} are exponentially suppressed, they still play a key role in the resolution of the identity in $\mathcal{H}^{\text{s.}}_\gamma$. Nevertheless, the terms corresponding to lowest weights in the twisted geometry coherent states \eqref{TGCS} will be neglected in the following analysis of this paper, since they always contribute exponentially suppressed small terms to the results in our discussion.
\subsubsection{The overlap function of the coherent states}
 Notice that the twisted geometry coherent state $\breve{\Psi}_{\gamma,\vec{\mathbb{H}}^o_e}$ on $\gamma$ is the product of the twisted geometry coherent state $\breve{\Psi}_{\mathbb{H}^o_e}(h_e)$
 on each edge $e\in\gamma$. Thus the the overlap function for $\breve{\Psi}_{\gamma,\vec{\mathbb{H}}^o_e}$ can be given by
\begin{equation}
i^t\left((\gamma,\vec{\mathbb{H}}^o_e),(\gamma, \vec{\mathbb{H}}'^o_e)\right):=\frac{|\langle \breve{\Psi}_{\gamma,\vec{\mathbb{H}}^o_e}|\breve{\Psi}_{\gamma,\vec{\mathbb{H}}'^o_e}\rangle|^2} {||\breve{\Psi}_{\gamma,\vec{\mathbb{H}}^o_e}||^2\cdot ||\breve{\Psi}_{\gamma,\vec{\mathbb{H}}'^o_e}||^2}=\prod_{e\in\gamma}i^t(\mathbb{H}^o_e, \mathbb{H}'^o_e)
\end{equation}
 with
 \begin{equation}
i^t(\mathbb{H}^o_e, \mathbb{H}'^o_e):=\frac{|\langle \breve{\Psi}_{\mathbb{H}^o_e}|\breve{\Psi}_{\mathbb{H}'^o_e}\rangle|^2}{||\breve{\Psi}_{\mathbb{H}^o_e}||^2 ||\breve{\Psi}_{\mathbb{H}'^o_e}||^2}
\end{equation}
being the overlap function for the coherent state $\breve{\Psi}_{\mathbb{H}^o_e}(h_e)$ on an edge $e$, where
\begin{equation}
||\breve{\Psi}_{\gamma,\vec{\mathbb{H}}^o_e}||^2
:=|\langle \breve{\Psi}_{\gamma,\vec{\mathbb{H}}^o_e}|\breve{\Psi}_{\gamma,\vec{\mathbb{H}}^o_e}\rangle|^2
\end{equation}
and
\begin{equation}
||\breve{\Psi}_{\mathbb{H}^o_e}||^2 :=|\langle \breve{\Psi}_{\mathbb{H}^o_e}|\breve{\Psi}_{\mathbb{H}'^o_e}\rangle|^2
\end{equation}
are the module squares of $\breve{\Psi}_{\gamma,\vec{\mathbb{H}}^o_e}$ and  $\breve{\Psi}_{\mathbb{H}^o_e}(h_e)$ respectively.
 In the following calculations and analysis, we will only consider $i^t(\mathbb{H}^o_e, \mathbb{H}'^o_e)$ without loss of generality to simplify our expressions.

We first find that
\begin{eqnarray}
&&||\breve{\Psi}_{\mathbb{H}^o_e}||^2 \stackrel{\text{large }\! \eta_e}{=}\sqrt{\frac{\pi}{2t}} e^{\frac{(\eta_e)^2+t^2(D-1)^2}{2t}}(\breve{\text{P}}(\frac{\eta_e}{2t}))^2\left(1+\mathcal{O}(e^{-\frac{1}{t}}) +\mathcal{O}(\frac{t}{\eta_e})\right)
\end{eqnarray}
with $\breve{\text{P}}(N)=\dim(\pi_N)$ is a polynomial of $N$. Notice that $\langle N_e, V'_e|N_e, V_e\rangle=0$ or $\langle N_e, -\tilde{V}_e|N_e,-\tilde{ V}'_e\rangle=0$ leads to $\langle \breve{\Psi}_{\mathbb{H}^o_e}|\breve{\Psi}_{\mathbb{H}'^o_e}\rangle =0$. Hence, we only consider the case of $\langle N_e, V'_e|N_e, V_e\rangle\neq0$ and $\langle N_e, -\tilde{V}_e|N_e,-\tilde{ V}'_e\rangle\neq0$ in the following part of this paper.
Then, we have
\begin{eqnarray}\label{overlapmmm}
&&\langle \breve{\Psi}_{\mathbb{H}^o_e}|\breve{\Psi}_{\mathbb{H}'^o_e}\rangle e^{-\frac{(\eta_e)^2+(\eta'_e)^2+2t^2(D-1)^2}{4t}}
\\\nonumber &{=}&e^{\mathbf{i}\frac{D-1}{2}(\xi^o_e-\xi'^o_e)}\sum_{N_e}(\dim(\pi_{N_e}))^2 \exp(-t(\frac{\eta_e}{2t}-d_{N_e})^2 -t(\frac{\eta'_e}{2t}-d_{N_e})^2)\\\nonumber
&&\cdot e^{\mathbf{i}N_e(\xi^o_e-\xi'^o_e+\varphi(u_e,u'_e) +\varphi(\tilde{u}_e,\tilde{u}'_e))}\exp(- N_e\widetilde{\Theta}_e)+\frac{1}{\sqrt{t}}\mathcal{O}(e^{-\frac{\eta'^2_e}{8t}})
\end{eqnarray}
for large $\eta'_e$, where $\widetilde{\Theta}_e:=\Theta(u_e,u'_e)+\Theta(\tilde{u}_e,\tilde{u}'_e)$ and we used the invention
\begin{equation}
\langle N_e, V'_e|N_e, V_e\rangle=\exp{(-N_e\Theta(u_e,u'_e))}e^{\mathbf{i}N_e\varphi(u_e,u'_e)},
\end{equation}
\begin{equation}
\langle N_e, -\tilde{V}_e|N_e,-\tilde{ V}'_e\rangle=\exp{(-N_e\Theta(\tilde{u}_e,\tilde{u}'_e))}e^{\mathbf{i}N_e\varphi(\tilde{u}_e,\tilde{u}'_e)},
\end{equation}
 where $\Theta(u_e,{u'}_e):=-\frac{\ln|\langle N_e, V'_e|N_e, V_e\rangle|}{N_e}\geq0$, $\Theta(\tilde{u}_e,\tilde{u}'_e):=-\frac{\ln|\langle N_e, -\tilde{V}_e|N_e,-\tilde{ V}'_e\rangle|}{N_e}\geq0$ with $\Theta(u_e,u'_e)=0, \Theta(\tilde{u}_e,\tilde{u}'_e)=0$ for $V_e=V'_e$, $\tilde{V}_e=\tilde{ V}'_e$ respectively, and $e^{\mathbf{i}N_e\varphi(u_e,u'_e)}:=\frac{\langle N_e, V'_e|N_e, V_e\rangle}{|\langle N_e, V'_e|N_e, V_e\rangle|}$, $e^{\mathbf{i}N_e\varphi(\tilde{u}_e,\tilde{u}'_e)}:=\frac{\langle N_e, -\tilde{V}_e|N_e, -\tilde{V}'_e\rangle}{|\langle N_e, -\tilde{V}_e|N_e, -\tilde{V}'_e\rangle|}$ with $\varphi(u_e,u'_e)=0, \varphi(\tilde{u}_e,\tilde{u}'_e)=0$ for $V_e=V'_e$, $\tilde{V}_e=\tilde{ V}'_e$ respectively. In order to study Eq. \eqref{overlapmmm}, the cases of $\widetilde{\Theta}_e\ll \eta_e+\eta'_e$ and $\widetilde{\Theta}_e\simeq\eta_e+\eta'_e$ or $\widetilde{\Theta}_e\gg \eta_e+\eta'_e$ are considered respectively.
\\ \\
\textbf{(i).} For the case of $\widetilde{\Theta}_e\ll \eta_e+\eta'_e$, the overlap function is expressed as
\begin{eqnarray}\label{overlap2}
&&i^t(\mathbb{H}^o_e, \mathbb{H}'^o_e):=\frac{|\langle \breve{\Psi}_{\mathbb{H}^o_e}|\breve{\Psi}_{\mathbb{H}'^o_e}\rangle|^2}{||\breve{\Psi}_{\mathbb{H}^o_e}||^2 ||\breve{\Psi}_{\mathbb{H}'^o_e}||^2}\\\nonumber
&{=}&\frac{ (f_{\text{Poly}}(\frac{\eta'_e}{t}, \frac{\eta_e}{t},\frac{\widetilde{\Theta}_e}{t}))^2}{(\breve{\text{P}}(\frac{\eta_e}{2t}))^2 (\breve{\text{P}}(\frac{\eta'_{e}}{2t}))^2} e^{-2t(\frac{\eta'_e}{2t}-\frac{\eta_e}{2t})^2 +4t(\frac{\eta'_e}{4t}-\frac{\eta_e}{4t}-\frac{\widetilde{\Theta}_e}{4t})^2} e^{-2(\frac{\eta_e}{2t}-\frac{D-1}{2})\widetilde{\Theta}_e}\exp(-\frac{(\xi^o_e-\xi'^o_e+\tilde{\varphi}_e)^2}{4t})\\\nonumber
&&\cdot\left(1+\mathcal{O}(\frac{t}{\eta'_e}) +\mathcal{O}(e^{-\frac{1}{t}})\right)
\end{eqnarray}
for large $\eta'_e$ and $\widetilde{\Theta}_e\ll \eta_e+\eta'_e$, where $f_{\text{Poly}}(\frac{\eta'_e}{t}, \frac{\eta_e}{t},\frac{\widetilde{\Theta}_e}{t})$ is a polynomial of the three variables $\frac{\eta'_e}{t}, \frac{\eta_e}{t},\frac{\widetilde{\Theta}_e}{t}$ which satisfies
\begin{equation}
f_{\text{Poly}}(\frac{\eta'_e}{t}, \frac{\eta_e}{t},\frac{\widetilde{\Theta}_e}{t}){=}(\breve{\text{P}}(\frac{\eta'_e}{4t}+\frac{\eta_e}{4t} -\frac{\widetilde{\Theta}_e}{4t}))^2(1+\mathcal{O}(\frac{t}{\eta'_e}))
\end{equation}
for large $\eta'_e$ and $\widetilde{\Theta}_e\ll \eta_e+\eta'_e$.
Then one can conclude the peakedness property of the overlap function in the case of $\widetilde{\Theta}_e\ll \eta_e+\eta'_e$. For the overlap function $i^t(\mathbb{H}^o_e, \mathbb{H}'^o_e)$ given by Eq.\eqref{overlap2}, one first find that it is sharply peaked at $\widetilde{\Theta}_e=0$ by the factor $e^{-2(\frac{\eta_e}{2t}-\frac{D-1}{2})\widetilde{\Theta}_e}$. Notice that $\tilde{\varphi}_e=0$ if $\widetilde{\Theta}_e=0$ by their definition, one can further conclude that the overlap function $i^t(\mathbb{H}^o_e, \mathbb{H}'^o_e)$ is sharply peaked at $\xi^o_e=\xi'^o_e$ and $\eta_e=\eta'_e$ by the factors $\exp(-\frac{(\xi^o_e-\xi'^o_e+\tilde{\varphi}_e)^2}{4t})$ and $e^{-2t(\frac{\eta'_e}{2t}-\frac{\eta_e}{2t})^2 +4t(\frac{\eta'_e}{4t}-\frac{\eta_e}{4t}-\frac{\widetilde{\Theta}_e}{4t})^2}$ respectively.
\\ \\
\textbf{(ii).} For the case of $\widetilde{\Theta}_e\simeq\eta_e+\eta'_e$ or $\widetilde{\Theta}_e\gg \eta_e+\eta'_e$,
 the overlap function is expressed as
\begin{eqnarray}\label{overlap1}
&&i^t(\mathbb{H}^o_e, \mathbb{H}'^o_e):=\frac{|\langle \breve{\Psi}_{\mathbb{H}^o_e}|\breve{\Psi}_{\mathbb{H}'^o_e}\rangle|^2}{||\breve{\Psi}_{\mathbb{H}^o_e}||^2 ||\breve{\Psi}_{\mathbb{H}'^o_e}||^2}\\\nonumber
&{\lesssim}&\frac{\left(\sqrt{\frac{2t}{\pi}}\left(e^{-t((\frac{\eta_e}{2t})^2+ (\frac{\eta'_e}{2t})^2)}+f(\eta_e,\eta'_e)e^{-t(\frac{\eta_e}{4t})^2}e^{- \widetilde{\Theta}_e}\right) + (\breve{\text{P}}(\frac{\eta_e}{4t}+\frac{\eta'_e}{4t}))^2e^{-\frac{t}{2}(\frac{\eta'_e}{2t}-\frac{\eta_e}{2t})^2} e^{- [\frac{\eta_e}{4t}]\widetilde{\Theta}_e}\right)^2}{ (\breve{\text{P}}(\frac{\eta_e}{2t}))^2 (\breve{\text{P}}(\frac{\eta'_e}{2t}))^2}
\end{eqnarray}
for large $\eta'_e$, where $f(\eta_e,\eta'_e)=[\eta_e/4t] \exp( -t(\frac{\eta'_e}{2t}-\frac{\eta_e}{4t}-\frac{D+1}{2})^2)(\breve{\text{P}}(\frac{\eta_e}{4t}))^2$.
Note that we considered $\widetilde{\Theta}_e\simeq\eta_e+\eta'_e$ or $\widetilde{\Theta}_e\gg \eta_e+\eta'_e$ here, it is obviously that the overlap function $i^t(\mathbb{H}^o_e, \mathbb{H}'^o_e)$ is suppressed exponentially by the factors $e^{-t((\frac{\eta_e}{2t})^2+ (\frac{\eta'_e}{2t})^2)}$, $e^{-t(\frac{\eta_e}{4t})^2}$ and $e^{- [\frac{\eta_e}{4t}]\widetilde{\Theta}_e}$  in Eq.\eqref{overlap1}.

Finally, let us  combine the analysis of the overlap function $i^t(\mathbb{H}^o_e, \mathbb{H}'^o_e)$ given by Eqs.\eqref{overlap2} and \eqref{overlap1}, one can concludes the peakedness property that the overlap function $i^t(\mathbb{H}^o_e, \mathbb{H}'^o_e)$ is sharply peaked at $\xi^o_e=\xi'^o_e$, $\eta_e=\eta'_e$ and $V_e=V'_e$, $\tilde{V}_e=\tilde{V}'_e$ for large $\eta'_e$.
\section{Ehrenfest property of twisted geometry coherent state}\label{sec3}
To establish the ``Ehrenfest Property'' of the twisted geometry coherent states, one needs to consider the expectation value of all elementary quantum operators in all dimensional LQG. In fact, for a given graph $\gamma$ and corresponding Hilbert space $\mathcal{H}_\gamma$, every polynomial of the elementary operators $\{\hat{h}_e,\hat{X}^{IJ}_e\}_{e\in\gamma}$ can be reduced to sums of monomials of the form
\begin{equation}
\hat{O}_\gamma=\prod_{e\in\gamma}\hat{O}_e,
\end{equation}
where the operator $\hat{O}_e$ is a certain polynomial of the elementary operators  $\hat{h}_e,\hat{X}^{IJ}_e$ on the edge $e$. The expectation value of $\hat{O}_\gamma$ with respect to the twisted geometry coherent states \eqref{TGCS} is given by
\begin{equation}
 \frac{\langle \breve{\Psi}_{\gamma,\vec{\mathbb{H}}^o_e}|\hat{O}_\gamma|\breve{\Psi}_{\gamma,\vec{\mathbb{H}}^o_e}\rangle} {||\breve{\Psi}_{\gamma,\vec{\mathbb{H}}^o_e}||^2}=\prod_{e\in\gamma} \frac{\langle \breve{\Psi}_{{\mathbb{H}}^o_e}|\hat{O}_e|\breve{\Psi}_{{\mathbb{H}}^o_e}\rangle} {||\breve{\Psi}_{{\mathbb{H}}^o_e}||^2}.
\end{equation}
As discussed in \cite{2000Gauge}, it is shown that in order to establish the ``Ehrenfest property'' it will be completely
sufficient to consider this problem for one copy of the edge only. In the following part of this paper, we will concentrate on the issues on a single edge $e$.
\subsection{ Expectation values of operator monomials}
In this section we will reduce the computation of expectation values of operator monomials to the computation of matrix elements of elementary operators between the twisted geometry coherent states. Recall the completeness relation \eqref{resoide}, which reads
 \begin{equation}
\mathbbm{1}_{\mathcal{H}^{\text{s.}}_e}=
\int_{\check{P}_e}d{\mathbb{H}}^o_e|\breve{\Psi}_{{\mathbb{H}}^o_e}\rangle\langle \breve{\Psi}_{{\mathbb{H}}^o_e}|. 
\end{equation}
Let us consider an operator monomial $\hat{O}_e=\hat{O}_{e,1}...\hat{O}_{e,n}$ where each of the $\hat{O}_{e,k}, k=1,...,n<\infty$ represents one of the elementary operators $\hat{h}_e, \hat{X}_e^{IJ}$. Then, by using \eqref{resoide}, we can write the expectation value of $\hat{O}_e$ as
\begin{eqnarray}\label{expectmono}
&& \frac{\langle \breve{\Psi}_{{\mathbb{H}}^o_e}|\hat{O}_e|\breve{\Psi}_{{\mathbb{H}}^o_e}\rangle} {||\breve{\Psi}_{{\mathbb{H}}^o_e}||^2}\\\nonumber
&=&\frac{1} {||\breve{\Psi}_{{\mathbb{H}}^o_e}||^2}\int_{\check{P}_e}d{\mathbb{H}}^o_{e,1}... \int_{\check{P}_e}d{\mathbb{H}}^o_{e,n-1}\prod_{k=1}^n \langle \breve{\Psi}_{{\mathbb{H}}^o_{e,k-1}}|\hat{O}_{e,k}|\breve{\Psi}_{{\mathbb{H}}^o_{e,k}}\rangle\\\nonumber
&=&\int_{\check{P}_e}d{\mathbb{H}}^o_{e,1}... \int_{\check{P}_e}d{\mathbb{H}}^o_{e,n-1}\left(\prod_{k=1}^{n-1}||\breve{\Psi}_{{\mathbb{H}}^o_{e,k}}||^2\right) \left(\prod_{k=1}^n \frac{\langle \breve{\Psi}_{{\mathbb{H}}^o_{e,k-1}}|\hat{O}_{e,k}|\breve{\Psi}_{{\mathbb{H}}^o_{e,k}}\rangle} {||\breve{\Psi}_{{\mathbb{H}}^o_{e,k-1}}||||\breve{\Psi}_{{\mathbb{H}}^o_{e,k}}||}\right),
\end{eqnarray}
where we have set ${\mathbb{H}}^o_{e,0}={\mathbb{H}}^o_{e,n}={\mathbb{H}}^o_{e}$. Notice that the quantity
\begin{equation}\
j^t(\mathbb{H}^o_{e},\mathbb{H}'^o_{e}):= \frac{\langle \breve{\Psi}_{{\mathbb{H}}^o_{e}}|\breve{\Psi}_{{\mathbb{H}}'^o_{e}}\rangle} {||\breve{\Psi}_{{\mathbb{H}}^o_{e}}||||\breve{\Psi}_{{\mathbb{H}}'^o_{e}}||}
\end{equation}
is exponentially small in the sense of a Gaussian needle of width $\sqrt{t}$ unless $\mathbb{H}^o_{e}=\mathbb{H}'^o_{e}$ (where it equals unity).  Thus, it is conceivable that
\begin{equation}\label{approx}
  \frac{\langle \breve{\Psi}_{{\mathbb{H}}^o_{e,k-1}}|\hat{O}_{e,k}|\breve{\Psi}_{{\mathbb{H}}^o_{e,k}}\rangle} {||\breve{\Psi}_{{\mathbb{H}}^o_{e,k-1}}||||\breve{\Psi}_{{\mathbb{H}}^o_{e,k}}||}\approx \frac{\langle \breve{\Psi}_{{\mathbb{H}}^o_{e,k}}|\hat{O}_{e,k}|\breve{\Psi}_{{\mathbb{H}}^o_{e,k}}\rangle} {||\breve{\Psi}_{{\mathbb{H}}^o_{e,k}}||||\breve{\Psi}_{{\mathbb{H}}^o_{e,k}}||} j^t(\mathbb{H}^o_{e,k-1},\mathbb{H}^o_{e,k}).
\end{equation}
By substituting Eq.\eqref{approx} into Eq.\eqref{expectmono}, we would have indeed shown that
 \begin{equation}\label{expectmono2}
 \frac{\langle \breve{\Psi}_{{\mathbb{H}}^o_e}|\hat{O}_e|\breve{\Psi}_{{\mathbb{H}}^o_e}\rangle} {||\breve{\Psi}_{{\mathbb{H}}^o_e}||^2}\approx\prod_{k=1}^n \frac{\langle \breve{\Psi}_{{\mathbb{H}}^o_e}|\hat{O}_{e,k}|\breve{\Psi}_{{\mathbb{H}}^o_e}\rangle} {||\breve{\Psi}_{{\mathbb{H}}^o_e}||^2}.
 \end{equation}
 Thus, in order to prove the desired result \eqref{expectmono2} it is sufficient to prove \eqref{approx} together with the precise meaning of $``\approx''$. In the following parts of this section, we will calculate and discuss the matrix elements of the elementary holonomy and flux operators in the twisted geometry coherent state basis, to gives a reliable proof of  \eqref{approx}.
 \subsection{Matrix elements of the elementary operators}
 Since the expectation values of holonomy and flux operators with respect to twisted geometry coherent state are well evaluated by their corresponding classical values up to $\mathcal{O}(t)$ \cite{Long:2021lmd}, we can prove \eqref{approx} by shown that
    \begin{eqnarray}\label{otf}
\left|\frac{\langle\breve{\Psi}_{\mathbb{H}^o_e}|\hat{O}_e |\breve{\Psi}_{\mathbb{H}'^o_e}\rangle}{||\breve{\Psi}_{\mathbb{H}'^o_e}||||\breve{\Psi}_{ \mathbb{H}^o_e}||} -O_e(\mathbb{H}'^o_e)\frac{\langle\breve{\Psi}_{\mathbb{H}^o_e} |\breve{\Psi}_{\mathbb{H}'^o_e}\rangle}{||\breve{\Psi}_{\mathbb{H}'^o_e}||||\breve{\Psi}_{ \mathbb{H}^o_e}||}\right|
&{\lesssim}&t\left|f_{O_e}(\mathbb{H}^o_e,\mathbb{H}'^o_e)\right|\cdot\left|\frac{\langle\breve{\Psi}_{\mathbb{H}^o_e} |\breve{\Psi}_{\mathbb{H}'^o_e}\rangle}{||\breve{\Psi}_{\mathbb{H}'^o_e}||||\breve{\Psi}_{ \mathbb{H}^o_e}||}\right|,
\end{eqnarray}
with $\hat{O}_e$ representing holonomy operator or flux operator here, $O_e(\mathbb{H}'^o_e)$ being the corresponding classical values of $\hat{O}_e$ given by $\mathbb{H}'^o_e$, and $f_{O_e}(\mathbb{H}^o_e,\mathbb{H}'^o_e)$ being a function whose growth is always suppressed by $\left|\frac{\langle\breve{\Psi}_{\mathbb{H}^o_e} |\breve{\Psi}_{\mathbb{H}'^o_e}\rangle}{||\breve{\Psi}_{\mathbb{H}'^o_e}||||\breve{\Psi}_{ \mathbb{H}^o_e}||}\right|$ exponentially as $|\eta_e-\eta'_e|$ ,$\widetilde{\Theta}_e$ and $|\xi^o_e-\xi'^o_e|$ going large. In the following part of this subsection, we will prove Eq.\eqref{otf} for holonomy and flux operators respectively.
\subsubsection{Matrix elements of the flux operator}

We consider the matrix elements of the flux operator in the twisted geometry coherent state basis, which are denoted by $\frac{\langle\breve{\Psi}_{\mathbb{H}^o_e}|\hat{X}_e^{IJ} |\breve{\Psi}_{\mathbb{H}'^o_e}\rangle}{||\breve{\Psi}_{\mathbb{H}'^o_e}||||\breve{\Psi}_{ \mathbb{H}^o_e}||}$. The numerator can be calculated as follows.
\begin{eqnarray}
&& \langle\breve{\Psi}_{\mathbb{H}^o_e}|\hat{X}_e^{IJ} |\breve{\Psi}_{\mathbb{H}'^o_e}\rangle e^{-\frac{(\eta_e)^2+(\eta'_e)^2+2t^2(D-1)^2}{4t}}\\\nonumber
&{=}&-\mathbf{i}\beta t\sum_{N_e}(\dim(\pi_{N_e}))^2 \exp(-t(\frac{\eta_e}{2t}-d_{N_e})^2 -t(\frac{\eta'_e}{2t}-d_{N_e})^2)\\\nonumber
&&\cdot e^{\mathbf{i}d_{N_e}(\xi^o_e-\xi'^o_e)}\langle N_e, V'_e|\tau^{IJ}|N_e, V_e\rangle\langle N_e, -\tilde{V}_e|N_e,-\tilde{ V}'_e\rangle+\beta \sqrt{t}\cdot\mathcal{O}(e^{-\frac{\eta'^2_e}{8t}}).
\end{eqnarray}
Without loss of generality, we set $V'_e=2\delta_{1}^{[I}\delta_2^{J]}$ to simplify the expressions. Then, the components of $\hat{X}_e^{IJ}$ can be decomposed as three sets and we can discuss them separately.\\
\\  \textbf{Set I: } \textit{Matrix elements of $\hat{X}_e^{12}$.}\\
 Similar to the calculation of the overlap function of twisted geometry coherent state, we defined $\widetilde{\Theta}_e:=\Theta(u_e,u'_e)+\Theta(\tilde{u}_e,\tilde{u}'_e)$ and used the invention
\begin{equation}
\langle N_e, V'_e|N_e, V_e\rangle=\exp{(-N_e\Theta(u_e,u'_e))}e^{\mathbf{i}N_e\varphi(u_e,u'_e)},
\end{equation}
\begin{equation}
\langle N_e, -\tilde{V}_e|N_e,-\tilde{ V}'_e\rangle=\exp{(-N_e\Theta(\tilde{u}_e,\tilde{u}'_e))}e^{\mathbf{i}N_e\varphi(\tilde{u}_e,\tilde{u}'_e)},
\end{equation}
 with $\Theta(u_e,{u'}_e):=-\frac{\ln|\langle N_e, V'_e|N_e, V_e\rangle|}{N_e}\geq0$, $\Theta(\tilde{u}_e,\tilde{u}'_e):=-\frac{\ln|\langle N_e, -\tilde{V}_e|N_e,-\tilde{ V}'_e\rangle|}{N_e}\geq0$, $e^{\mathbf{i}N_e\varphi(u_e,u'_e)}:=\frac{\langle N_e, V'_e|N_e, V_e\rangle}{|\langle N_e, V'_e|N_e, V_e\rangle|}$ and $e^{\mathbf{i}N_e\varphi(\tilde{u}_e,\tilde{u}'_e)}:=\frac{\langle N_e, -\tilde{V}_e|N_e, -\tilde{V}'_e\rangle}{|\langle N_e, -\tilde{V}_e|N_e, -\tilde{V}'_e\rangle|}$ are variables independent with $N_e$. Then, we have
\begin{eqnarray}\label{F12}
&& \langle\breve{\Psi}_{\mathbb{H}^o_e}|\hat{X}_e^{12} |\breve{\Psi}_{\mathbb{H}'^o_e}\rangle e^{-\frac{(\eta_e)^2+(\eta'_e)^2+2t^2(D-1)^2}{4t}}\\\nonumber
&{=}& \beta t \sum_{N_e}N_e(\dim(\pi_{N_e}))^2 \exp(-t(\frac{\eta_e}{2t}-d_{N_e})^2 -t(\frac{\eta'_e}{2t}-d_{N_e})^2)\\\nonumber
&&\cdot e^{\mathbf{i}d_{N_e}(\xi^o_e-\xi'^o_e)}\langle N_e, V'_e|N_e, V_e\rangle\langle N_e, -\tilde{V}_e|N_e,-\tilde{ V}'_e\rangle+\beta \sqrt{t}\cdot\mathcal{O}(e^{-\frac{\eta'^2_e}{8t}})\\\nonumber
&{=}&\beta t e^{\mathbf{i}\frac{D-1}{2}(\xi^o_e-\xi'^o_e)}\sum_{N_e}N_e(\dim(\pi_{N_e}))^2\exp(-t(\frac{\eta_e}{2t}-d_{N_e})^2 -t(\frac{\eta'_e}{2t}-d_{N_e})^2)\\\nonumber
&&\cdot e^{\mathbf{i}N_e(\xi^o_e-\xi'^o_e+\varphi(u_e,u'_e) +\varphi(\tilde{u}_e,\tilde{u}'_e))}\exp(- N_e\widetilde{\Theta}_e)+\beta \sqrt{t}\cdot\mathcal{O}(e^{-\frac{\eta'^2_e}{8t}})\\\nonumber
\end{eqnarray}
for large $\eta'_e$, where $\tilde{\varphi}_e:=\varphi(u_e,u'_e) +\varphi(\tilde{u}_e,\tilde{u}'_e)$ and $\widetilde{\Theta}_e:=\Theta(u_e,u'_e)+\Theta(\tilde{u}_e,\tilde{u}'_e)$. The calculation of Eq.\eqref{F12} follows the similar procedures of the calculation of $\langle\breve{\Psi}_{\mathbb{H}^o_e} |\breve{\Psi}_{\mathbb{H}'^o_e}\rangle$ in \cite{Long:2021lmd}.  Let us consider the cases of $\widetilde{\Theta}_e\ll \eta_e+\eta'_e$ and $\widetilde{\Theta}_e\simeq\eta_e+\eta'_e$ or $\widetilde{\Theta}_e\gg \eta_e+\eta'_e$ separately.
\\ \\
\textbf{(i).} For the case $\widetilde{\Theta}_e\ll \eta_e+\eta'_e$, Eq.\eqref{F12} reads
\begin{eqnarray}\label{F122}
&& \langle\breve{\Psi}_{\mathbb{H}^o_e}|\hat{X}_e^{12} |\breve{\Psi}_{\mathbb{H}'^o_e}\rangle e^{-\frac{(\eta_e)^2+(\eta'_e)^2+2t^2(D-1)^2}{4t}}\\\nonumber
&{=}&\beta te^{\mathbf{i}\frac{D-1}{2}(\xi^o_e-\xi'^o_e)}\sum_{N_e}N_e(\dim(\pi_{N_e}))^2\exp(-t(\frac{\eta_e}{2t}-d_{N_e})^2 -t(\frac{\eta'_e}{2t}-d_{N_e})^2)\\\nonumber
&&\cdot e^{\mathbf{i}N_e(\xi^o_e-\xi'^o_e+\varphi(u_e,u'_e) +\varphi(\tilde{u}_e,\tilde{u}'_e))}\exp(- N_e\widetilde{\Theta}_e)+\beta \sqrt{t}\cdot\mathcal{O}(e^{-\frac{\eta'^2_e}{8t}})\\\nonumber
&=&\beta te^{\mathbf{i}\frac{D-1}{2}(\xi^o_e-\xi'^o_e)}e^{-t(\frac{\eta'_e}{2t}-\frac{\eta_e}{2t})^2 +2t(\frac{\eta_e}{4t}-\frac{\eta'_e}{4t}-\frac{\widetilde{\Theta}_e}{4t})^2} e^{\mathbf{i}(\frac{\eta_e}{4t}-\frac{D-1}{2}+\frac{\eta'_e}{4t}-\frac{\widetilde{\Theta}_e}{4t})(\xi^o_e- \xi'^o_e+\tilde{\varphi}_e)} \\\nonumber
&&\cdot \exp(-(\frac{\eta'_e}{2t}-\frac{D-1}{2})\widetilde{\Theta}_e) \sum_{[\tilde{k}_e]}\widetilde{\text{P}}(\tilde{k}_e)\left(\exp(-2t\tilde{k}_e^2)e^{\mathbf{i}\tilde{k}_e(\xi^o_e-\xi'^o_e +\tilde{\varphi}_e)}\right)+\beta \sqrt{t}\cdot\mathcal{O}(e^{-\frac{\eta'^2_e}{8t}})\\\nonumber
&{=}&\beta t\frac{\sqrt{\pi}}{\sqrt{2 t}}e^{\mathbf{i}\frac{D-1}{2}(\xi^o_e-\xi'^o_e)}e^{-t(\frac{\eta'_e}{2t}-\frac{\eta_e}{2t})^2 +2t(\frac{\eta_e}{4t}-\frac{\eta'_e}{4t}-\frac{\widetilde{\Theta}_e}{4t})^2} e^{\mathbf{i}(\frac{\eta_e}{4t}-\frac{D-1}{2}+\frac{\eta'_e}{4t}-\frac{\widetilde{\Theta}_e}{4t}) (\xi^o_e-\xi'^o_e+\tilde{\varphi}_e)}e^{-(\frac{\eta'_e}{2t}-\frac{D-1}{2})\widetilde{\Theta}_e} \\\nonumber
&&\cdot\sum_{n=-\infty}^{\infty} \check{\text{P}}(2\pi n-(\xi^o_e-\xi'^o_e+\tilde{\varphi}_e))\exp(-\frac{(2\pi n-(\xi^o_e-\xi'^o_e+\tilde{\varphi}_e))^2}{8t})e^{\mathbf{i}2\pi n\text{mod}(\tilde{k}_e,1)}(1+\mathcal{O}(\frac{t}{\eta'_e}))\\\nonumber
&&+\beta \sqrt{t}\cdot\mathcal{O}(e^{-\frac{\eta'^2_e}{8t}})
\end{eqnarray}
for large $\eta'_e$, 
where $\tilde{k}_e:=d_{N_e}- \frac{\eta'_e}{4t}-\frac{\eta_e}{4t}+\frac{\widetilde{\Theta}_e}{4t}=[\tilde{k}_e]+\text{mod}(\tilde{k}_e,1)$ with $[\tilde{k}_e]$ being the maximum integer less than or equal to $\tilde{k}_e$ and $\text{mod}(\tilde{k}_e,1)$ being the corresponding remainder, and $\widetilde{\text{P}}(\tilde{k}_e)$ is a polynomial of $\tilde{k}_e$ defined by $\widetilde{\text{P}}(\tilde{k}_e)=N_e(\dim(\pi_{N_e}))^2$. Besides, $ \check{\text{P}}(x)$ is also a polynomial which is given by
 \begin{equation}
\check{\text{P}}(x)=\left((\mathbf{i})^n a_n\frac{d^n}{dx^n} \exp(-\frac{x^2}{8t})+(\mathbf{i})^{n-1} a_{n-1}\frac{d^{n-1}}{dx^{n-1}} \exp(-\frac{x^2}{8t})+...+a_0 \exp(-\frac{x^2}{8t})\right) \exp(\frac{x^2}{8t})
\end{equation}
with $a_n,a_{n-1},...,a_0$ is given by the expanding  $\widetilde{\text{P}}(\tilde{k}_e)=a_n\tilde{k}_e^n+a_{n-1}\tilde{k}_e^{n-1}+...+a_0$, wherein $a_0=(\frac{\eta'_e}{4t}+\frac{\eta_e}{4t}-\frac{\widetilde{\Theta}_e}{4t})(\breve{\text{P}}(\frac{\eta'_e}{4t} +\frac{\eta_e}{4t}-\frac{\widetilde{\Theta}_e}{4t}))^2$,  $\frac{a_0}{a_{m}}\simeq(\frac{\eta_e}{4t})^m$, $0\leq m\leq n$ for  large $\eta'_e$ by the definition of  $\widetilde{\text{P}}(\tilde{k}_e)$.
 In addition, we used the Poisson summation formula in third step of Eq.\eqref{F122}, which  converts a slowly converging series into a rapidly converging series which in our case almost only the term with $n=0$ need to be relevant. Following this analysis, we can further give
 \begin{eqnarray}\label{F122}
&& \langle\breve{\Psi}_{\mathbb{H}^o_e}|\hat{X}_e^{12} |\breve{\Psi}_{\mathbb{H}'^o_e}\rangle\\\nonumber
&{=}&\beta t\frac{\sqrt{\pi}}{\sqrt{2 t}}e^{\mathbf{i}\frac{D-1}{2}(\xi^o_e-\xi'^o_e)}e^{\frac{(\eta_e)^2+(\eta'_e)^2+2t^2(D-1)^2}{4t}}e^{-t(\frac{\eta'_e}{2t}-\frac{\eta_e}{2t})^2 +2t(\frac{\eta_e}{4t}-\frac{\eta'_e}{4t}-\frac{\widetilde{\Theta}_e}{4t})^2} e^{\mathbf{i}(\frac{\eta_e}{4t}-\frac{D-1}{2}+\frac{\eta'_e}{4t}-\frac{\widetilde{\Theta}_e}{4t}) (\xi^o_e-\xi'^o_e+\tilde{\varphi}_e)} \\\nonumber
&&\cdot e^{-(\frac{\eta'_e}{2t}-\frac{D-1}{2})\widetilde{\Theta}_e} (\frac{\eta'_e}{4t}+\frac{\eta_e}{4t}-\frac{\widetilde{\Theta}_e}{4t})(\breve{\text{P}}(\frac{\eta'_e}{4t}+\frac{\eta_e}{4t}-\frac{\widetilde{\Theta}_e}{4t}))^2 \exp(-\frac{(\xi^o_e-\xi'^o_e+\tilde{\varphi}_e)^2}{8t})(1+\mathcal{O}(\frac{t}{\eta'_e})+\mathcal{O}(e^{-\frac{1}{t}}))\\\nonumber
&=&\beta(\frac{\eta'_e}{4}+\frac{\eta_e}{4}-\frac{\widetilde{\Theta}_e}{4})\langle\breve{\Psi}_{\mathbb{H}^o_e} |\breve{\Psi}_{\mathbb{H}'^o_e}\rangle(1+\mathcal{O}(\frac{t}{\eta'_e})+\mathcal{O}(e^{-\frac{1}{t}}))
\end{eqnarray}
for large $\eta'_e$. Notice that the classical evaluation of operator $\hat{X}^{12}$ is given by ${X}^{12}(\mathbb{H}'^o_e)=\frac{\eta'_e}{2}$. Hence one can estimate the matrix elements of $\hat{X}_{12}$ with respect to the twisted geometry coherent states by
\begin{eqnarray}\label{F12esti}
&&\frac{\langle\breve{\Psi}_{\mathbb{H}^o_e}|\hat{X}_e^{12} |\breve{\Psi}_{\mathbb{H}'^o_e}\rangle}{||\breve{\Psi}_{\mathbb{H}'^o_e}||||\breve{\Psi}_{ \mathbb{H}^o_e}||} -\beta\frac{\eta'_e}{2}\frac{\langle\breve{\Psi}_{\mathbb{H}^o_e} |\breve{\Psi}_{\mathbb{H}'^o_e}\rangle}{||\breve{\Psi}_{\mathbb{H}'^o_e}||||\breve{\Psi}_{ \mathbb{H}^o_e}||}\\\nonumber
  &\stackrel{\text{large}\ \eta'_e}{=}&\beta\left((\frac{\eta_e}{4}-\frac{\eta'_e}{4} -\frac{\widetilde{\Theta}_e}{4})+(\frac{\eta_e}{4}+\frac{\eta'_e}{4} -\frac{\widetilde{\Theta}_e}{4})(\mathcal{O}(\frac{t}{\eta'_e})+\mathcal{O}(e^{-\frac{1}{t}}))\right)\frac{\langle\breve{\Psi}_{\mathbb{H}^o_e} |\breve{\Psi}_{\mathbb{H}'^o_e}\rangle}{||\breve{\Psi}_{\mathbb{H}'^o_e}||||\breve{\Psi}_{ \mathbb{H}^o_e}||}.
\end{eqnarray}
Recall the overlap function $i^t(\mathbb{H}^o_e, \mathbb{H}'^o_e)$ is sharply peaked at $\eta_e=\eta'_e$ and $\widetilde{\Theta}_e=0$, and it decays
exponentially  for $\eta_e\neq\eta'_e$ or $\widetilde{\Theta}_e\neq0$. Hence we can conclude that the right hand side of Eq.\eqref{F12esti} is bounded by a correction term which tends to zero in the limit $t\to 0$ for large $\eta'_e$.
\\ \\
\textbf{(ii).} For the case $\widetilde{\Theta}_e\simeq\eta_e+\eta'_e$ or $\widetilde{\Theta}_e\gg \eta_e+\eta'_e$. Similar to the analysis of the overlap function in \cite{Long:2021lmd}, we have
\begin{eqnarray}\label{F1212000}
&& \langle\breve{\Psi}_{\mathbb{H}^o_e}|\hat{X}_e^{12} |\breve{\Psi}_{\mathbb{H}'^o_e}\rangle e^{-\frac{(\eta_e)^2+(\eta'_e)^2+2t^2(D-1)^2}{4t}}\\\nonumber
&{=}&\beta te^{\mathbf{i}\frac{D-1}{2}(\xi^o_e-\xi'^o_e)}\sum_{N_e}N_e(\dim(\pi_{N_e}))^2\exp(-t(\frac{\eta_e}{2t}-d_{N_e})^2 -t(\frac{\eta'_e}{2t}-d_{N_e})^2)\\\nonumber
&&\cdot e^{\mathbf{i}N_e(\xi^o_e-\xi'^o_e+\varphi(u_e,u'_e) +\varphi(\tilde{u}_e,\tilde{u}'_e))}\exp(- N_e\widetilde{\Theta}_e)+\beta \sqrt{t}\cdot\mathcal{O}(e^{-\frac{\eta'^2_e}{8t}})\\\nonumber
&{<}&
\beta t[\eta'_e/4t] \exp(-t(\frac{\eta'_e}{4t}-\frac{D+1}{2})^2 -t(\frac{\eta_e}{2t}-\frac{\eta'_e}{4t}-\frac{D+1}{2})^2)(\frac{\eta'_e}{4t})(\breve{\text{P}}(\frac{\eta'_e}{4t}))^2 \exp(- \widetilde{\Theta}_e)\\\nonumber
&&+\beta t\sum_{N_e=[\frac{\eta'_e}{4t}]+1}^{+\infty} N_e(\dim(\pi_{N_e}))^2\left(\exp(-t(\frac{\eta'_e}{2t}-d_{N_e})^2 -t(\frac{\eta_e}{2t}-d_{N_e})^2) \exp(- [\frac{\eta'_e}{4t}]\widetilde{\Theta}_e)\right)+\beta \sqrt{t}\cdot\mathcal{O}(e^{-\frac{\eta'^2_e}{8t}})
\\\nonumber
&{\simeq}&\beta t[\eta'_e/4t] \exp(-t(\frac{\eta'_e}{4t}-\frac{D+1}{2})^2 -t(\frac{\eta_e}{2t}-\frac{\eta'_e}{4t}-\frac{D+1}{2})^2)(\frac{\eta'_e}{4t})(\breve{\text{P}}(\frac{\eta'_e}{4t}))^2 \exp(- \widetilde{\Theta}_e)\\\nonumber
&&+\beta t\sqrt{\frac{\pi}{2t}} (\frac{\eta_e}{4t}+\frac{\eta'_e}{4t})(\breve{\text{P}}(\frac{\eta_e}{4t}+\frac{\eta'_e}{4t}))^2 e^{-\frac{t}{2}(\frac{\eta_e}{2t}-\frac{\eta'_e}{2t})^2}  \exp(- [\frac{\eta'_e}{4t}]\widetilde{\Theta}_e)+\beta \sqrt{t}\cdot\mathcal{O}(e^{-\frac{\eta'^2_e}{8t}})\\\nonumber
\end{eqnarray}
for $\eta'_e$ being large, wherein the sign ``<'' means that the module of its left side less than that of its right side. Then, in this case the matrix elements of $\hat{X}_{12}$ is estimated by
 \begin{eqnarray}\label{F121212}
0&<&\left|\frac{\langle\breve{\Psi}_{\mathbb{H}^o_e}|\hat{X}_e^{12} |\breve{\Psi}_{\mathbb{H}'^o_e}\rangle}{||\breve{\Psi}_{\mathbb{H}'^o_e}||||\breve{\Psi}_{ \mathbb{H}^o_e}||}\right|\\\nonumber
&{\lesssim}&\beta t\frac{\sqrt{\frac{2t}{\pi}}f_1(\eta_e,\eta'_e)e^{-t(\frac{\eta'_e}{4t})^2}e^{- \widetilde{\Theta}_e} + (\frac{\eta_e}{4t}+\frac{\eta'_e}{4t})(\breve{\text{P}}(\frac{\eta_e}{4t}+\frac{\eta'_e}{4t}))^2e^{-\frac{t}{2}(\frac{\eta'_e}{2t}-\frac{\eta_e}{2t})^2} e^{- [\frac{\eta'_e}{4t}]\widetilde{\Theta}_e}}{ \breve{\text{P}}(\frac{\eta_e}{2t}) \breve{\text{P}}(\frac{\eta'_e}{2t})}
\end{eqnarray}
for large $\eta'_e$,
where $f_1(\frac{\eta_e}{t},\frac{\eta'_e}{t}):=[\eta'_e/4t] \exp( -t(\frac{\eta_e}{2t}-\frac{\eta'_e}{4t}-\frac{D+1}{2})^2)(\frac{\eta'_e}{4t})(\breve{\text{P}}(\frac{\eta'_e}{4t}))^2$. Notice that $\widetilde{\Theta}_e\simeq\eta_e+\eta'_e$ or $\widetilde{\Theta}_e\gg \eta_e+\eta'_e$ in this case, hence we can conclude that $\left|\frac{\langle\breve{\Psi}_{\mathbb{H}^o_e}|\hat{X}_e^{12} |\breve{\Psi}_{\mathbb{H}'^o_e}\rangle}{||\breve{\Psi}_{\mathbb{H}'^o_e}||||\breve{\Psi}_{ \mathbb{H}^o_e}||}\right|$ is always suppressed exponentially by the factors $e^{-t(\frac{\eta'_e}{4t})^2}$ and $e^{-\frac{t}{2}(\frac{\eta'_e}{2t}-\frac{\eta_e}{2t})^2} e^{- [\frac{\eta'_e}{4t}]\widetilde{\Theta}_e}$ in Eq.\eqref{F121212}.

It is ready to combine the results in the cases for $\widetilde{\Theta}_e\ll \eta_e+\eta'_e$ and $\widetilde{\Theta}_e\simeq\eta_e+\eta'_e$ or $\widetilde{\Theta}_e\gg \eta_e+\eta'_e$. In general, the matrix elements of $\hat{X}_{12}$ with respect to the twisted geometry coherent state is estimated by
\begin{eqnarray}\label{F12final}
\left|\frac{\langle\breve{\Psi}_{\mathbb{H}^o_e}|\hat{X}_e^{12} |\breve{\Psi}_{\mathbb{H}'^o_e}\rangle}{||\breve{\Psi}_{\mathbb{H}'^o_e}||||\breve{\Psi}_{ \mathbb{H}^o_e}||} -\frac{\eta'_e}{2}\frac{\langle\breve{\Psi}_{\mathbb{H}^o_e} |\breve{\Psi}_{\mathbb{H}'^o_e}\rangle}{||\breve{\Psi}_{\mathbb{H}'^o_e}||||\breve{\Psi}_{ \mathbb{H}^o_e}||}\right|
&\stackrel{\text{large}\ \eta'_e}{\lesssim}&t\left|f_{12}(\mathbb{H}^o_e,\mathbb{H}'^o_e)\right|\cdot\left|\frac{\langle\breve{\Psi}_{\mathbb{H}^o_e} |\breve{\Psi}_{\mathbb{H}'^o_e}\rangle}{||\breve{\Psi}_{\mathbb{H}'^o_e}||||\breve{\Psi}_{ \mathbb{H}^o_e}||}\right|,
\end{eqnarray}
where $f_{12}(\mathbb{H}^o_e,\mathbb{H}'^o_e)$ is a function grows no faster than exponentials as $|\eta_e-\eta'_e|$ ,$\widetilde{\Theta}_e$ and $|\xi^o_e-\xi'^o_e|$ going large for large $\eta'_e$.\\
\\  \textbf{Set II:}\textit{Matrix elements of $\hat{X}_e^{IJ}$ with $I,J\in \{3,...,D+1\}$.}\\ \\
Notice that $\langle N_e, V'_e|\tau^{IJ}|N_e, V_e\rangle=0$ with $V'_e=2\delta_{1}^{[I}\delta_2^{J]}$ and  $I,J\in \{3,...,D+1\}$, it is straightforward to get that
\begin{eqnarray}
&& \langle\breve{\Psi}_{\mathbb{H}^o_e}|\hat{X}_e^{IJ} |\breve{\Psi}_{\mathbb{H}'^o_e}\rangle
=0 ,\quad \text{for}\  I, J\in \{3,...,D+1\}.
\end{eqnarray}
\\ \\ \textbf{Set III:} \textit{Matrix elements of $\hat{X}_e^{1J}$ and $\hat{X}_e^{2J}$ with $J\in \{3,...,D+1\}$.}\\ \\
Let us consider the rest components $\langle\breve{\Psi}_{\mathbb{H}^o_e}|\hat{X}_e^{1J} |\breve{\Psi}_{\mathbb{H}'^o_e}\rangle$ and $\langle\breve{\Psi}_{\mathbb{H}^o_e}|\hat{X}_e^{2J} |\breve{\Psi}_{\mathbb{H}'^o_e}\rangle$ with $J\in \{3,...,D+1\}$.
We first have
\begin{eqnarray}
\langle N_e, V'_e|\tau^{1J}|N_e, V_e\rangle=N_e  \langle 1,V'_e|\tau_{1J}|1, V_e\rangle\langle (N_e-1), V'_e|(N_e-1), V_e\rangle, \quad \text{for}\   J\in \{3,...,D+1\}.
\end{eqnarray}
Here we note that $\langle 1,V'_e|\tau_{1J}|1, V_e\rangle$ with $J\neq 1,2$ as functions of $V_e$  vanish at $V_e=V'_e$ and the growth of their modules are restricted by their derivatives evaluated by Eq.\eqref{deri} as $V_e$ being transformed by $e^{t\tau_{KL}}\in SO(D+1)$. Then, by checking how ${\Theta}({u}_e,{u}'_e)=-\ln|\langle 1,V'_e|1, V_e\rangle|$ grows as $V_e$ being transformed by $e^{t\tau_{KL}}\in SO(D+1)$, we can conclude that $\langle 1,V'_e|\tau_{1J}|1, V_e\rangle$ with $J\neq 1,2$ grows no faster than exponentials as ${\Theta}({u}_e,{u}'_e)=-\ln|\langle 1,V'_e|1, V_e\rangle|$ going large.
Then, for large $\eta'_e$ we can give
\begin{eqnarray}\label{X1J}
 && \langle\breve{\Psi}_{\mathbb{H}^o_e}|\hat{X}_e^{1J} |\breve{\Psi}_{\mathbb{H}'^o_e}\rangle e^{-\frac{(\eta_e)^2+(\eta'_e)^2+2t^2(D-1)^2}{4t}}\\\nonumber
  &=& -\mathbf{i}\beta t\langle 1,V'_e|\tau_{1J}|1, V_e\rangle \sum_{N_e}N_e(\dim(\pi_{N_e}))^2 \exp(-t(\frac{\eta_e}{2t}-d_{N_e})^2 -t(\frac{\eta'_e}{2t}-d_{N_e})^2)\\\nonumber
&&\cdot e^{\mathbf{i}d_{N_e}(\xi^o_e-\xi'^o_e)}\langle (N_e-1), V'_e|(N_e-1), V_e\rangle\langle N_e, -\tilde{V}_e|N_e,-\tilde{ V}'_e\rangle+\beta\sqrt{t}\mathcal{O}(e^{-(\eta'_e)^2/(8t)})\\\nonumber
&=&  -\mathbf{i}\beta t e^{\mathbf{i}\frac{D-1}{2}(\xi^o_e-\xi'^o_e)}\frac{\langle 1,V'_e|\tau_{1J}|1, V_e\rangle}{\langle 1,V'_e|1, V_e\rangle} \sum_{N_e}N_e(\dim(\pi_{N_e}))^2\exp(-t(\frac{\eta_e}{2t}-d_{N_e})^2 -t(\frac{\eta'_e}{2t}-d_{N_e})^2)\\\nonumber
&&\cdot e^{\mathbf{i}N_e(\xi^o_e-\xi'^o_e+\varphi(u_e,u'_e) +\varphi(\tilde{u}_e,\tilde{u}'_e))}\exp(- N_e\widetilde{\Theta}_e)+\beta\sqrt{t}\mathcal{O}(e^{-(\eta'_e)^2/(8t)}),
\end{eqnarray}
where we only consider the case of $\langle 1,V'_e|1, V_e\rangle\neq0$ since one have $\langle 1,V'_e|\tau_{1J}|1, V_e\rangle=0$ and $\langle\breve{\Psi}_{\mathbb{H}^o_e}|\hat{X}_e^{1J} |\breve{\Psi}_{\mathbb{H}'^o_e}\rangle=0$ if $\langle 1,V'_e|1, V_e\rangle=0$.
By comparing Eq.\eqref{X1J} with Eq.\eqref{F12}, one have
\begin{eqnarray}
 \langle\breve{\Psi}_{\mathbb{H}^o_e}|\hat{X}_e^{1J} |\breve{\Psi}_{\mathbb{H}'^o_e}\rangle \stackrel{\text{large}\ \eta'_e}{=}-\mathbf{i}\frac{\langle 1,V'_e|\tau_{1J}|1, V_e\rangle}{\langle 1,V'_e|1, V_e\rangle}  \langle\breve{\Psi}_{\mathbb{H}^o_e}|\hat{X}^{12}_e|\breve{\Psi}_{\mathbb{H}'^o_e}\rangle+ e^{\frac{(\eta_e)^2+(\eta'_e)^2+2t^2(D-1)^2}{4t}}\beta\sqrt{t}\mathcal{O}(e^{-(\eta'_e)^2/(8t)})
\end{eqnarray}
and further
\begin{eqnarray}
\left|\frac{ \langle\breve{\Psi}_{\mathbb{H}^o_e}|\hat{X}_e^{1J} |\breve{\Psi}_{\mathbb{H}'^o_e}\rangle}{||\breve{\Psi}_{\mathbb{H}'^o_e}||||\breve{\Psi}_{ \mathbb{H}^o_e}||} \right| \stackrel{\text{large}\ \eta'_e}{=}\left|\frac{\langle 1,V'_e|\tau_{1J}|1, V_e\rangle}{\langle 1,V'_e|1, V_e\rangle}  \right| \frac{\langle\breve{\Psi}_{\mathbb{H}^o_e}|\hat{X}^{12}_e|\breve{\Psi}_{\mathbb{H}'^o_e}\rangle}{||\breve{\Psi}_{\mathbb{H}'^o_e}||||\breve{\Psi}_{ \mathbb{H}^o_e}||}+ \beta{t}\mathcal{O}(e^{-(\eta'_e)^2/(8t)}),
\end{eqnarray}
wherein one should note that $|\langle 1,V'_e|1, V_e\rangle|=e^{-{\Theta}(u_e,u'_e)}$, $0\leq|\langle 1,V'_e|\tau_{1J}|1, V_e\rangle|\leq 1$ with $\langle 1,V'_e|\tau_{1J}|1, V_e\rangle=0$ at $V'_e=V_e$, and $\langle 1,V'_e|\tau_{1J}|1, V_e\rangle$ with $J\neq 1,2$ grow no faster than exponentials as ${\Theta}({u}_e,{u}'_e)=-\ln|\langle 1,V'_e|1, V_e\rangle|$ going large.
Then let us recall Eq.\eqref{F12} and notice that $\frac{\langle\breve{\Psi}_{\mathbb{H}^o_e} |\breve{\Psi}_{\mathbb{H}'^o_e}\rangle}{||\breve{\Psi}_{\mathbb{H}'^o_e}||||\breve{\Psi}_{ \mathbb{H}^o_e}||}$ is unity at $\mathbb{H}'^o_e=\mathbb{H}^o_e$ and decays exponentially fast to $0$ as $|\eta_e-\eta'_e|$, $\widetilde{\Theta}_e$ and $|\xi^o_e-\xi'^o_e|$ going large  for large $\eta'_e$, we can conclude that
\begin{eqnarray}
\left|\frac{\langle\breve{\Psi}_{\mathbb{H}^o_e}|\hat{X}_e^{1J} |\breve{\Psi}_{\mathbb{H}'^o_e}\rangle}{||\breve{\Psi}_{\mathbb{H}'^o_e}||||\breve{\Psi}_{ \mathbb{H}^o_e}||}\right|
&\stackrel{\text{large}\ \eta'_e}{\lesssim}&t\left| f_{1J}(\mathbb{H}^o_e,\mathbb{H}'^o_e)\right|\cdot\left|\frac{\langle\breve{\Psi}_{\mathbb{H}^o_e} |\breve{\Psi}_{\mathbb{H}'^o_e}\rangle}{||\breve{\Psi}_{\mathbb{H}'^o_e}||||\breve{\Psi}_{ \mathbb{H}^o_e}||}\right|,\quad \text{for}\   J\in \{3,...,D+1\},
\end{eqnarray}
where $f_{1J}(\mathbb{H}^o_e,\mathbb{H}'^o_e)$ is a function of $\mathbb{H}^o_e,\mathbb{H}'^o_e$ which vanishes for $\Theta(u_e,u'_e)=0$ and whose growth is always suppressed by $\left|\frac{\langle\breve{\Psi}_{\mathbb{H}^o_e} |\breve{\Psi}_{\mathbb{H}'^o_e}\rangle}{||\breve{\Psi}_{\mathbb{H}'^o_e}||||\breve{\Psi}_{ \mathbb{H}^o_e}||}\right|$ exponentially as $|\eta_e-\eta'_e|$, $\widetilde{\Theta}_e$ and $|\xi^o_e-\xi'^o_e|$ going large  for large $\eta'_e$.
Similar discussion can be given for $\langle\breve{\Psi}_{\mathbb{H}'^o_e}|\hat{X}_e^{2J} |\breve{\Psi}_{\mathbb{H}^o_e}\rangle$ and we reach that
\begin{eqnarray}
\frac{\langle\breve{\Psi}_{\mathbb{H}^o_e}|\hat{X}_e^{2J} |\breve{\Psi}_{\mathbb{H}'^o_e}\rangle}{||\breve{\Psi}_{\mathbb{H}'^o_e}||||\breve{\Psi}_{ \mathbb{H}^o_e}||}
&\stackrel{\text{large}\ \eta'_e}{\lesssim}&t\left| f_{2J}(\mathbb{H}^o_e,\mathbb{H}'^o_e)\right|\cdot\left|\frac{\langle\breve{\Psi}_{\mathbb{H}^o_e} |\breve{\Psi}_{\mathbb{H}'^o_e}\rangle}{||\breve{\Psi}_{\mathbb{H}'^o_e}||||\breve{\Psi}_{ \mathbb{H}^o_e}||}\right|,\quad \text{for}\   J\in \{3,...,D+1\},
\end{eqnarray}
where $f_{2J}(\mathbb{H}^o_e,\mathbb{H}'^o_e)$ is a function of $\mathbb{H}^o_e,\mathbb{H}'^o_e$ which vanishes for $\Theta(u_e,u'_e)=0$ and whose growth is always suppressed by $\left|\frac{\langle\breve{\Psi}_{\mathbb{H}^o_e} |\breve{\Psi}_{\mathbb{H}'^o_e}\rangle}{||\breve{\Psi}_{\mathbb{H}'^o_e}||||\breve{\Psi}_{ \mathbb{H}^o_e}||}\right|$ exponentially as $|\eta_e-\eta'_e|$, $\widetilde{\Theta}_e$ and $|\xi^o_e-\xi'^o_e|$ going large  for large $\eta'_e$.

Finally, collecting all the results in this subsection we arrive at the first main result of this paper.
\begin{itemize}
  \item The matrix elements of the flux operators with respect to the twisted geometry coherent states can be estimated by
      \begin{eqnarray}\label{Ffinal}
\left|\frac{\langle\breve{\Psi}_{\mathbb{H}^o_e}|\hat{X}_e^{IJ} |\breve{\Psi}_{\mathbb{H}'^o_e}\rangle}{||\breve{\Psi}_{\mathbb{H}'^o_e}||||\breve{\Psi}_{ \mathbb{H}^o_e}||} -\frac{\eta'_e}{2}V'^{IJ}\frac{\langle\breve{\Psi}_{\mathbb{H}^o_e} |\breve{\Psi}_{\mathbb{H}'^o_e}\rangle}{||\breve{\Psi}_{\mathbb{H}'^o_e}||||\breve{\Psi}_{ \mathbb{H}^o_e}||}\right|
&\stackrel{\text{large}\ \eta'_e}{\lesssim}&t\left|f_{X}(\mathbb{H}^o_e,\mathbb{H}'^o_e)\right|\cdot\left| \frac{\langle\breve{\Psi}_{\mathbb{H}^o_e} |\breve{\Psi}_{\mathbb{H}'^o_e}\rangle}{||\breve{\Psi}_{\mathbb{H}'^o_e}||||\breve{\Psi}_{ \mathbb{H}^o_e}||}\right|,
\end{eqnarray}
where $f_{X}(\mathbb{H}^o_e,\mathbb{H}'^o_e)$ is a function of $\mathbb{H}^o_e,\mathbb{H}'^o_e$  whose growth is always suppressed by $\left|\frac{\langle\breve{\Psi}_{\mathbb{H}^o_e} |\breve{\Psi}_{\mathbb{H}'^o_e}\rangle}{||\breve{\Psi}_{\mathbb{H}'^o_e}||||\breve{\Psi}_{ \mathbb{H}^o_e}||}\right|$ exponentially as $|\eta_e-\eta'_e|$, $\widetilde{\Theta}_e$ and $|\xi^o_e-\xi'^o_e|$ going large  for large $\eta'_e$.
\end{itemize}
\subsubsection{Matrix elements of the holonomy operator}
 The matrix elements of holonomy operator in twisted geometry coherent basis will be considered for the cases that $(D+1)$ is even or odd separately. Let us first consider the case of $(D+1)$ being even.
Notice that each one of the matrix element of the classical holonomy in the definition representation space of $SO(D+1)$ corresponds to a holonomy operator which acts on the twisted geometry coherent state by multiplying. In order to give a specific holonomy operator, one need to consider an orthonormal and complete basis $\{|1,V_{\imath\jmath}\rangle|(\imath,\jmath)\in\{(1,2),(2,1),(3,4),(4,3),...,(D,D+1),(D+1,D)\}\}$ of the definition representation space of $SO(D+1)$, where $V_{\imath\jmath}$ are the elements in a set of bi-vectors $\{V_{\imath\jmath}=2\delta_\imath^{[I}\delta_\jmath^{J]}|(\imath,\jmath)\in \{(1,2),(2,1),(3,4),(4,3),...,(D,D+1),(D+1,D)\}\}$ in $\mathbb{R}^{D+1}$,  and the interpretation of these notations of this basis are explained in Appendix \ref{app0} explicitly. Then, the matrix elements of the classical holonomy $h_e$ in the basis $\{|1,V_{\imath\jmath}\rangle$ of definition representation space of $SO(D+1)$ can be promoted as holonomy operators  as
 \begin{equation}
 \langle 1, V_{\imath\jmath}|h_e|1,V_{\imath'\jmath'}\rangle \mapsto  (\widehat{h_e})_{\imath\jmath,\imath'\jmath'}.
 \end{equation}
 For a given twisted geometry coherent state $\breve{\Psi}_{\gamma,\vec{\mathbb{H}}^o_e}$, the label $\vec{\mathbb{H}}^o_e$ assigns classical labels $u_e=u(V_e)$ and $\tilde{u}_e=\tilde{u}(\tilde{V}_e)$ to each edge $e$. Then, in order to adapt the holonomy operators to the state $\breve{\Psi}_{\gamma,\vec{\mathbb{H}}^o_e}$, let us consider two orthonormal and complete basis $\{\tilde{u}_e|1,V_{\imath\jmath}\rangle\}$ and $\{u_e|1,V_{\imath\jmath}\rangle\}$ of the definition representation space of $SO(D+1)$. These two  orthonormal and complete basis select the matrix elements $({u_e^{-1}h_e\tilde{u}_e})_{\imath\jmath,\imath'\jmath'}:=\langle 1, V_{\imath\jmath}|u_e^{-1}h_e\tilde{u}_e|1,V_{\imath'\jmath'}\rangle$ of the classical holomomy $h_e$, which can be promoted as the holonomy operator $(\widehat{u_e^{-1}h_e\tilde{u}_e})_{\imath\jmath,\imath'\jmath'}$ acting on the coherent state $\breve{\Psi}_{\gamma,\vec{\mathbb{H}}^o_e}$ by multiplying. We show the details of the action of the holonomy operator $(\widehat{u_e^{-1}h_e\tilde{u}_e})_{\imath\jmath,\imath'\jmath'}$ on the adapting states in Appendix \ref{app0}.  Now, we can start to calculate the matrix elements of the holonomy operator $(\widehat{u_e^{-1}h_e\tilde{u}_e})_{\imath\jmath,\imath'\jmath'}$ in the twisted geometry coherent state basis, which read
 \begin{equation}
 \frac{\langle\breve{\Psi}_{\mathbb{H}^o_e}|(\widehat{u'^{-1}_e h_e\tilde{u}'_e})_{\imath\jmath,\imath'\jmath'}  |\breve{\Psi}_{\mathbb{H}'^o_e}\rangle}{||\breve{\Psi}_{\mathbb{H}'^o_e}||||\breve{\Psi}_{ \mathbb{H}^o_e}||}.
 \end{equation}

We first consider the simplest component $(\widehat{u_e^{-1}h_e\tilde{u}_e})_{12,12} $ of holonomy operator, whose matrix elements in twisted geometry coherent basis is given by
\begin{eqnarray}\label{expectah1212}
&& \langle\breve{\Psi}_{\mathbb{H}^o_e}|(\widehat{u'^{-1}_eh_e\tilde{u}'_e})_{12,12}  |\breve{\Psi}_{\mathbb{H}'^o_e}\rangle e^{-\frac{(\eta_e)^2+(\eta'_e)^2+2t^2(D-1)^2}{4t}}\\\nonumber
&\stackrel{\text{large}\ \eta'_e }{=}&e^{\mathbf{i}\xi'^o_e}\sum_{N_e}(\dim(\pi_{N_e}))^{3/2}(\dim(\pi_{N_e+1}))^{1/2} \exp(-t(\frac{\eta_e}{2t} -d_{N_e+1})^2 -t(\frac{\eta'_e}{2t}-d_{N_e})^2)\\\nonumber
&&\cdot e^{\mathbf{i}d_{N_e+1}(\xi^o_e-\xi'^o_e)}\langle N_e+1, V'_e|N_e+1, V_e\rangle\langle N_e+1, -\tilde{V}_e|N_e+1,-\tilde{ V}'_e\rangle+\frac{1}{\sqrt{t}}\mathcal{O}(e^{-\frac{(\eta'_e)^2}{8t}}),
\end{eqnarray}
wherein we used that $|1, V_{12}\rangle\otimes |N_e,V_{12}\rangle =| N_e+1, V_{12}\rangle$.
 The calculation of Eq.\eqref{expectah1212} is proceeded in Appendix \ref{app1} and we get
\begin{eqnarray}\label{h1212fina}
\left|\frac{\langle\breve{\Psi}_{\mathbb{H}^o_e}|(\widehat{u'^{-1}_e h_e\tilde{u}'_e})_{12,12}  |\breve{\Psi}_{\mathbb{H}'^o_e}\rangle}{||\breve{\Psi}_{\mathbb{H}'^o_e}||||\breve{\Psi}_{ \mathbb{H}^o_e}||} -e^{\mathbf{i}\xi'^o_e}\frac{\langle\breve{\Psi}_{\mathbb{H}^o_e} |\breve{\Psi}_{\mathbb{H}'^o_e}\rangle}{||\breve{\Psi}_{\mathbb{H}'^o_e}||||\breve{\Psi}_{ \mathbb{H}^o_e}||}\right|
&\stackrel{\text{large}\ \eta'_e}{\lesssim}&t\left|\tilde{f}_{12}(\mathbb{H}^o_e,\mathbb{H}'^o_e)\right|\cdot\left|\frac{\langle\breve{\Psi}_{\mathbb{H}^o_e} |\breve{\Psi}_{\mathbb{H}'^o_e}\rangle}{||\breve{\Psi}_{\mathbb{H}'^o_e}||||\breve{\Psi}_{ \mathbb{H}^o_e}||}\right|,
\end{eqnarray}
where $\tilde{f}_{12}(\mathbb{H}^o_e,\mathbb{H}'^o_e)$ is a function whose growth is always suppressed by $\left|\frac{\langle\breve{\Psi}_{\mathbb{H}^o_e} |\breve{\Psi}_{\mathbb{H}'^o_e}\rangle}{||\breve{\Psi}_{\mathbb{H}'^o_e}||||\breve{\Psi}_{ \mathbb{H}^o_e}||}\right|$ exponentially as $|\eta_e-\eta'_e|$, $\widetilde{\Theta}_e$ and $|\xi^o_e-\xi'^o_e|$ going large  for large $\eta'_e$.

Similarly, the matrix elements of $(\widehat{u'^{-1}_eh_e\tilde{u}'_e})_{21,21} $ in twisted geometry coherent basis is given by
\begin{eqnarray}\label{h212121}
 &&\langle\breve{\Psi}_{\mathbb{H}^o_e}|(\widehat{u'^{-1}_eh_e\tilde{u}'_e})_{21,21}  |\breve{\Psi}_{\mathbb{H}'^o_e}\rangle \\\nonumber &\stackrel{\text{large}\ \eta'_e}{=}&\text{FRHS\ of  Eq.}\eqref{h212121}+\text{SRHS\ of  Eq.}\eqref{h212121}+\text{TRHS\ of  Eq.}\eqref{h212121}\\\nonumber
 &&+\frac{1}{\sqrt{t}}e^{\frac{(\eta_e)^2+(\eta'_e)^2+2t^2(D-1)^2}{4t}}\mathcal{O}(e^{-\frac{(\eta'_e)^2}{8t}}),
\end{eqnarray}
where we defined
\begin{eqnarray}\label{h21F}
 &&\text{FRHS\ of  Eq.}\eqref{h212121}\\\nonumber
 &:=& e^{\frac{(\eta_e)^2+(\eta'_e)^2+2t^2(D-1)^2}{4t}}e^{-\mathbf{i}\xi'^o_e}\sum_{N_e}(\dim(\pi_{N_e}))^{3/2} (\dim(\pi_{N_e-1}))^{1/2} \exp(-t(\frac{\eta_e}{2t} -d_{N_e-1})^2 -t(\frac{\eta'_e}{2t}-d_{N_e})^2)\\\nonumber
&&\cdot e^{\mathbf{i}d_{N_e-1}(\xi^o_e-\xi'^o_e)}\langle N_e-1, V'_e|N_e-1, V_e\rangle\langle N_e-1, -\tilde{V}_e|N_e-1,-\tilde{ V}'_e\rangle
\end{eqnarray}
\begin{eqnarray}\label{h21S}
 &&\text{SRHS\ of  Eq.}\eqref{h212121}\\\nonumber
 &:=&- e^{\frac{(\eta_e)^2+(\eta'_e)^2+2t^2(D-1)^2}{4t}}e^{-\mathbf{i}\xi'^o_e}\sum_{N_e}(\dim(\pi_{N_e}))^{3/2} (\dim(\pi_{N_e-1}))^{1/2} \exp(-t(\frac{\eta_e}{2t} -d_{N_e-1})^2 -t(\frac{\eta'_e}{2t}-d_{N_e})^2)\\\nonumber
&&\cdot e^{\mathbf{i}d_{N_e-1}(\xi^o_e-\xi'^o_e)}(1-|\alpha_1(N_e)|^2 )\langle N_e-1, V'_e|N_e-1, V_e\rangle\langle N_e-1, -\tilde{V}_e|N_e-1,-\tilde{ V}'_e\rangle
\end{eqnarray}
and
\begin{eqnarray}\label{h21T}
 &&\text{TRHS\ of  Eq.}\eqref{h212121}\\\nonumber
 &:=&e^{\frac{(\eta_e)^2+(\eta'_e)^2+2t^2(D-1)^2}{4t}}e^{\mathbf{i}\xi'^o_e}\sum_{N_e}\Big((\dim(\pi_{N_e}))^{3/2} (\dim(\pi_{N_e+1}))^{1/2} \exp(-t(\frac{\eta_e}{2t} -d_{N_e+1})^2 -t(\frac{\eta'_e}{2t}-d_{N_e})^2)\\\nonumber
&&\cdot e^{\mathbf{i}d_{N_e+1}(\xi^o_e-\xi'^o_e)}|\alpha_2(N_e)|^2\langle N_e+1,...| u'^{-1}_eu_e|N_e+1,V_{12}\rangle\langle N_e+1,V_{12}|\tilde{u}^{-1}_e\tilde{u}'_e|N_e+1,...\rangle,
\end{eqnarray}
 wherein
 \begin{eqnarray}\label{V21V12}
|1, V_{21}\rangle\otimes |N_e,V_{12}\rangle =\alpha_1(N_e)| N_e-1, V_{12}\rangle +\alpha_2(N_e)|N_e+1,...\rangle+\alpha_3(N_e)|\text{not\ simple}\rangle,
\end{eqnarray}
with $|\alpha_1(N)|^2=\frac{N(2N+D-3)}{(D+N-2)(2N+D-1)}$, $|\alpha_2(N)|^2\leq1-|\alpha_1(N)|^2$, see more details in Appendix \ref{app0}. As shown in Appendix \ref{app1}, the three terms in the right-hand side (RHS) of ``='' in Eq.\eqref{h212121} are evaluated by Eqs.\eqref{H21212133}, \eqref{H212121},  \eqref{STFFh21212121} and \eqref{TTFFh21212121}.
Combining these results we get
  \begin{eqnarray}\label{h2121fina2}
\left|\frac{\langle\breve{\Psi}_{\mathbb{H}^o_e}|(\widehat{u'^{-1}_eh_e\tilde{u}'_e})_{21,21}  |\breve{\Psi}_{\mathbb{H}'^o_e}\rangle}{||\breve{\Psi}_{\mathbb{H}'^o_e}||||\breve{\Psi}_{ \mathbb{H}^o_e}||} -e^{-\mathbf{i}\xi'^o_e}\frac{\langle\breve{\Psi}_{\mathbb{H}^o_e} |\breve{\Psi}_{\mathbb{H}'^o_e}\rangle}{||\breve{\Psi}_{\mathbb{H}'^o_e}||||\breve{\Psi}_{ \mathbb{H}^o_e}||}\right|
&\stackrel{\text{large}\ \eta'_e}{\lesssim}&t\left|\tilde{f}_{21}(\mathbb{H}^o_e,\mathbb{H}'^o_e)\right|\cdot\left|\frac{\langle\breve{\Psi}_{\mathbb{H}^o_e} |\breve{\Psi}_{\mathbb{H}'^o_e}\rangle}{||\breve{\Psi}_{\mathbb{H}'^o_e}||||\breve{\Psi}_{ \mathbb{H}^o_e}||}\right|,
\end{eqnarray}
where $\tilde{f}_{21}(\mathbb{H}^o_e,\mathbb{H}'^o_e)$ is a function whose growth is always suppressed by $\left|\frac{\langle\breve{\Psi}_{\mathbb{H}'^o_e} |\breve{\Psi}_{\mathbb{H}^o_e}\rangle}{||\breve{\Psi}_{\mathbb{H}'^o_e}||||\breve{\Psi}_{ \mathbb{H}^o_e}||}\right|$ exponentially as $|\eta_e-\eta'_e|$, $\widetilde{\Theta}_e$ and $|\xi^o_e-\xi'^o_e|$ going large  for large $\eta'_e$.

   For the off-diagonal component $(\widehat{u'^{-1}_eh_e\tilde{u}'_e})_{12,21}$ of holonomy operators, we have
\begin{eqnarray}\label{h122100}
&& \langle\breve{\Psi}_{\mathbb{H}^o_e}|(\widehat{u'^{-1}_eh_e\tilde{u}'_e})_{12,21}  |\breve{\Psi}_{\mathbb{H}'^o_e}\rangle\\\nonumber
&\stackrel{\text{large}\ \eta'_e}{=}&e^{\frac{(\eta_e)^2+(\eta'_e)^2+2t^2(D-1)^2}{4t}}e^{\mathbf{i}\xi'^o_e}\sum_{N_e}(\dim(\pi_{N_e}))^{\frac{3}{2}} (\dim(\pi_{N_e+1}))^{\frac{1}{2}} \exp(-t(\frac{\eta_e}{2t} -d_{N_e+1})^2 -t(\frac{\eta'_e}{2t}-d_{N_e})^2)\\\nonumber
&&\cdot e^{\mathbf{i}d_{N_e+1}(\xi^o_e-\xi'^o_e)}\alpha_2(N_e)\langle N_e+1, V'_e|N_e+1, V_e\rangle\langle N_e+1, -\tilde{V}_e|N_e+1,...\rangle+\frac{1}{\sqrt{t}}\mathcal{O}(e^{-(\eta'_e)^2/(8t)})
\end{eqnarray}
where $\alpha_2(N_e)$ satisfies $0\leq|\alpha_2(N_e)|\leq \sqrt{1-|\alpha_1(N_e)|^2} $. The explicit calculation of Eq.\eqref{h122100} is given in Appendix \ref{app1}. Then, based on Eq.\eqref{TTFFh12211221}, we can evaluate Eq.\eqref{h122100} by
\begin{equation}\label{h1221}
\frac{\left|\langle\breve{\Psi}_{\mathbb{H}^o_e}|(\widehat{u'^{-1}_eh_e\tilde{u}'_e})_{12,21}  |\breve{\Psi}_{\mathbb{H}'^o_e}\rangle\right|}{||\breve{\Psi}_{\mathbb{H}'^o_e}||||\breve{\Psi}_{ \mathbb{H}^o_e}||}
\stackrel{\text{large}\ \eta'_e}{\lesssim}t\left|\tilde{f}'(\mathbb{H}^o_e,\mathbb{H}'^o_e)\right|\frac{\left|\langle\breve{\Psi}_{\mathbb{H}^o_e} |\breve{\Psi}_{\mathbb{H}'^o_e}\rangle\right|}{||\breve{\Psi}_{\mathbb{H}'^o_e}||||\breve{\Psi}_{ \mathbb{H}^o_e}||},
\end{equation}
where $\tilde{f}'(\mathbb{H}^o_e,\mathbb{H}'^o_e)$ is a function whose growth is always suppressed by $\frac{\left|\langle\breve{\Psi}_{\mathbb{H}^o_e} |\breve{\Psi}_{\mathbb{H}'^o_e}\rangle\right|}{||\breve{\Psi}_{\mathbb{H}'^o_e}||||\breve{\Psi}_{ \mathbb{H}^o_e}||}$ exponentially as $|\eta_e-\eta'_e|, \Theta(u_e,u'_e)$ and $|\xi^o_e-\xi'^o_e|$  going large  for large $\eta'_e$. Following similar calculations we can also give
\begin{equation}\label{h2112}
\frac{\left|\langle\breve{\Psi}_{\mathbb{H}^o_e}|(\widehat{u'^{-1}_eh_e\tilde{u}'_e})_{21,12}  |\breve{\Psi}_{\mathbb{H}'^o_e}\rangle\right|}{||\breve{\Psi}_{\mathbb{H}'^o_e}||||\breve{\Psi}_{ \mathbb{H}^o_e}||}
\stackrel{\text{large}\ \eta'_e}{\lesssim}t\left|\tilde{f}''(\mathbb{H}^o_e,\mathbb{H}'^o_e)\right|\frac{\left|\langle\breve{\Psi}_{\mathbb{H}^o_e} |\breve{\Psi}_{\mathbb{H}'^o_e}\rangle\right|}{||\breve{\Psi}_{\mathbb{H}'^o_e}||||\breve{\Psi}_{ \mathbb{H}^o_e}||},
\end{equation}
where $\tilde{f}''(\mathbb{H}^o_e,\mathbb{H}'^o_e)$ is a function whose growth is always suppressed by $\frac{\left|\langle\breve{\Psi}_{\mathbb{H}^o_e} |\breve{\Psi}_{\mathbb{H}'^o_e}\rangle\right|}{||\breve{\Psi}_{\mathbb{H}'^o_e}||||\breve{\Psi}_{ \mathbb{H}^o_e}||}$ exponentially as $|\eta_e-\eta'_e|, \Theta(u_e,u'_e)$ and $|\xi^o_e-\xi'^o_e|$  going large  for large $\eta'_e$.

Then, let us consider the components $(\widehat{u'^{-1}_eh_e\tilde{u}'_e})_{\imath\jmath,\imath'\jmath'}$ of holonomy operator with $(\imath,\jmath),(\imath',\jmath')\in\{(3,4),(4,3),...,(D,D+1),(D+1,D)\}$. Similar to the calculations of Eqs.\eqref{expectah1212}, \eqref{h212121}, and \eqref{h122100} in Appendix \ref{app1}, the matrix elements of $(\widehat{u'^{-1}_eh_e\tilde{u}'_e})_{\imath\jmath,\imath'\jmath'}$ in twisted geometry coherent state basis can be evaluated as
\begin{eqnarray}\label{h3456}
&& \langle\breve{\Psi}_{\mathbb{H}^o_e}|(\widehat{u'^{-1}_eh_e\tilde{u}'_e})_{\imath\jmath,\imath'\jmath'}  |\breve{\Psi}_{\mathbb{H}'^o_e}\rangle e^{-\frac{(\eta_e)^2+(\eta'_e)^2+2t^2(D-1)^2}{4t}}\\\nonumber
&{=}&-e^{\mathbf{i}\xi'^o_e}\sum_{N_e}\Big((\dim(\pi_{N_e}))^{3/2} (\dim(\pi_{N_e+1}))^{1/2} \exp(-t(\frac{\eta_e}{2t} -d_{N_e+1})^2 -t(\frac{\eta'_e}{2t}-d_{N_e})^2)\\\nonumber
&&\cdot e^{\mathbf{i}d_{N_e+1}(\xi^o_e-\xi'^o_e)}\frac{1}{(N_e+1)^2}\langle N_e+1,V_{12}|u^{-1}_eu'_e(\tau^{1\imath}\pm\tau^{2\jmath})|N_e+1,V_{12}\rangle\\\nonumber
&&\cdot\langle N_e+1,V_{12}|(\tau^{\imath'1}\pm\tau^{\jmath'2})\tilde{u}'^{-1}_e\tilde{u}_e|N_e+1,V_{12}\rangle\Big) +\frac{1}{\sqrt{t}}\mathcal{O}(e^{-\frac{(\eta'_e)^2}{8t}})\\\nonumber
&{=}&-e^{\mathbf{i}\xi'^o_e} \sum_{N_e}\Big((\dim(\pi_{N_e}))^{3/2} (\dim(\pi_{N_e+1}))^{1/2} \exp(-t(\frac{\eta_e}{2t} -d_{N_e+1})^2 -t(\frac{\eta'_e}{2t}-d_{N_e})^2)\\\nonumber
&&\cdot e^{\mathbf{i}d_{N_e+1}(\xi^o_e-\xi'^o_e)} \mathcal{T}_{\pm\imath\jmath,\pm\imath'\jmath'}(u^{-1}_eu'_e,\tilde{u}'^{-1}_e\tilde{u}_e)\langle N_e+1,V_{12}|u^{-1}_eu'_e|N_e+1,V_{12}\rangle\langle N_e+1,V_{12}|\tilde{u}'^{-1}_e\tilde{u}_e|N_e+1,V_{12}\rangle\Big)\\\nonumber
&&+\frac{1}{\sqrt{t}}\mathcal{O}(e^{-\frac{(\eta'_e)^2}{8t}})\\\nonumber
&=& \text{FRHS\ of  Eq.}\eqref{h3456}+\frac{1}{\sqrt{t}}\mathcal{O}(e^{-\frac{(\eta'_e)^2}{8t}})
\end{eqnarray}
for large $\eta'_e$ and $(\imath,\jmath),(\imath',\jmath')\in\{(3,4),(4,3),...,(D,D+1),(D+1,D)\}$, where we used Eq.\eqref{holotype3} and defined
\begin{eqnarray}
&&\text{FRHS\ of  Eq.}\eqref{h3456}\\\nonumber
&:=& -e^{\mathbf{i}\xi'^o_e} \sum_{N_e}\Big((\dim(\pi_{N_e}))^{3/2} (\dim(\pi_{N_e+1}))^{1/2} \exp(-t(\frac{\eta_e}{2t} -d_{N_e+1})^2 -t(\frac{\eta'_e}{2t}-d_{N_e})^2)\\\nonumber
&&\cdot e^{\mathbf{i}d_{N_e+1}(\xi^o_e-\xi'^o_e)} \mathcal{T}_{\pm\imath\jmath,\pm\imath'\jmath'}(u^{-1}_eu'_e,\tilde{u}'^{-1}_e\tilde{u}_e)\langle N_e+1,V_{12}|u^{-1}_eu'_e|N_e+1,V_{12}\rangle\langle N_e+1,V_{12}|\tilde{u}'^{-1}_e\tilde{u}_e|N_e+1,V_{12}\rangle\Big)
 \end{eqnarray}
 with
 \begin{eqnarray}
&&\mathcal{T}_{\pm\imath\jmath,\pm\imath'\jmath'}(u^{-1}_eu'_e,\tilde{u}'^{-1}_e\tilde{u}_e)\\\nonumber
&:=&\frac{\langle 1,V_{12}|\tilde{u}'^{-1}_e\tilde{u}_e(\tau^{1\imath'}\pm\tau^{2\jmath'})|1,V_{12}\rangle}{\langle 1,V_{12}|\tilde{u}'^{-1}_e\tilde{u}_e|1,V_{12}\rangle}\frac{\langle 1,V_{12}|(\tau^{1\imath}\pm\tau^{2\jmath})u^{-1}_eu'_e|1,V_{12}\rangle}{\langle 1,V_{12}|u^{-1}_eu'_e|1,V_{12}\rangle}\\\nonumber
 &=&e^{\widetilde{\Theta}_e}e^{\mathbf{i}\tilde{\varphi}_e}\widetilde{ \mathcal{T}}_{\pm\imath\jmath,\pm\imath'\jmath'}(u^{-1}_eu'_e,\tilde{u}'^{-1}_e\tilde{u}_e),
 \end{eqnarray}
 and
 \begin{eqnarray}
&&\widetilde{ \mathcal{T}}_{\pm\imath\jmath,\pm\imath'\jmath'}(u^{-1}_eu'_e,\tilde{u}'^{-1}_e\tilde{u}_e)\\\nonumber
&:=&{\langle 1,V_{12}|\tilde{u}'^{-1}_e\tilde{u}_e(\tau^{1\imath'}\pm\tau^{2\jmath'})|1,V_{12}\rangle}{\langle 1,V_{12}|(\tau^{1\imath}\pm\tau^{2\jmath})u^{-1}_eu'_e|1,V_{12}\rangle},
 \end{eqnarray}
where we used the notation that $\tau^{1\imath}\pm\tau^{2\jmath}$ takes $\tau^{1\imath}+\tau^{2\jmath}$ if $\imath<\jmath$, and $\tau^{1\imath}-\tau^{2\jmath}$ if $\imath>\jmath$. Note that $\mathcal{T}_{\pm\imath\jmath,\pm\imath'\jmath'}(u^{-1}_eu'_e,\tilde{u}'^{-1}_e\tilde{u}_e)$ and $ \widetilde{\mathcal{T}}_{\pm\imath\jmath,\pm\imath'\jmath'}(u^{-1}_eu'_e,\tilde{u}'^{-1}_e\tilde{u}_e)$ satisfy
\begin{equation}\label{Tij1}
\mathcal{T}_{\pm\imath\jmath,\pm\imath'\jmath'}(u^{-1}_eu'_e,\tilde{u}'^{-1}_e\tilde{u}_e)=0,\ \ \widetilde{\mathcal{T}}_{\pm\imath\jmath,\pm\imath'\jmath'}(u^{-1}_eu'_e,\tilde{u}'^{-1}_e\tilde{u}_e)=0, \ \   \text{if}\ \  \widetilde{\Theta}_e=0
\end{equation}
 and
 \begin{equation}\label{Tij2}
\left|\mathcal{T}_{\pm\imath\jmath,\pm\imath'\jmath'}(u^{-1}_eu'_e,\tilde{u}'^{-1}_e\tilde{u}_e)\right|\leq 4e^{\widetilde{\Theta}_e},\quad \left|\widetilde{\mathcal{T}}_{\pm\imath\jmath,\pm\imath'\jmath'}(u^{-1}_eu'_e,\tilde{u}'^{-1}_e\tilde{u}_e)\right|\leq 4,
 \end{equation}
for $(\imath,\jmath),(\imath',\jmath')\in\{(3,4),(4,3),...,(D,D+1),(D+1,D)\}$.
Moreover, by recalling Eq.\eqref{deri}, we also have that $\left|\widetilde{\mathcal{T}}_{\pm\imath\jmath,\pm\imath'\jmath'}(u^{-1}_eu'_e,\tilde{u}'^{-1}_e\tilde{u}_e)\right|$ with $(\imath,\jmath),(\imath',\jmath')\in\{(3,4),(4,3),...,(D,D+1),(D+1,D)\}$ grows no faster than exponentials as ${\Theta}({u}_e,{u}'_e)$ or ${\Theta}(\tilde{u}_e,\tilde{u}'_e)$ going large.
Then, similar to the analysis for Eqs.\eqref{h21F}, \eqref{h21S} and \eqref{h21T} in Appendix \ref{app1}, we have
\begin{eqnarray}\label{TTFFhijij0}
&&\left|\text{FRHS\ of  Eq.}\eqref{h3456}\right|\\\nonumber
&\stackrel{\text{large}\ \eta'_e}{=}&e^{-\frac{(\eta_e)^2+(\eta'_e)^2+2t^2(D-1)^2}{4t}}\left|\langle\breve{\Psi}_{\mathbb{H}^o_e} |\breve{\Psi}_{\mathbb{H}'^o_e}\rangle \right|\cdot\left| \mathcal{T}_{\pm\imath\jmath,\pm\imath'\jmath'}(u^{-1}_eu'_e,\tilde{u}'^{-1}_e\tilde{u}_e)\right| (1+\mathcal{O}(t)+\mathcal{O}(e^{-1/t})).
 \end{eqnarray}
Finally, let us combine Eqs.\eqref{Tij1}, \eqref{Tij2}, and \eqref{TTFFhijij0}, and notice that $\frac{\langle\breve{\Psi}_{\mathbb{H}^o_e} |\breve{\Psi}_{\mathbb{H}'^o_e}\rangle}{||\breve{\Psi}_{\mathbb{H}'^o_e}||||\breve{\Psi}_{ \mathbb{H}^o_e}||}$ is unity at $\mathbb{H}'^o_e=\mathbb{H}^o_e$ and decaying exponentially fast to $0$ for $\mathbb{H}'^o_e\neq\mathbb{H}^o_e$, we get
\begin{eqnarray}\label{hijij}
&&\left| \frac{\langle\breve{\Psi}_{\mathbb{H}^o_e}|(\widehat{u'^{-1}_eh_e\tilde{u}'_e})_{\imath\jmath,\imath'\jmath'}  |\breve{\Psi}_{\mathbb{H}'^o_e}\rangle}{||\breve{\Psi}_{\mathbb{H}'^o_e}||||\breve{\Psi}_{ \mathbb{H}^o_e}||}\right|\stackrel{\text{large}\ \eta'_e}{\lesssim}t|\tilde{f}'''(\mathbb{H}^o_e,\mathbb{H}'^o_e)|\left|\frac{\langle\breve{\Psi}_{\mathbb{H}^o_e} |\breve{\Psi}_{\mathbb{H}'^o_e}\rangle}{||\breve{\Psi}_{\mathbb{H}'^o_e}||||\breve{\Psi}_{ \mathbb{H}^o_e}||}\right|
\end{eqnarray}
with $(\imath,\jmath),(\imath',\jmath')\in\{(3,4),(4,3),...,(D,D+1),(D+1,D)\}$,
where $\tilde{f}'''(\mathbb{H}^o_e,\mathbb{H}'^o_e)$ is a function whose growth is always suppressed by $\frac{\left|\langle\breve{\Psi}_{\mathbb{H}^o_e} |\breve{\Psi}_{\mathbb{H}'^o_e}\rangle\right|}{||\breve{\Psi}_{\mathbb{H}'^o_e}||||\breve{\Psi}_{ \mathbb{H}^o_e}||}$ exponentially as $|\eta_e-\eta'_e|, \Theta(u_e,u'_e)$ and $|\xi^o_e-\xi'^o_e|$  going large for large $\eta'_e$.

The rest components of holonomy operators are $(\widehat{u'^{-1}_eh_e\tilde{u}'_e})_{12,\imath\jmath}$ and $(\widehat{u'^{-1}_eh_e\tilde{u}'_e})_{21,\imath\jmath}$ and their transpositions $(\widehat{u'^{-1}_eh_e\tilde{u}'_e})_{\imath\jmath,12}$ and $(\widehat{u'^{-1}_eh_e\tilde{u}'_e})_{\imath\jmath,21}$ with $(\imath,\jmath)\in\{(3,4),(4,3),...,(D,D+1),(D+1,D)\}$. By using the similar techniques utilized in the calculations of Eqs.\eqref{h122100} and \eqref{h3456}, the matrix elements of these components of holonomy operators in twisted geometry coherent state basis can be evaluated and the results combining with Eqs.\eqref{h1212fina}, \eqref{h2121fina2}, \eqref{h1221} and \eqref{hijij} give the second main result of this paper as follows.

\begin{itemize}
  \item The matrix elements of the holonomy operators with respect to the twisted geometry coherent states can be estimated by
      \begin{eqnarray}\label{Hfinal}
\left|\frac{\langle\breve{\Psi}_{\mathbb{H}^o_e}|\widehat{u'^{-1}_eh_e\tilde{u}'_e} |\breve{\Psi}_{\mathbb{H}'^o_e}\rangle}{||\breve{\Psi}_{\mathbb{H}'^o_e}||||\breve{\Psi}_{ \mathbb{H}^o_e}||} -u'^{-1}_eh'^s_e\tilde{u}'_e\frac{\langle\breve{\Psi}_{\mathbb{H}^o_e} |\breve{\Psi}_{\mathbb{H}'^o_e}\rangle}{||\breve{\Psi}_{\mathbb{H}'^o_e}||||\breve{\Psi}_{ \mathbb{H}^o_e}||}\right|
\stackrel{\text{large}\ \eta'_e}{\lesssim}t\left|f_{h}(\mathbb{H}^o_e,\mathbb{H}'^o_e)\right|\cdot\left|\frac{\langle\breve{\Psi}_{\mathbb{H}^o_e} |\breve{\Psi}_{\mathbb{H}'^o_e}\rangle}{||\breve{\Psi}_{\mathbb{H}'^o_e}||||\breve{\Psi}_{ \mathbb{H}^o_e}||}\right|,
\end{eqnarray}
where $f_{h}(\mathbb{H}^o_e,\mathbb{H}'^o_e)$ is a function whose growth is always suppressed by $\frac{\left|\langle\breve{\Psi}_{\mathbb{H}^o_e} |\breve{\Psi}_{\mathbb{H}'^o_e}\rangle\right|}{||\breve{\Psi}_{\mathbb{H}'^o_e}||||\breve{\Psi}_{ \mathbb{H}^o_e}||}$ exponentially as $|\eta_e-\eta'_e|$ ,$\widetilde{\Theta}_e$ and $|\xi^o_e-\xi'^o_e|$ going large for large $\eta'_e$, and $u'^{-1}_eh'^s_e\tilde{u}'_e$ is defined by $\mathbb{H}'^o_e$ and its matrix elements in the definition representation space of $SO(D+1)$ is given as
\begin{eqnarray}\label{hs1}
&&(u'^{-1}_eh'^s_e\tilde{u}'_e)_{12,12}=e^{\mathbf{i}\xi'^o_e},\quad (u'^{-1}_eh'^s_e\tilde{u}'_e)_{21,21}=e^{-\mathbf{i}\xi'^o_e},\\\nonumber
&&(u'^{-1}_eh'^s_e\tilde{u}'_e)_{12,21}=(u'^{-1}_eh'^o_e\tilde{u}'_e)_{21,12}=0
\end{eqnarray}
and
\begin{eqnarray}\label{hs22}
&&(u'^{-1}_eh'^s_e\tilde{u}'_e)_{12,\imath\jmath} =(u'^{-1}_eh'^s_e\tilde{u}'_e)_{21,\imath\jmath}=0,\quad \text{for}\  (\imath,\jmath)\neq (1,2)\ \text{or}\  (2,1),\\\nonumber  &&(u'^{-1}_eh'^s_e\tilde{u}'_e)_{\imath\jmath,12}=(u'^{-1}_eh'^s_e\tilde{u}'_e)_{\imath\jmath,21}=0,
 \quad \text{for}\  (\imath,\jmath)\neq (1,2)\ \text{or}\  (2,1),\\\nonumber
&&(u'^{-1}_eh'^s_e\tilde{u}'_e)_{\imath\jmath,\imath'\jmath'}=0, \quad \text{for}\  (\imath,\jmath)\neq (1,2)\ \text{or}\  (2,1),
\end{eqnarray}
with $(u'^{-1}_eh'^s_e\tilde{u}'_e)_{\imath\jmath,\imath'\jmath'}:=\langle 1, V_{\imath\jmath}|u'^{-1}_eh'^s_e\tilde{u}'_e|1,V_{\imath'\jmath'}\rangle$.
\end{itemize}
Here we would like to emphasis that  the matrix elements of the holonomy operators $\hat{h}_e$ with respect to the twisted geometry coherent states are not estimated by the classical holonomies $h'_e=u'_ee^{\bar{\xi}'^\mu_e\bar{\tau}_\mu}e^{\xi'^o_e\tau_o}\tilde{u}'^{-1}_e$, but by the simplicity resolved holonomies $h'^s_e$ which are independent with the gauge component $e^{\bar{\xi}'^\mu_e\bar{\tau}_\mu}$.

We still need to discuss the case of $(D+1)$ being odd. In this case the above results still hold except that the equation \eqref{hijij} holds for $(\imath,\jmath)\   \text{and}\  (\imath',\jmath')\in \{(3,4),(4,3),...,(D-1,D),(D,D-1)\}$. Besides, there are extra holonomy operators $(\widehat{u_e^{-1}h_e\tilde{u}_e})_{\imath\jmath,(D+1)}$ and $(\widehat{u_e^{-1}h_e\tilde{u}_e})_{(D+1),\imath\jmath}$ with $(\imath,\jmath)\in \{(1,2),(2,1),(3,4),...,(D-1,D),(D,D-1)\}$ in this case, which are defined by
\begin{equation}
(\widehat{u_e^{-1}h_e\tilde{u}_e})_{\imath\jmath,(D+1)}:=\widehat{\langle1, V_{\imath\jmath}|u_e^{-1}h_e\tilde{u}_e|1,\delta_{D+1}\rangle}
\end{equation}
and
\begin{equation}
(\widehat{u_e^{-1}h_e\tilde{u}_e})_{(D+1),\imath\jmath}:=\widehat{\langle1,\delta_{D+1}|u_e^{-1}h_e\tilde{u}_e|1, V_{\imath\jmath}\rangle}
\end{equation}
respectively, where $|1,\delta_{D+1}\rangle$ is defined in Appendix \ref{app0}. Let us consider $(\widehat{u_e^{-1}h_e\tilde{u}_e})_{(D+1),\imath\jmath}$ as an example. Notice Eq.\eqref{holoopD} in Appendix \ref{app0}, we have
\begin{eqnarray}\label{h34D+1}
&& \langle\breve{\Psi}_{\mathbb{H}^o_e}|(\widehat{u'^{-1}_eh_e\tilde{u}'_e})_{D+1,\imath'\jmath'}  |\breve{\Psi}_{\mathbb{H}'^o_e}\rangle e^{-\frac{(\eta_e)^2+(\eta'_e)^2+2t^2(D-1)^2}{4t}}\\\nonumber
&{=}&-e^{\mathbf{i}\xi'^o_e} \sum_{N_e}\Big((\dim(\pi_{N_e}))^{3/2} (\dim(\pi_{N_e+1}))^{1/2} \exp(-t(\frac{\eta_e}{2t} -d_{N_e+1})^2 -t(\frac{\eta'_e}{2t}-d_{N_e})^2)\\\nonumber
&&\cdot e^{\mathbf{i}d_{N_e+1}(\xi^o_e-\xi'^o_e)}\frac{\sqrt{2}}{(N_e+1)}\langle N_e+1,V_{12}|u_e^{-1}u'_e\tau^{1,D+1}|N_e+1,V_{12}\rangle\\\nonumber
&&\cdot\langle N_e+1,V_{12}|\tilde{u}_e^{-1}\tilde{u}'_e|1, V_{\imath'\jmath'};N_e,V_{12}\rangle\Big) +\frac{1}{\sqrt{t}}\mathcal{O}(e^{-\frac{(\eta'_e)^2}{8t}})
\end{eqnarray}
for large $\eta'_e$. Note that $(\imath',\jmath')\in \{(1,2),(2,1),(3,4),...,(D-1,D),(D,D-1)\}$ in \text{Eq.}\eqref{h34D+1},
by using Eqs.\eqref{ccgg1}, \eqref{CG1}, \eqref{ccgg3} and \eqref{ccgg4}, the \text{Eq.}\eqref{h34D+1} can be further calculated case by case. The operator $(\widehat{u_e^{-1}h_e\tilde{u}_e})_{\imath\jmath,(D+1)}$ can be discussed similarly.
 Finally, we can conclude that Eqs.\eqref{Hfinal}, \eqref{hs1} and \eqref{hs22} still hold for the case of $(D+1)$ being odd if the equations
\begin{eqnarray}\label{hs33}
&&(u'^{-1}_eh'^s_e\tilde{u}'_e)_{\imath\jmath,(D+1)}:=\langle 1, V_{\imath\jmath}|u'^{-1}_eh'^s_e\tilde{u}'_e|1,\delta_{D+1}\rangle=0,\\\nonumber
&&(u'^{-1}_eh'^s_e\tilde{u}'_e)_{(D+1),\imath\jmath}:=\langle 1, \delta_{D+1} |u'^{-1}_eh'^s_e\tilde{u}'_e|1,V_{\imath\jmath}\rangle=0
\end{eqnarray}
with $(\imath,\jmath)\in \{(1,2),(2,1),(3,4),...,(D-1,D),(D,D-1)\}$
being added to Eq.\eqref{hs22}.
\subsection{Expectation values of non-polynomial operators}
Similar to that of the heat-kernel coherent state in $SU(2)$ LQG \cite{2000Gauge}, the expectation values of non-polynomial operators with respect to twisted geometry coherent state in all dimensional LQG can be studied by reformulating it as the Hamburger moment problem.\\
 \\
 \textbf{Theorem (Hamburger)}  Given a sequence of real numbers $a_n\in\mathbb{R}$, $n=0,1,2,...$ A sufficient and necessary condition for the existence of a positive, finite measure $d\rho(x)$ on $\mathbb{R}$ such that the $a_n$ are its moments, that is,
\begin{equation}
a_n=\int_{\mathbb{R}}d\rho(x)x^n
\end{equation}
 is that for arbitrary natural number $0\leq M<\infty$ and arbitrary complex numbers $z_i$, $i=0,...,M$ it holds that
 \begin{equation}\label{HT}
 \sum_{i,j=0}^M\bar{z}_iz_ja_{i+j}\geq0.
 \end{equation}
 The measure is faithful if equality in \eqref{HT} occurs only for $z_i=0$. Moreover, the measure $\rho$ is unique if there exist constants $\alpha,\beta>0$ such that $|a_n|\leq\alpha\beta^n(n!)$ for all $n$.\\
 \\ The proof of this theorem can be found in Refs. \cite{1975Methods}\cite{2000Gauge}.
 In this section, we consider the operators whose arbitrary powers are densely defined on a common domain. Then, by using the Hamburger theorem, we can extend the Theorem 3.6, Corollary 3.1, and Theorem 3.7 in the Ref.\cite{2000Gauge} which consider the heat-kernel coherent state in $SU(2)$ LQG, to the case of twisted geometry coherent state in all dimensional LQG. The result of this extension leads to the following three corollaries for the expectation values of non-polynomial operators with respect to twisted geometry coherent state.
\\ \\
\textbf{Corollary (i)} Consider a self-adjoint operator $\hat{O}=O((\hat{X}_e^{IJ}, \hat{h}_e)_{e\in\gamma})$ on $\mathcal{H}_\gamma$ which is constructed from $\{(\hat{X}_e^{IJ}, \hat{h}_e)|e\in\gamma\}$. Let $O=O(({X}_e^{IJ}, {h}_e)_{e\in\gamma})$ be its real valued classical counterpart. Define its real valued and simplicity resolved classical counterpart $O^{s}(\vec{\mathbb{H}}^o)=O(({X}_e^{IJ}(\mathbb{H}^o_e), {h}^{s}_e(\mathbb{H}^o_e))_{e\in\gamma})$ of $\hat{O}$ by replacing $({X}_e^{IJ}, {h}_e)$ with $({X}_e^{IJ}(\mathbb{H}^o_e), {h}^{s}_e(\mathbb{H}^o_e))$ in the expression of $O=O(({X}_e^{IJ}, {h}_e)_{e\in\gamma})$, where ${X}_e^{IJ}(\mathbb{H}^o_e)=\frac{1}{2}\eta_eV_e^{IJ}$ and ${h}^{s}_e(\mathbb{H}^o_e)$ is given by Eqs.\eqref{hs1}, \eqref{hs22} and \eqref{hs33}. Suppose that for every $n\in\mathbb{N}$
\begin{equation}
\lim_{t\rightarrow0}\langle\hat{O}^n\rangle^t_{\gamma,\vec{\mathbb{H}}^o}=(O^{s}(\vec{\mathbb{H}}^o))^n,
\end{equation}
where $\langle\hat{o}\rangle^t_{\gamma,\vec{\mathbb{H}}^o}:=\langle\phi^t_{\gamma,\vec{\mathbb{H}}^o}|\hat{o} |\phi^t_{\gamma,\vec{\mathbb{H}}^o}\rangle$ with $\vec{\mathbb{H}}^o:=\{...,\mathbb{H}_e^o,...\}_{e\in\gamma}$ and  $\phi^t_{\gamma,\vec{\mathbb{H}}^o}:=\breve{\Psi} _{\gamma,\vec{\mathbb{H}}_e^o}/{||\breve{\Psi}_{\gamma,\vec{\mathbb{H}}_e^o}||}$ being the normalized formulation of $\breve{\Psi} _{\gamma,\vec{\mathbb{H}}_e^o}$.
Then for arbitrary  Borel measurable function $f$ on $\mathbb{R}$ such that $\langle f(\hat{O})^\dagger f(\hat{O})\rangle^t_{\gamma,\vec{\mathbb{H}}^o}<\infty$ we have
\begin{equation}
\lim_{t\rightarrow0}\langle f(\hat{O})\rangle^t_{\gamma,\vec{\mathbb{H}}^o}=f(O^{s}(\vec{\mathbb{H}}^o)).
\end{equation}
\\
\textbf{Proof:} This corollary is a direct generalization of the Theorem 3.6 in Ref.\cite{2000Gauge} which consider the heat-kernel coherent state in $SU(2)$ LQG. Let us give the main idea of this proof as follows. Denoted by $E(x)$, $x\in\mathbb{R}$ the spectral projection of $\hat{O}$. Then, by assumption and the spectral theorem we have
\begin{equation}
\lim_{t\rightarrow0}\int_{\mathbb{R}}d\langle\phi^t_{\gamma,\vec{\mathbb{H}}^o}|E(x) |\phi^t_{\gamma,\vec{\mathbb{H}}^o}\rangle x^n=(O^{s}(\vec{\mathbb{H}}^o))^n.
\end{equation}
Define $a_n:=(O^{s}(\vec{\mathbb{H}}^o))^n$, it obviously  satisfies all the criteria of the Hamburger theorem and we conclude that there exists a measure $d\rho_{\vec{\mathbb{H}}^o}(x)$ on $\mathbb{R}$ satisfying
\begin{equation}\label{measure}
\int_{\mathbb{R}}d\rho_{\vec{\mathbb{H}}^o}(x) x^n=(O^{s}(\vec{\mathbb{H}}^o))^n.
\end{equation}
It is obviously that the Dirac measure $d\rho_{\vec{\mathbb{H}}^o}(x) =\delta_{\mathbb{R}}(x,O^s(\vec{\mathbb{H}}^o))dx$ satisfies \eqref{measure} and it satisfies the uniqueness part of the criterion by choosing $\alpha=1, \beta=|O^s(\vec{\mathbb{H}}^o)|$ in the Hamburger theorem. Hence we can conclude that  the Dirac measure is the unique solution of this moment problem and it follows that the spectral measure $d\rho^t_{\vec{\mathbb{H}}^o}(x):= d\langle\phi^t_{\gamma,\vec{\mathbb{H}}^o}|E(x) |\phi^t_{\gamma,\vec{\mathbb{H}}^o}\rangle$ approaches the Dirac measure in the limit $t\rightarrow0$. Now, for arbitrary  Borel measurable function $f$ on $\mathbb{R}$ such that $\langle f(\hat{O})^\dagger f(\hat{O})\rangle^t_{\gamma,\vec{\mathbb{H}}^o}<\infty$, the spectral theorem applies and one can get
 \begin{equation}
 \langle f(\hat{O})\rangle^t_{\gamma,\vec{\mathbb{H}}^o}=\int_{\mathbb{R}}d\rho^t_{\vec{\mathbb{H}}^o}(x)f(x),
 \end{equation}
 and then
 \begin{equation}
\lim_{t\rightarrow0}\langle f(\hat{O})\rangle^t_{\gamma,\vec{\mathbb{H}}^o}=\lim_{t\rightarrow0} \int_{\mathbb{R}}d\rho^t_{\vec{\mathbb{H}}^o}(x)f(x)=\int_{\mathbb{R}}dx\delta_{\mathbb{R}}(x,O^s(\vec{\mathbb{H}}^o))f(x) =f(O^s(\vec{\mathbb{H}}^o)).
\end{equation}
This finish the proof. $\square$
\\ \\
\textbf{Corollary (ii)} Consider the self-adjoint, not necessarily commuting, operators
\begin{equation}
\hat{O}_1=O_1((\hat{X}_e^{IJ}, \hat{h}_e)_{e\in\gamma}), ...,\hat{O}_m=O_m((\hat{X}_e^{IJ}, \hat{h}_e)_{e\in\gamma})
\end{equation}
on $\mathcal{H}_\gamma$ which is constructed from $\{(\hat{X}_e^{IJ}, \hat{h}_e)|e\in\gamma\}$. Let $O^{s}_1(\vec{\mathbb{H}}^o)=O_1(({X}_e^{IJ}(\mathbb{H}^o_e), {h}^{s}_e(\mathbb{H}^o_e))_{e\in\gamma})$,...,$O^{s}_m(\vec{\mathbb{H}}^o)=O_m( ({X}_e^{IJ}(\mathbb{H}^o_e), {h}^{s}_e(\mathbb{H}^o_e))_{e\in\gamma})$ be their real valued and simplicity resolved classical counterpart. Suppose that for every $n_k\in\mathbb{N}$
\begin{equation}
\lim_{t\rightarrow0}\langle\prod_{k=1}^m\hat{O}_k^{n_k}\rangle^t_{\gamma,\vec{\mathbb{H}}^o} =\prod_{k=1}^mO^{s}_k(\vec{\mathbb{H}}^o)^{n_k}.
\end{equation}
Then for arbitrary  Borel measurable function $f$ on $\mathbb{R}^m$ such that $\langle f(\{\hat{O}_k\}_{k=1}^m)^\dagger f(\{\hat{O}_k\}_{k=1}^m)\rangle^t_{\gamma,\vec{\mathbb{H}}^o}<\infty$ we have
\begin{equation}
\lim_{t\rightarrow0}\langle f(\{\hat{O}_k\}_{k=1}^m)\rangle^t_{\gamma,\vec{\mathbb{H}}^o}=f(\{{O}^{s}_k(\vec{\mathbb{H}}^o)\}_{k=1}^m).
\end{equation}
\\
\textbf{Proof:} Similar to the proof of \textbf{Corollary (i)}, this corollary can be proven directly by using the spectral theorem
\begin{equation}
\lim_{t\rightarrow0}\int_{\mathbb{R}}d^m\langle\phi^t_{\gamma,\vec{\mathbb{H}}^o}|E_1(x_1)...E_m(x_m) |\phi^t_{\gamma,\vec{\mathbb{H}}^o}\rangle \prod_{k=1}^mx_k^{n_k}=\prod_{k=1}^mO^{s}_k(\vec{\mathbb{H}}^o)^{n_k}
\end{equation}
and the uniqueness part of Hamburger theorem.
\\
\\
\textbf{Corollary (iii)} Consider the self-adjoint, not necessarily commuting, operators $\hat{O}_1=O_1((\hat{X}_e^{IJ}, \hat{h}_e)_{e\in\gamma})$ and $\hat{O}_2=O_2((\hat{X}_e^{IJ}, \hat{h}_e)_{e\in\gamma})$ on $\mathcal{H}_\gamma$ which is constructed from $\{(\hat{X}_e^{IJ}, \hat{h}_e)|e\in\gamma\}$. Let $O_1=O_1(({X}_e^{IJ}, {h}_e)_{e\in\gamma})$ and $O_2=O_2(({X}_e^{IJ}, {h}_e)_{e\in\gamma})$ be their real valued classical counterpart, and $O^{s}_1(\vec{\mathbb{H}}^o)=O_1(({X}_e^{IJ}(\mathbb{H}^o_e), {h}^{s}_e(\mathbb{H}^o_e))_{e\in\gamma})$ and $O^{s}_2(\vec{\mathbb{H}}^o)=O_2(({X}_e^{IJ}(\mathbb{H}^o_e), {h}^{s}_e(\mathbb{H}^o_e))_{e\in\gamma})$ be their real valued and simplicity resolved classical counterpart. Suppose that $\hat{O}_1$ is positive semi-definite and
\begin{equation}
\lim_{t\rightarrow0}\frac{\langle[\hat{O}_1,\hat{O}_2]\rangle^t_{\gamma,\vec{\mathbb{H}}^o} }{\mathbf{i}t} =O_{\{1,2\}}(({X}_e^{IJ}(\mathbb{H}^o_e), {h}^{s}_e(\mathbb{H}^o_e))_{e\in\gamma})
\end{equation}
with $O_{\{1,2\}}(({X}_e^{IJ}(\mathbb{H}^o_e), {h}^{s}_e(\mathbb{H}^o_e))_{e\in\gamma})$ being given by replacing $({X}_e^{IJ},{h}_e)$ with $({X}_e^{IJ}(\mathbb{H}^o_e), {h}^{s}_e(\mathbb{H}^o_e))$ in the expression of $O_{\{1,2\}}(({X}_e^{IJ}, {h}_e)_{e\in\gamma})$, where $O_{\{1,2\}}(({X}_e^{IJ}, {h}_e)_{e\in\gamma}):=\{O_1(({X}_e^{IJ}, {h}_e)_{e\in\gamma}),O_2(({X}_e^{IJ}, {h}_e)_{e\in\gamma})\}$ is the Poisson bracket between $O_1(({X}_e^{IJ}, {h}_e)_{e\in\gamma})$ and $O_2(({X}_e^{IJ}, {h}_e)_{e\in\gamma})$.
Then for arbitrary  real number $r$ we have
\begin{equation}
\lim_{t\rightarrow0}\frac{\langle[(\hat{O}_1)^r,\hat{O}_2]\rangle^t_{\gamma,\vec{\mathbb{H}}^o} }{\mathbf{i}t} =O_{\{1(r),2\}}(({X}_e^{IJ}(\mathbb{H}^o_e), {h}^{s}_e(\mathbb{H}^o_e))_{e\in\gamma})
\end{equation}
with $O_{\{1(r),2\}}(({X}_e^{IJ}(\mathbb{H}^o_e), {h}^{s}_e(\mathbb{H}^o_e))_{e\in\gamma})$ being given by replacing $({X}_e^{IJ},{h}_e)$ with $({X}_e^{IJ}(\mathbb{H}^o_e), {h}^{s}_e(\mathbb{H}^o_e))$ in the expression of $O_{\{1(r),2\}}(({X}_e^{IJ}, {h}_e)_{e\in\gamma})$, where $O_{\{1(r),2\}}(({X}_e^{IJ}, {h}_e)_{e\in\gamma}):=\{\left(O_1(({X}_e^{IJ}, {h}_e)_{e\in\gamma})\right)^r,O_2(({X}_e^{IJ}, {h}_e)_{e\in\gamma})\}$ is the Poisson bracket between $\left(O_1(({X}_e^{IJ}, {h}_e)_{e\in\gamma})\right)^r$ and $O_2(({X}_e^{IJ}, {h}_e)_{e\in\gamma})$.
\\
\\
\textbf{Proof:} Following the proof of Theorem 3.7 in Ref.\cite{2000Gauge}, this corollary can be proven similarly by using the completeness relation of the twisted geometry coherent states and applying the \textbf{Corollary (ii)}.
\section{Conclusion}\label{sec4}
The ``Ehrenfest property'' of the twisted geometry coherent state in all dimensional LQG is studied in this paper. Based on the completeness relation and the peakedness property of the twisted geometry coherent state, it is shown that in order to establish ``Ehrenfest property'' for polynomials of elementary operators, it is completely sufficient
to prove that the matrix elements of holonomy and flux operators in the twisted geometry coherent state basis
are estimated by their corresponding  classical values up to first order of $t$. Then, with the Clebsh-Gordan coefficients related to the states in the simple representation space of $SO(D+1)$ being given, we complete this proof by using the properties of the Perelomov type coherent states of $SO(D+1)$ and the Gaussian functions. Besides, it is shown that the expectation values of non-polynomial operators with respect to twisted geometry coherent state in all dimensional LQG  can be reformulated as the Hamburger moment problem. By extending the similar researches for the heat-kernel coherent state in $SU(2)$ LQG, we show that the ``Ehrenfest property'' for non-polynomial operators can be established at zeroth order of $t$. Therefore,  the twisted geometry coherent state provides us a reliable candidate for the study of the effective dynamics of all dimensional LQG.

\section*{Acknowledgments}
This work is supported by the project funded by China Postdoctoral Science Foundation  with Grant No. 2021M691072, and the National Natural Science Foundation of China (NSFC) with Grants No. 12047519, No. 11875006 and No. 11961131013.
\appendix
\section{The matrix elements of holonomy operator in spin-network basis}\label{app0}
In order to study the matrix elements of holonomy operator in spin-network basis in all dimensional LQG,
one need to consider the Clebsh-Gordan coefficient related to the states in the simple representation space of $SO(D+1)$.  Recall that the space $\mathfrak{H}^N_{D+1}$ of the sphere harmonic function on $S^D$ with degree $N$ is the simple representation space of $SO(D+1)$ labelled by $N$, and it has the orthonormal basis $\{\Xi^{N,\mathbf{M}}(\bm{x})\}$ (or $\{|N,\mathbf{M}\rangle\}$ in Dirac bracket formulation). Then, the Clebsh-Gordan coefficient can be given by
\begin{equation}\label{CGGENERAL}
\langle N', \mathbf{M}';N'', \mathbf{M}''|N, \mathbf{M}\rangle \langle N, \mathbf{0}|N', \mathbf{0};N'', \mathbf{0}\rangle =\dim(\pi_{N})\int_{SO(D+1)}dg \overline{D_{(\mathbf{M},\mathbf{0})}^N(g)}D_{(\mathbf{M'},\mathbf{0})}^{N'} (g) D_{(\mathbf{M}'', \mathbf{0})}^{N''}(g),
\end{equation}
where $\dim(\pi_{N})=\frac{(D+N-2)!(2N+D-1)}{(D-1)!N!}$, $|N', \mathbf{M}';N'', \mathbf{M}''\rangle:=|N', \mathbf{M}'\rangle\otimes|N'', \mathbf{M}''\rangle$ and
\begin{equation}
D_{(\mathbf{M},\mathbf{0})}^N(g):=\langle N,\mathbf{M}|g|N, \mathbf{0}\rangle
\end{equation}
is the matrix element function on $SO(D+1)$ selected by $|N,\mathbf{M}\rangle $ and $|N, \mathbf{0}\rangle$.
Based on Eq.\eqref{CGGENERAL}, it is easy to see that
\begin{equation}
 |\langle N+1, \mathbf{0}|N, \mathbf{0};1, \mathbf{0}\rangle|^2 =\dim(\pi_{N+1})\int_{SO(D+1)}dg \overline{D_{(\mathbf{0},\mathbf{0})}^{N+1}(g)}D_{(\mathbf{0},\mathbf{0})}^{N} (g) D_{(\mathbf{0}, \mathbf{0})}^{1}(g).
\end{equation}
Moreover, let us note that \cite{vilenkin2013representation}
\begin{equation}
D_{(\mathbf{0},\mathbf{0})}^{N} (g) =D_{(\mathbf{0},\mathbf{0})}^{N} (\theta)=\frac{(D-2)!N!}{(D+N-2)!}C_N^{\frac{D-1}{2}}(\cos\theta) ,\quad C_1^{\frac{D-1}{2}}(\cos\theta)=(D-1)\cos\theta,
\end{equation}
and
\begin{equation}
C_{N+1}^{\frac{D-1}{2}}(\cos\theta) =\frac{2N+D-1}{N+1}\cos\theta C_{N}^{\frac{D-1}{2}}(\cos\theta) -\frac{N+D-2}{N+1} C_{N-1}^{\frac{D-1}{2}}(\cos\theta).
\end{equation}
Then, we can calculate
\begin{equation}\label{0001}
 |\langle N+1, \mathbf{0}|N, \mathbf{0};1, \mathbf{0}\rangle|^2 =\frac{D+N-1}{2N+D-1},
\end{equation}
and similarly we have
\begin{equation}\label{0002}
 |\langle N-1, \mathbf{0}|N, \mathbf{0};1, \mathbf{0}\rangle|^2 =\frac{N}{2N+D-1},
\end{equation}
Futhermore, the relation between the function $D_{(\mathbf{M},\mathbf{0})}^N(g)$ and $\Xi^{N,\mathbf{M}}(\bm{x})$ can be found in Ref.\cite{vilenkin2013representation} and it leads
\begin{eqnarray}\label{NNNNNN}
&&\langle N', \mathbf{M}';N'', \mathbf{M}''|N, \mathbf{M}\rangle \langle N, \mathbf{0}|N', \mathbf{0};N'', \mathbf{0}\rangle\\\nonumber& =&\sqrt{\frac{\dim(\pi_{N})}{\dim(\pi_{N'})\dim(\pi_{N''})}}
\int_{SO(D+1)} d\bm{x}\overline{\Xi^{N,\mathbf{M}}(\bm{x})} {\Xi^{N',\mathbf{M}''}(\bm{x}) }{\Xi^{N'',\mathbf{M}''}(\bm{x})}.
\end{eqnarray}

Now let us turn to consider the functions which are involved in the main part of this article.
The normalized harmonic function $c_N(x_\imath+\mathbf{i}x_\jmath)$ can be denoted by $|N,V_{\imath\jmath}\rangle$ with $\imath, \jmath=1,...,D+1$, $\bm{x}=(x_1,...,x_{D+1})\in S^D$, $V_{\imath\jmath}:=2\delta_\imath^{[I}\delta_\jmath^{J]}$ and $c_N$ being the normalization factor.
A harmonic function basis of the definition representation space of $SO(D+1)$ can be given by,
\begin{equation}
(x_1+\mathbf{i}x_2),\quad (x_1-\mathbf{i}x_2), \quad (x_3+\mathbf{i}x_4),..., (x_{D}+\mathbf{i}x_{D+1}),\quad (x_{D}-\mathbf{i}x_{D+1})
\end{equation}
for $D+1$ being even, and
\begin{equation}
(x_1+\mathbf{i}x_2),\quad (x_1-\mathbf{i}x_2), \quad (x_3+\mathbf{i}x_4),..., (x_{D-1}+\mathbf{i}x_{D}),\quad (x_{D-1}-\mathbf{i}x_{D}),\quad x_{D+1}
\end{equation}
for $D+1$ being odd.
 These functions can be expressed as following normalized states by using Dirac bracket notation, which reads
\begin{equation}
\{|1, V_{\imath\jmath}\rangle\},\quad (\imath,\jmath)\in\{(1,2),(2,1),(3,4),(4,3),...,(D,D+1),(D+1,D)\}
\end{equation}
for $D+1$ being even, and
\begin{equation}
\{|1, V_{\imath\jmath}\rangle,\   |1,\delta_{D+1}\rangle\}, \quad (\imath,\jmath)\in\{(1,2),(2,1),(3,4),(4,3),...,(D-1,D),(D,D-1)\}
\end{equation}
for $D+1$ being odd.
First, one can check that
\begin{equation}\label{ccgg1}
|1, V_{12};N,V_{12}\rangle:=|1, V_{12}\rangle\otimes |N,V_{12}\rangle=| N+1, V_{12}\rangle.
\end{equation}
Then,  one can further calculate some of the other Clebsh-Gordan coefficients. Following Eqs.\eqref{NNNNNN} and \eqref{ccgg1}, let us consider
\begin{eqnarray}\label{N1212}
&&\langle N, V_{12};1, V_{12}|N+1, V_{12}\rangle \langle N+1, \mathbf{0}|N, \mathbf{0};1, \mathbf{0}\rangle=\langle N+1, \mathbf{0}|N, \mathbf{0};1, \mathbf{0}\rangle\\\nonumber
& =&\sqrt{\frac{\dim(\pi_{N+1})}{(D+1)\cdot\dim(\pi_{N}) }}
\int_{SO(D+1)} d\bm{x}\overline{\Xi^{N+1,V_{12}}(\bm{x})} {\Xi^{N,V_{12}}(\bm{x}) }{\Xi^{1,V_{12}}(\bm{x})},
\end{eqnarray}
and
\begin{eqnarray}\label{N1221}
&&\langle N+1, V_{12};1, V_{21}|N, V_{12}\rangle \langle N, \mathbf{0}|N+1, \mathbf{0};1, \mathbf{0}\rangle\\\nonumber
& =&\sqrt{\frac{\dim(\pi_{N})}{(D+1)\cdot\dim(\pi_{N+1}) }}
\int_{SO(D+1)} d\bm{x}\overline{\Xi^{N,V_{12}}(\bm{x})} {\Xi^{N+1,V_{12}}(\bm{x}) }{\Xi^{1,V_{21}}(\bm{x})}.
\end{eqnarray}
By substituting \eqref{0001} into \eqref{N1212} we get
\begin{equation}\label{cccN}
\frac{c_Nc_{1}}{c_{N+1}}=\langle N+1, \mathbf{0}|N, \mathbf{0};1, \mathbf{0}\rangle\sqrt{\frac{(D+1)\cdot\dim(\pi_{N}) }{\dim(\pi_{N+1})}}=e^{-\mathbf{i}\alpha}\sqrt{\frac{(N+1)(D+1)}{2N+D+1}}.
\end{equation}
Notice that $\Xi^{1,V_{21}}(\bm{x})=\overline{\Xi^{1,V_{12}}(\bm{x})}$, and then Eqs.\eqref{N1212} and \eqref{N1221} give
\begin{equation}\label{CGooooo}
\langle N, V_{12}|N+1, V_{12};1, V_{21}\rangle=\frac{\dim(\pi_{N})}{\dim(\pi_{N+1}) }\frac{\langle N+1, \mathbf{0}|N, \mathbf{0};1, \mathbf{0}\rangle}{\langle N+1, \mathbf{0};1, \mathbf{0}|N, \mathbf{0}\rangle}.
\end{equation}
Let us denote
\begin{eqnarray}\label{CG1}
|1, V_{21};N,V_{12}\rangle& :=&|1, V_{21}\rangle\otimes |N,V_{12}\rangle\\\nonumber
 &=&\alpha_1(N)| N-1, V_{12}\rangle +\alpha_2(N)|N+1,...\rangle+\alpha_3(N)|\text{not\ simple}\rangle,
\end{eqnarray}
where $|N+1,...\rangle$ is a state in the simple representation space $\mathfrak{H}^{N+1}_{D+1}$ and $|\text{not\ simple}\rangle$ is a state not belonging to any simple representation. By using Eq.\eqref{CGooooo}, one can get
\begin{eqnarray}
|\alpha_1(N)|^2&=&\left|\frac{\dim(\pi_{N-1})}{\dim(\pi_{N}) }\frac{\langle N, \mathbf{0}|N-1, \mathbf{0};1, \mathbf{0}\rangle}{\langle N, \mathbf{0};1, \mathbf{0}|N-1, \mathbf{0}\rangle}\right|^2
\\\nonumber
&=&\frac{N(2N+D-3)}{(D+N-2)(2N+D-1)}
\end{eqnarray}
and $|\alpha_2(N)|^2\leq1-|\alpha_1(N)|^2$.

We are also interested in the special Clebsh-Gordan coefficient $\langle1, V_{\imath\jmath};N, V_{12}|N', V'\rangle$ with $(\imath,\jmath)\in \{(3,4),(4,3),...\}$, which can be given by
\begin{eqnarray}\label{propto}
&&\langle1, V_{\imath\jmath};N, V_{12}|N', V'\rangle \langle N', \mathbf{0}|1, \mathbf{0};N, \mathbf{0}\rangle\\\nonumber& =&\sqrt{\frac{\dim(\pi_{N'})}{(D+1)\cdot\dim(\pi_{N})}}
\int_{SO(D+1)} d\bm{x}\overline{\Xi^{N',V'}(\bm{x})} \Xi^{1,V_{\imath\jmath}}(\bm{x}) {\Xi^{N,V_{12}}(\bm{x})}.
\end{eqnarray}
Notice that we have the relation
\begin{equation}
\Xi^{1,V_{\imath\jmath}}(\bm{x}) {\Xi^{N,V_{12}}(\bm{x})}=-\frac{c_Nc_{1}}{c_{N+1}}\frac{1}{N+1}(\tau^{1\imath}+\tau^{2\jmath}){\Xi^{N+1,V_{12}}(\bm{x})},\quad \text{for}\ \imath<\jmath, \  \text{and} \  \imath,\jmath\neq1,2,
\end{equation}
\begin{equation}
\Xi^{1,V_{\imath\jmath}}(\bm{x}) {\Xi^{N,V_{12}}(\bm{x})}=-\frac{c_Nc_{1}}{c_{N+1}}\frac{1}{N+1}(\tau^{1\imath}-\tau^{2\jmath}){\Xi^{N+1,V_{12}}(\bm{x})},\quad \text{for}\ \imath>\jmath, \  \text{and} \  \imath,\jmath\neq1,2.
\end{equation}
Then, by using above equations and substituting \eqref{0001} and \eqref{cccN} into equation \eqref{propto}, we can check that
\begin{eqnarray}\label{ccgg3}
&&\langle1, V_{\imath\jmath};N, V_{12}|N', V'\rangle  \\\nonumber &=&-\frac{1}{(N+1)}
\langle N',V'|(\tau^{1\imath}+\tau^{2\jmath})|N+1,V_{12}\rangle,\  \text{for}\ \imath<\jmath, \  \text{and} \  \imath,\jmath\neq1,2,
\end{eqnarray}
and
\begin{eqnarray}\label{ccgg4}
&&\langle1, V_{\imath\jmath};N, V_{12}|N', V'\rangle  \\\nonumber &=&-\frac{1}{(N+1)}
\langle N',V'|(\tau^{1\imath}-\tau^{2\jmath})|N+1,V_{12}\rangle,\  \text{for}\ \imath>\jmath, \  \text{and} \  \imath,\jmath\neq1,2.
\end{eqnarray}
Similarly, we can also consider
\begin{eqnarray}
&&\langle1,\delta_{D+1};N, V_{12}|N', V'\rangle \langle N', \mathbf{0}|1, \mathbf{0};N, \mathbf{0}\rangle\\\nonumber& =&\sqrt{\frac{\dim(\pi_{N'})}{(D+1)\cdot\dim(\pi_{N})}}
\int_{SO(D+1)} d\bm{x}\overline{\Xi^{N',V'}(\bm{x})} \Xi^{1,\delta_{D+1}}(\bm{x}) {\Xi^{N,V_{12}}(\bm{x})},
\end{eqnarray}
where $\Xi^{1,\delta_{D+1}}(\bm{x}) =\sqrt{D+1}x_{D+1}$. Note that
\begin{equation}
\Xi^{1,\delta_{D+1}}(\bm{x}) {\Xi^{N,V_{12}}(\bm{x})}=-\frac{\sqrt{2}c_1c_N}{c_{N+1}}\frac{1}{N+1}\tau^{1,D+1} {\Xi^{N+1,V_{12}}(\bm{x})},
\end{equation}
with $c_1=\sqrt{\frac{D+1}{2}}.$ Then, we have
\begin{eqnarray}\label{ccgg5}
\langle1,\delta_{D+1};N, V_{12}|N', V'\rangle =-\frac{\sqrt{2}}{(N+1)}
\langle N',V'|\tau^{1,D+1} |N+1,V_{12}\rangle.
\end{eqnarray}

Now, we can consider the holonomy operator $\hat{h}$ acting on the matrix element function $\Xi^{N}_{u^{-1},\tilde{u}}(h):=\langle N,V_{12}|u^{-1}h\tilde{u}|N,V_{12}\rangle$. For the case of $D+1$ being even, the action of the holonomy operator corresponding to the holonomy component $\langle 1, V_{\imath\jmath}|u^{-1}h\tilde{u}|1, V_{\imath'\jmath'}\rangle$ is given by
\begin{eqnarray}
\widehat{\langle 1, V_{\imath\jmath}|u^{-1}{h}\tilde{u}|1, V_{\imath'\jmath'}\rangle} \circ\Xi^{N}_{u^{-1},\tilde{u}}(h)&:=&\langle 1, V_{\imath\jmath}|u^{-1}h\tilde{u}|1, V_{\imath'\jmath'}\rangle \cdot \Xi^{N}_{u^{-1},\tilde{u}}(h)\\\nonumber
&=&\langle 1, V_{\imath\jmath}|u^{-1}h\tilde{u}|1, V_{\imath'\jmath'}\rangle \cdot \langle N,V_{12}|u^{-1}h\tilde{u}|N,V_{12}\rangle\\\nonumber
&=& \langle N,V_{12};1, V_{\imath\jmath}|u^{-1}h\tilde{u}|N,V_{12};1, V_{\imath'\jmath'}\rangle
\end{eqnarray}
with $(\imath,\jmath),(\imath',\jmath')\in \{(1,2),(2,1),...\}$.
Based on this action, the matrix elements of the operator $\widehat{\langle 1,V_{\imath\jmath}|u^{-1}h\tilde{u}|1, V_{\imath'\jmath'}\rangle}$ in the basis spanned by the states $\left|N,u^{-1},\tilde{u}\right\rangle$ corresponding to $\Xi^{N}_{u^{-1},\tilde{u}}(h)$ can be given by
\begin{eqnarray}\label{holoop}
&& \left\langle N',u'^{-1},\tilde{u}'\left|\widehat{\langle 1, V_{\imath\jmath}|u^{-1}h\tilde{u}|1, V_{\imath'\jmath'}\rangle}\right|N,u^{-1},\tilde{u}\right\rangle
  \\\nonumber&:=&\int_{SO(D+1)}dh \overline{\langle N',V_{12}|u'^{-1}h\tilde{u}'|N',V_{12}\rangle} \cdot \langle 1, V_{\imath\jmath}|u^{-1}h\tilde{u}|1,V_{\imath'\jmath'}\rangle \cdot \langle N,V_{12}|u^{-1}h\tilde{u}|N,V_{12}\rangle \\\nonumber
   &=&\frac{1}{\dim(\pi_{N'})}\langle1, V_{\imath\jmath};N,V_{12}| u^{-1}u'|N',V_{12}\rangle\cdot \langle N',V_{12}|\tilde{u}'^{-1}\tilde{u}|1, V_{\imath'\jmath'};N,V_{12}\rangle,
\end{eqnarray}
where $\dim(\pi_N)=\frac{(D+N-2)!(2N+D-1)}{(D-1)!N!}$. Then, by using Eqs.\eqref{ccgg1}, \eqref{CG1}, \eqref{ccgg3}, \eqref{ccgg4} and \eqref{ccgg5}, the \text{Eq.}\eqref{holoop} can be further calculated and the results are obvious. For instance, we have
\begin{eqnarray}
 && \text{Eq.}\eqref{holoop}\\\nonumber
  &{=}& \frac{1}{\dim(\pi_{N+1})}\langle N+1,V_{12}| u^{-1}u'|N',V_{12}\rangle\langle N',V_{12}|\tilde{u}'^{-1}\tilde{u}|N+1,V_{12}\rangle,\quad \text{if}\ (\imath,\jmath)=(1,2), (\imath',\jmath') =(1,2),
\end{eqnarray}
\begin{eqnarray}
 && \text{Eq.}\eqref{holoop}\\\nonumber
  &{=}& \frac{1}{\dim(\pi_{N-1})}|\alpha_1(N)|^2\langle N-1,V_{12}| u^{-1}u'|N',V_{12}\rangle\langle N',V_{12}|\tilde{u}'^{-1}\tilde{u}|N-1,V_{12}\rangle\\\nonumber
 && + \frac{1}{\dim(\pi_{N+1})}|\alpha_2(N)|^2\langle N+1,...| u^{-1}u'|N',V_{12}\rangle\langle N',V_{12}|\tilde{u}'^{-1}\tilde{u}|N+1,...\rangle,\\\nonumber
  &&\quad \quad \quad \quad \quad \quad \quad \quad\quad \quad \quad \quad\quad \quad\quad \quad \quad \quad \quad \quad \quad \quad \quad\quad \quad \quad\text{if}\ (\imath,\jmath)=(2,1), (\imath',\jmath') =(2,1),
\end{eqnarray}
and
 \begin{eqnarray}\label{holotype3}
 && \text{Eq.}\eqref{holoop}\\\nonumber
  &=&- \frac{1}{\dim(\pi_{N+1})}\frac{1}{(N+1)^2}\langle N',V_{12}|u'^{-1}u(\tau^{1\imath}\pm\tau^{2\jmath})|N+1,V_{12}\rangle\langle N+1,V_{12}|(\tau^{1\imath'}\pm\tau^{2\jmath'})\tilde{u}^{-1}\tilde{u}'|N',V_{12}\rangle,\\\nonumber
  &&\quad \quad \quad \quad \quad \quad \quad \quad\quad \quad\quad  \quad \quad\quad \quad \quad \quad \quad \quad \quad \quad \quad\quad \quad\quad  \quad \quad\quad  \text{if}\  \imath,\jmath,\imath',\jmath',\neq1,2,
\end{eqnarray}
where $\tau^{1\imath}\pm\tau^{2\jmath}$ takes $\tau^{1\imath}+\tau^{2\jmath}$ if $\imath<\jmath$, and $\tau^{1\imath}-\tau^{2\jmath}$ if $\imath>\jmath$.

For the case of $D+1$ being odd, the previous discussion of the operator $\widehat{\langle 1, V_{\imath\jmath}|u^{-1}{h}\tilde{u}|1, V_{\imath'\jmath'}\rangle} $ with $(\imath,\jmath),(\imath',\jmath')\in \{(1,2),(2,1),...\}$ still holds. Besides, we have the extra holonomy operator $\widehat{\langle 1, \delta_{D+1}|u^{-1}{h}\tilde{u}|1, V_{\imath'\jmath'}\rangle}$ and $\widehat{\langle 1, V_{\imath\jmath}|u^{-1}{h}\tilde{u}|1, \delta_{D+1}\rangle}$. Let us consider $\widehat{\langle 1, \delta_{D+1}|u^{-1}{h}\tilde{u}|1, V_{\imath'\jmath'}\rangle}$ as an example, whose action on $\Xi^{N}_{u^{-1},\tilde{u}}(h)$ is given by
\begin{eqnarray}
\widehat{\langle 1, \delta_{D+1}|u^{-1}{h}\tilde{u}|1, V_{\imath'\jmath'}\rangle} \circ\Xi^{N}_{u^{-1},\tilde{u}}(h)&:=&\langle 1, \delta_{D+1}|u^{-1}{h}\tilde{u}|1, V_{\imath'\jmath'}\rangle \cdot \Xi^{N}_{u^{-1},\tilde{u}}(h)\\\nonumber
&=&\langle 1, \delta_{D+1}|u^{-1}{h}\tilde{u}|1, V_{\imath'\jmath'}\rangle \cdot \langle N,V_{12}|u^{-1}h\tilde{u}|N,V_{12}\rangle\\\nonumber
&=& \langle N,V_{12};1, \delta_{D+1}|u^{-1}h\tilde{u}|N,V_{12};1, V_{\imath'\jmath'}\rangle
\end{eqnarray}
with $(\imath,\jmath),(\imath',\jmath')\in \{(1,2),(2,1),...\}$. Then, similar to \text{Eq.}\eqref{holoop}, we have
\begin{eqnarray}\label{holoopD}
&& \left\langle N',u'^{-1},\tilde{u}'\left|\widehat{\langle 1, \delta_{D+1}|u^{-1}h\tilde{u}|1, V_{\imath'\jmath'}\rangle}\right|N,u^{-1},\tilde{u}\right\rangle
  \\\nonumber
  &:=&\int_{SO(D+1)}dh \overline{\langle N',V_{12}|u'^{-1}h\tilde{u}'|N',V_{12}\rangle} \cdot \langle 1, \delta_{D+1}|u^{-1}h\tilde{u}|1,V_{\imath'\jmath'}\rangle \cdot \langle N,V_{12}|u^{-1}h\tilde{u}|N,V_{12}\rangle \\\nonumber
   &=&\frac{1}{\dim(\pi_{N'})}\langle1, \delta_{D+1};N,V_{12}| u^{-1}u'|N',V_{12}\rangle\cdot \langle N',V_{12}|\tilde{u}'^{-1}\tilde{u}|1, V_{\imath'\jmath'};N,V_{12}\rangle\\\nonumber
   &=&-\frac{1}{\dim(\pi_{N+1})}\frac{\sqrt{2}}{(N+1)}\delta_{N',N+1}
\langle N+1,V_{12}|u'^{-1}u\tau^{1,D+1}|N+1,V_{12}\rangle\cdot \langle N+1,V_{12}|\tilde{u}'^{-1}\tilde{u}|1, V_{\imath'\jmath'};N,V_{12}\rangle.
\end{eqnarray}
By using Eqs.\eqref{ccgg1}, \eqref{CG1}, \eqref{ccgg3} and \eqref{ccgg4}, the \text{Eq.}\eqref{holoopD} can be further calculated and the results are obvious.
\section{Poisson Summation Formula and relevant calculations}\label{app1}
The Poisson summation formula is used multiple times in this article. Let us introduce it and show the details of its application here. Let $f$ be an function in $L_1(\mathbb{R}, dx)$ such that the series
\begin{equation}
\phi(y)=\sum_{n=-\infty}^{\infty}f(y+ns)
\end{equation}
is absolutely and uniformly convergent for $y\in [0, s], s > 0$. Then
\begin{equation}
  \sum_{n=-\infty}^{\infty}f(ns)=\frac{2\pi}{s}\sum_{n=-\infty}^\infty\tilde{f}(\frac{2\pi n}{s}),
\end{equation}
where $  \tilde{f}(k):=\int_{\mathbf{R}}\frac{dx}{2\pi}e^{-\mathbf{i}kx}f(x)$ is the Fourier transform of $f$. The proof of this theorem can be found in \cite{Poissonsummation}. In this paper, the application of the Poisson summation formula is involved in the calculation of the expression
\begin{equation}
\sum_{N=0}^{\infty}F(N)e^{\mathbf{i}N\xi}\exp(-2t(\alpha_\eta-N)^2),
\end{equation}
where we defined $\alpha_\eta:=c_1\frac{\eta}{t}+c_2>0$, where $c_1$ and $c_2$ are constants satisfying $c_1>0$ and $|c_2|\ll c_1\frac{\eta}{t}$.
Now let us consider three cases of $F(x)$ separately. \\
\textbf{Case I:} $F(x)=x^\ell$ is a polynomial with $\ell\in\mathbb{N}$.  Consider the following calculations.
\begin{eqnarray}\label{xell}
&&\sum_{N=0}^{\infty}F(N)e^{\mathbf{i}N\xi}\exp(-2t({N}-\alpha_\eta)^2)\\\nonumber
&=&e^{\mathbf{i}\alpha_\eta\xi}\sum_{k={-\alpha_\eta}}^{\infty}(\alpha_\eta+k)^\ell\exp(-2tk^2)e^{\mathbf{i}k\xi}\\\nonumber
&=&(\alpha_\eta)^\ell e^{\mathbf{i}\alpha_\eta\xi} \sum_{[k]=[-\alpha_\eta]}^{\infty}(1+\frac{([k]+r)}{\alpha_\eta})^\ell\exp(-2t([k]+r)^2)e^{\mathbf{i}([k]+r)\xi},
\end{eqnarray}
where  $k:={N}-\alpha_\eta$, $r=-\alpha_\eta-[-\alpha_\eta]$ and $[k]$ represents the maximal integer no greater than $k$. Note that we have
\begin{eqnarray}
&&\sum_{m=-\infty}^{\infty}(m+r)^{\ell'} e^{(-2t(m+r)^2)}e^{\mathbf{i}(m+r)\xi} =\sum_{m=-\infty}^{\infty}e^{2\pi\mathbf{i}mr}\sqrt{\frac{\pi}{2t}}\check{P}(2\pi m-\xi)e^{-\frac{( 2\pi m-\xi)^2}{8t}},\  \ell'\in\mathbb{N},
\end{eqnarray}
here we defined $\check{P}(x):=(\mathbf{i})^{\ell'}(\frac{d^{\ell'}} {dx^{\ell'}}e^{-\frac{( x)^2}{8t}})e^{\frac{( x)^2}{8t}}$.
Then by setting $[k]=m$ in Eq.\eqref{xell}, we have
\begin{eqnarray}
\sum_{N=0}^{\infty}F(N)\exp(-2t(\alpha_\eta-N)^2)e^{\mathbf{i}N\xi}
\xlongequal[ ]{\text{ large}\ \! \eta}\sqrt{\frac{\pi}{2t}}(\alpha_\eta)^\ell e^{\mathbf{i}\alpha_\eta\xi} e^{-\frac{(\xi)^2}{8t}} \left(1+\mathcal{O}(\frac{t}{\eta}) +\mathcal{O}(e^{-\frac{\pi^2}{2t}})\right).
\end{eqnarray}
\\ \textbf{Case II:} $F(x)=(P(x))^{1/4}$, where $P(x)$ is a polynomial of $x$ with $P(x)>0$ if $x>0$. Let us focus on the case of $\alpha_\eta>0, x>0$ involved in this paper. We can reformulate $F(x)$ as $F(x)=(P(\alpha_\eta))^{1/4}f(z)$ with $f(z):=(1+z)^{1/4}$ and $z:=\frac{P(x)}{P(\alpha_\eta)}-1>-1$. Then by Taylor's theorem, we have
\begin{equation}
f(z)=1+\sum_{n=1}^{\infty}\left(\begin{array}{lr}
q \\
 n
\end{array}
\right)z^n,\quad  \left(\begin{array}{lr}
q \\
 n
\end{array}
\right)=(-1)^{n+1}\frac{q(1-q)...(n-1+q)}{n!}
\end{equation}
with $q=1/4$ here.
To proceed next step of the calculation, we introduce a lemma as follows.\\
\\ \textbf{Lemma.} For each $l\geq0$ there exist $0<\beta_{l}<\infty$ such that
\begin{equation}
f_{2l+1}(z)-\beta_lz^{2l+2}\leq f(z)\leq f_{2l+1}(z),
\end{equation}
where $f_{l}(z)=1+\sum_{n=1}^{l}\left(\begin{array}{lr}
q \\
 n
\end{array}
\right)z^n,\quad  \left(\begin{array}{lr}
q \\
 n
\end{array}
\right)$ denotes the partial Taylor series of $f(z)=(1+z)^q$, $0<q\leq 1/4$, up to order $z^k$.
\\
The proof of this lemma can be find in \cite{Giesel_2007}. \\
\\
Now let us set $x=N$ in $P(x)$ so that $z=\frac{P(N)}{P(\alpha_\eta)}-1$. Then by using the results of \textbf{Case I}, we can give
 \begin{eqnarray}
\sqrt{\frac{2t}{\pi}}\sum_{N=0}^{\infty}z^{\ell'}\exp(-2t(\alpha_\eta-N)^2)e^{\mathbf{i}N\xi} e^{\frac{(\xi)^2}{8t}}e^{-\mathbf{i}\alpha_\eta\xi}
\xlongequal[
 ]{\text{ large}\ \! \eta}\mathcal{O}(\frac{t}{\eta}) +\mathcal{O}(e^{-\frac{\pi^2}{2t}}),\ \text{for} \ \ell'\in\mathbb{N}_+.
\end{eqnarray}
Further by using the above \textbf{Lemma}, we get
\begin{eqnarray}
\sqrt{\frac{2t}{\pi}}\sum_{N=0}^{\infty}F(N)\exp(-2t(\alpha_\eta-N)^2)e^{\mathbf{i}N\xi}
\xlongequal[]{\text{ large}\ \! \eta}(P(\alpha_\eta))^{1/4}e^{\mathbf{i}\alpha_\eta\xi} e^{-\frac{(\xi)^2}{8t}} (1+\mathcal{O}(\frac{t}{\eta}) +\mathcal{O}(e^{-\frac{\pi^2}{2t}})).
\end{eqnarray}
\\ \textbf{Case III:} $F(x)=(P_1(x))^{1/4}(\frac{P_2(x)}{P_3(x)})^{\frac{1}{m}}$, $m\in\mathbb{N}_+$, where $P_1(x)$ is a polynomial of $x$ with degree larger than $4$ and it satisfies $P_1(x)\geq0$, $\frac{d}{dx}({P_1(x)})>0$ for $x\geq0$, $P_2(x)$ and $P_3(x)$ are both polynomials of $x$ which satisfy $P_3(x)>P_2(x)\geq0$ for $x\geq0$, $\frac{d}{dx}(\frac{P_2(x)}{P_3(x)})<0$ for $x\geq1$ and the degree of $P_3(x)$ larger than that of $P_2(x)$. 
We can evaluate that
\begin{eqnarray}
 &&\sqrt{\frac{2t}{\pi}}\left|\sum_{N=0}^{\infty}F(N)\exp(-2t(\alpha_\eta-N)^2) e^{\mathbf{i}N\xi}\right|\\\nonumber
   &<&\sqrt{\frac{2t}{\pi}}\sum_{N=0}^{\infty}F(N)\exp(-2t(\alpha_\eta-N)^2)\\\nonumber
   &<& \sqrt{\frac{2t}{\pi}}\left((P_1(\frac{\alpha_\eta}{2}))^{1/4}\sum_{N=0}^{[\frac{\alpha_\eta}{2}]} \exp(-2t(\alpha_\eta-N)^2)+ \sum_{N=[\frac{\alpha_\eta}{2}]+1}^{\infty}(P_1(N))^{1/4}\frac{P_2(\frac{\alpha_\eta}{2}+1)} {P_3(\frac{\alpha_\eta}{2}+1)}\exp(-2t(\alpha_\eta-N)^2)\right) \\\nonumber
   &{\lesssim}&   \sqrt{\frac{2t}{\pi}}([\frac{\alpha_\eta}{2}]+1)(P_1(\alpha_\eta))^{1/4}\exp(-\frac{1}{2}t(\alpha_\eta)^2)+ P_1(\alpha_\eta)^{1/4}\left(\frac{P_2(\frac{\alpha_\eta}{2}+1)} {P_3(\frac{\alpha_\eta}{2}+1)}\right)^{\frac{1}{m}}(1+\mathcal{O}(\frac{t}{\eta}) +\mathcal{O}(e^{-\frac{\pi^2}{2t}}))\\\nonumber
   &{\simeq}&(P_1(\alpha_\eta))^{1/4}(\alpha_\eta\sqrt{t}\mathcal{O}(e^{-\frac{\eta^2}{8t}})+\mathcal{O}((\frac{t}{\eta})^{\frac{1}{m}}))
\end{eqnarray}
for large $\eta$.

Based on the three cases discussed above, we can further calculate the specific equations \eqref{expectah1212}, \eqref{h212121}, and \eqref{h122100} appeared in the main part of this paper. \\
\\ \textbf{Calculation of Eq.\eqref{expectah1212}:}
We also discuss two cases separately. In the first case, we consider $\widetilde{\Theta}_e\gg \eta_e+\eta'_e$ or $\widetilde{\Theta}_e\simeq \eta_e+\eta'_e$. Similar to the analysis of Eqs.\eqref{F1212000} and \eqref{F121212}, we have
\begin{eqnarray}\label{H1212000}
&& \langle\breve{\Psi}_{\mathbb{H}^o_e}|(\widehat{u'^{-1}_e h_e\tilde{u}'_e})_{12,12}  |\breve{\Psi}_{\mathbb{H}'^o_e}\rangle e^{-\frac{(\eta_e)^2+(\eta'_e)^2+2t^2(D-1)^2}{4t}}\\\nonumber
&{=}&e^{\mathbf{i}\xi'^o_e} \sum_{N_e}(\dim(\pi_{N_e}))^{3/2}(\dim(\pi_{N_e+1}))^{1/2} \exp(-t(\frac{\eta_e}{2t} -d_{N_e+1})^2 -t(\frac{\eta'_e}{2t}-d_{N_e})^2)\\\nonumber
&&\cdot e^{\mathbf{i}d_{N_e+1}(\xi^o_e-\xi'^o_e)}\langle N_e+1, V'_e|N_e+1, V_e\rangle\langle N_e+1, -\tilde{V}_e|N_e+1,-\tilde{ V}'_e\rangle+\frac{1}{\sqrt{t}}\mathcal{O}(e^{-\frac{(\eta'_e)^2}{8t}})\\\nonumber
&{<}&
([\eta'_e/4t]+1) \exp(-t(\frac{\eta'_e}{4t}-\frac{D+1}{2})^2 -t(\frac{\eta_e}{2t}-\frac{\eta'_e}{4t}-\frac{D-1}{2})^2)(\breve{\text{P}}(\frac{\eta'_e}{4t}+1))^{1/2}(\breve{\text{P}}(\frac{\eta'_e}{4t}))^{3/2} \\\nonumber
&&+\sum_{N_e=[\frac{\eta'_e}{4t}]+1}^{+\infty} (\dim(\pi_{N_e}))^{3/2}(\dim(\pi_{N_e+1}))^{1/2}\left(\exp(-t(\frac{\eta'_e}{2t}-d_{N_e+1})^2 -t(\frac{\eta_e}{2t}-d_{N_e})^2) e^{- [\frac{\eta'_e}{4t}]\widetilde{\Theta}_e}\right)\\\nonumber
&&+ \frac{1}{\sqrt{t}}\cdot\mathcal{O}(e^{-\frac{\eta'^2_e}{8t}})
\\\nonumber
&{\simeq}&([\eta'_e/4t]+1) \exp(-t(\frac{\eta'_e}{4t}-\frac{D+1}{2})^2 -t(\frac{\eta_e}{2t}-\frac{\eta'_e}{4t}-\frac{D-1}{2})^2)(\breve{\text{P}}(\frac{\eta'_e}{4t}+1))^{1/2} (\breve{\text{P}}(\frac{\eta'_e}{4t}))^{3/2} \\\nonumber
&&+\sqrt{\frac{\pi}{2t}} (\breve{\text{P}}(\frac{\eta_e}{4t}+\frac{\eta'_e}{4t}))^{3/2}  (\breve{\text{P}}(\frac{\eta_e}{4t}+\frac{\eta'_e}{4t}+1))^{1/2} e^{-\frac{t}{2}(\frac{\eta_e}{2t}-\frac{\eta'_e}{2t})^2}  \exp(- [\frac{\eta'_e}{4t}]\widetilde{\Theta}_e)+ \frac{1}{\sqrt{t}}\cdot\mathcal{O}(e^{-\frac{\eta'^2_e}{8t}})\\\nonumber
\end{eqnarray}
for $\eta'_e$ being large, where ``$<$'' represents that the module of its left-hand side less than that the module of  its right-hand side. Then, in this case the matrix elements of $(\widehat{u_e^{-1}h_e\tilde{u}_e})_{12,12} $ is estimated by
 \begin{eqnarray}\label{H121212}
0&<&\left|\frac{\langle\breve{\Psi}_{\mathbb{H}^o_e}|(\widehat{u'^{-1}_e h_e\tilde{u}'_e})_{12,12}  |\breve{\Psi}_{\mathbb{H}'^o_e}\rangle }{||\breve{\Psi}_{\mathbb{H}'^o_e}||||\breve{\Psi}_{ \mathbb{H}^o_e}||}\right|\\\nonumber
&{\lesssim}&\frac{\sqrt{\frac{2t}{\pi}}\tilde{f}_1(\frac{\eta_e}{t},\frac{\eta'_e}{t})e^{-t(\frac{\eta'_e}{4t}-\frac{D+1}{2})^2}+ (\breve{\text{P}}(\frac{\eta_e}{4t}+\frac{\eta'_e}{4t}))^{3/2}  (\breve{\text{P}}(\frac{\eta_e}{4t}+\frac{\eta'_e}{4t}+1))^{1/2}e^{-\frac{t}{2}(\frac{\eta'_e}{2t}-\frac{\eta_e}{2t})^2} e^{- [\frac{\eta'_e}{4t}]\widetilde{\Theta}_e}}{ \breve{\text{P}}(\frac{\eta_e}{2t}) \breve{\text{P}}(\frac{\eta'_e}{2t})}
\end{eqnarray}
for large $\eta'_e$,
where $\tilde{f}_1(\frac{\eta_e}{t},\frac{\eta'_e}{t}):=([\eta'_e/4t]+1) \exp( -t(\frac{\eta_e}{2t}-\frac{\eta'_e}{4t}-\frac{D-1}{2})^2)(\breve{\text{P}}(\frac{\eta'_e}{4t}+1))^{1/2} (\breve{\text{P}}(\frac{\eta'_e}{4t}))^{3/2}$. Notice $\widetilde{\Theta}_e\simeq\eta_e+\eta'_e$ or $\widetilde{\Theta}_e\gg \eta_e+\eta'_e$ in this case, hence we can conclude that $\left|\frac{\langle\breve{\Psi}_{\mathbb{H}'^o_e}|(\widehat{u'^{-1}_e h_e\tilde{u}'_e})_{12,12} |\breve{\Psi}_{\mathbb{H}^o_e}\rangle}{||\breve{\Psi}_{\mathbb{H}'^o_e}||||\breve{\Psi}_{ \mathbb{H}^o_e}||}\right|$ is always suppressed exponentially by the factors $e^{-t(\frac{\eta'_e}{4t}-\frac{D+1}{2})^2}$ and $e^{-\frac{t}{2}(\frac{\eta'_e}{2t}-\frac{\eta_e}{2t})^2} e^{- [\frac{\eta'_e}{4t}]\widetilde{\Theta}_e}$ in Eq.\eqref{H121212}.
In the second case, we consider $\widetilde{\Theta}_e\ll \eta_e+\eta'_e$.
\begin{eqnarray}\label{H1212120}
&&\langle\breve{\Psi}_{\mathbb{H}^o_e}|(\widehat{u'^{-1}_e h_e\tilde{u}'_e})_{12,12}  |\breve{\Psi}_{\mathbb{H}'^o_e}\rangle e^{-\frac{(\eta_e)^2+(\eta'_e)^2+2t^2(D-1)^2}{4t}}\\\nonumber
&\stackrel{\text{large}\ \eta'_e}{=}&e^{\mathbf{i}\xi'^o_e} \sum_{N_e}(\dim(\pi_{N_e}))^{3/2}(\dim(\pi_{N_e+1}))^{1/2} \exp(-t(\frac{\eta_e}{2t} -d_{N_e+1})^2 -t(\frac{\eta'_e}{2t}-d_{N_e})^2)\\\nonumber
&&\cdot e^{\mathbf{i}d_{N_e+1}(\xi^o_e-\xi'^o_e)}\langle N_e+1, V'_e|N_e+1, V_e\rangle\langle N_e+1, -\tilde{V}_e|N_e+1,-\tilde{ V}'_e\rangle+\frac{1}{\sqrt{t}}\mathcal{O}(e^{-\frac{(\eta'_e)^2}{8t}})\\\nonumber
&=&e^{\mathbf{i}\xi'^o_e}e^{\mathbf{i}\frac{D+1}{2}(\xi^o_e-\xi'^o_e)} \sum_{N_e}(\dim(\pi_{N_e}))^{3/2}(\dim(\pi_{N_e+1}))^{1/2}\exp(-t(\frac{\eta_e}{2t}-1-d_{N_e})^2 -t(\frac{\eta'_e}{2t}-d_{N_e})^2)\\\nonumber
&&\cdot e^{\mathbf{i}(N_e+1)(\xi^o_e-\xi'^o_e+\varphi(u_e,u'_e) +\varphi(\tilde{u}_e,\tilde{u}'_e))}e^{-\widetilde{\Theta}_e} \exp(- N_e\widetilde{\Theta}_e)+\frac{1}{\sqrt{t}}\mathcal{O}(e^{-\frac{(\eta'_e)^2}{8t}})\\\nonumber
&=&e^{\mathbf{i}\xi'^o_e}e^{\mathbf{i}\frac{D+1}{2}(\xi^o_e-\xi'^o_e)} e^{\mathbf{i}(\xi^o_e-\xi'^o_e+\tilde{\varphi}_e)}e^{-t(\frac{\eta'_e}{2t}-\frac{\eta_e}{2t}+1)^2 +2t(\frac{\eta_e}{4t}-\frac{1}{2}-\frac{\eta'_e}{4t}-\frac{\widetilde{\Theta}_e}{4t})^2} e^{\mathbf{i}(\frac{\eta_e}{4t}-\frac{1}{2}-\frac{D-1}{2}+\frac{\eta'_e}{4t}-\frac{\widetilde{\Theta}_e}{4t})(\xi^o_e- \xi'^o_e+\tilde{\varphi}_e)} \\\nonumber
&&\cdot e^{-\widetilde{\Theta}_e}\exp(-(\frac{\eta'_e}{2t}-\frac{D-1}{2})\widetilde{\Theta}_e) \sum_{[\tilde{k}_e]}(\widetilde{\text{P}}(\tilde{k}_e))^{1/4}\left(\exp(-2t\tilde{k}_e^2)e^{\mathbf{i}\tilde{k}_e(\xi^o_e-\xi'^o_e +\tilde{\varphi}_e)}\right)+\frac{1}{\sqrt{t}}\mathcal{O}(e^{-\frac{(\eta'_e)^2}{8t}}),
\end{eqnarray}
here we defined $\tilde{k}_e:=d_{N_e}- \frac{\eta'_e}{4t}-\frac{\eta_e}{4t}+\frac{1}{2}+\frac{\widetilde{\Theta}_e}{4t}=[\tilde{k}_e]+\text{mod}(\tilde{k}_e,1)$ with $[\tilde{k}_e]$ being the maximum integer less than or equal to $\tilde{k}_e$ and $\text{mod}(\tilde{k}_e,1)$ being the corresponding remainder, and $\widetilde{\text{P}}(\tilde{k}_e)$ is a polynomial of $\tilde{k}_e$ defined by $(\widetilde{\text{P}}(\tilde{k}_e))^{1/4}=(\dim(\pi_{N_e}))^{3/2}(\dim(\pi_{N_e+1}))^{1/2}$.
 By applying the \textbf{Case II} discussed above, we can immediately get
\begin{eqnarray}\label{H121212122}
&&\langle\breve{\Psi}_{\mathbb{H}^o_e}|(\widehat{u'^{-1}_e h_e\tilde{u}'_e})_{12,12}  |\breve{\Psi}_{\mathbb{H}'^o_e}\rangle e^{-\frac{(\eta_e)^2+(\eta'_e)^2+2t^2(D-1)^2}{4t}}\\\nonumber
&\stackrel{\text{large}\ \eta'_e}{=}&e^{\mathbf{i}\xi'^o_e}e^{\mathbf{i}\frac{D+1}{2}(\xi^o_e-\xi'^o_e)}e^{\mathbf{i}(\xi^o_e-\xi'^o_e+\tilde{\varphi}_e)} e^{-t(\frac{\eta'_e}{2t}-\frac{\eta_e}{2t}+1)^2 +2t(\frac{\eta_e}{4t}-\frac{1}{2}-\frac{\eta'_e}{4t}-\frac{\widetilde{\Theta}_e}{4t})^2} e^{\mathbf{i}(\frac{\eta_e}{4t}-\frac{1}{2}-\frac{D-1}{2}+\frac{\eta'_e}{4t}-\frac{\widetilde{\Theta}_e}{4t})(\xi^o_e- \xi'^o_e+\tilde{\varphi}_e)} \\\nonumber
&&\cdot\exp(-\widetilde{\Theta}_e) \exp(-(\frac{\eta'_e}{2t}-\frac{D-1}{2})\widetilde{\Theta}_e) \sqrt{\frac{\pi}{2t}} (\widetilde{\text{P}}(0))^{1/4}e^{-\frac{(\xi^o_e-\xi'^o_e +\tilde{\varphi}_e)^2}{8t}}( 1+\mathcal{O}(\frac{t}{\eta'_e})+\mathcal{O}(e^{-1/t}))\\\nonumber
&&+\frac{1}{\sqrt{t}}\mathcal{O}(e^{-\frac{(\eta'_e)^2}{8t}})
\end{eqnarray}
and then
\begin{eqnarray}\label{H121212133}
&&\frac{\langle\breve{\Psi}_{\mathbb{H}^o_e}|(\widehat{u'^{-1}_e h_e\tilde{u}'_e})_{12,12}  |\breve{\Psi}_{\mathbb{H}'^o_e}\rangle }{||\breve{\Psi}_{\mathbb{H}'^o_e}||||\breve{\Psi}_{ \mathbb{H}^o_e}||}\\\nonumber
&\stackrel{\text{large}\ \eta'_e}{=}&e^{\mathbf{i}\xi'^o_e}e^{-\widetilde{\Theta}_e/2}e^{-t/2}e^{\mathbf{i}\frac{D+1}{2}(\xi^o_e-\xi'^o_e)}e^{\frac{\mathbf{i}}{2}(\xi^o_e-\xi'^o_e+\tilde{\varphi}_e)} e^{(\eta_e/2-\eta'_e/2)}\frac{\langle\breve{\Psi}_{\mathbb{H}^o_e} |\breve{\Psi}_{\mathbb{H}'^o_e}\rangle}{||\breve{\Psi}_{\mathbb{H}'^o_e}||||\breve{\Psi}_{ \mathbb{H}^o_e}||}( 1+\mathcal{O}(\frac{t}{\eta'_e})+\mathcal{O}(e^{-1/t})).
\end{eqnarray}
Combine the results \eqref{H121212} and \eqref{H121212133} of these two cases, we reach Eq.\eqref{h1212fina} immediately.\\
\\
\textbf{Calculation of Eq.\eqref{h212121}:}  The first term in the right-hand side (FRHS) of ``='' in Eq.\eqref{h212121} is given as
\begin{eqnarray}\label{FTFh21}
&&\text{FRHS\ of  Eq.}\eqref{h212121}\\\nonumber
&=& e^{\frac{(\eta_e)^2+(\eta'_e)^2+2t^2(D-1)^2}{4t}}e^{-\mathbf{i}\xi'^o_e}\sum_{N_e}(\dim(\pi_{N_e}))^{3/2} (\dim(\pi_{N_e-1}))^{1/2} \exp(-t(\frac{\eta_e}{2t} -d_{N_e-1})^2 -t(\frac{\eta'_e}{2t}-d_{N_e})^2)\\\nonumber
&&\cdot e^{\mathbf{i}d_{N_e-1}(\xi^o_e-\xi'^o_e)}\langle N_e-1, V'_e|N_e-1, V_e\rangle\langle N_e-1, -\tilde{V}_e|N_e-1,-\tilde{ V}'_e\rangle.
\end{eqnarray}
By following a similar analysis of Eqs. \eqref{H1212000} and \eqref{H1212120}, we can immediately give
\begin{eqnarray}\label{H21212133}
&&\frac{\text{FRHS\ of  Eq.}\eqref{h212121} }{||\breve{\Psi}_{\mathbb{H}'^o_e}||||\breve{\Psi}_{ \mathbb{H}^o_e}||}\\\nonumber
&\stackrel{\text{large}\ \eta'_e}{=}&e^{-\mathbf{i}\xi'^o_e}e^{\widetilde{\Theta}_e/2}e^{-t/2}e^{\mathbf{i}\frac{D+1}{2}(\xi^o_e-\xi'^o_e)}e^{-\frac{\mathbf{i}}{2}(\xi^o_e-\xi'^o_e+\tilde{\varphi}_e)} e^{-(\eta_e/2-\eta'_e/2)}\frac{\langle\breve{\Psi}_{\mathbb{H}^o_e} |\breve{\Psi}_{\mathbb{H}'^o_e}\rangle}{||\breve{\Psi}_{\mathbb{H}'^o_e}||||\breve{\Psi}_{ \mathbb{H}^o_e}||}( 1+\mathcal{O}(\frac{t}{\eta'_e})+\mathcal{O}(e^{-1/t}))
\end{eqnarray}
for large $\widetilde{\Theta}_e\ll \eta_e+\eta'_e$, and
 \begin{eqnarray}\label{H212121}
0&<&\left|\frac{\text{FRHS\ of  Eq.}\eqref{h212121}\rangle }{||\breve{\Psi}_{\mathbb{H}'^o_e}||||\breve{\Psi}_{ \mathbb{H}^o_e}||}\right|\\\nonumber
&\stackrel{\text{large}\ \eta'_e}{\lesssim}&\frac{\sqrt{\frac{2t}{\pi}}\tilde{f}'_1(\frac{\eta_e}{t},\frac{\eta'_e}{t})e^{-t(\frac{\eta'_e}{4t}-\frac{D+1}{2})^2}+ (\breve{\text{P}}(\frac{\eta_e}{4t}+\frac{\eta'_e}{4t}))^{3/2}  (\breve{\text{P}}(\frac{\eta_e}{4t}+\frac{\eta'_e}{4t}-1))^{1/2}e^{-\frac{t}{2}(\frac{\eta'_e}{2t}-\frac{\eta_e}{2t})^2} e^{- [\frac{\eta'_e}{4t}]\widetilde{\Theta}_e}}{ \breve{\text{P}}(\frac{\eta_e}{2t}) \breve{\text{P}}(\frac{\eta'_e}{2t})}
\end{eqnarray}
for $\widetilde{\Theta}_e\simeq\eta_e+\eta'_e$ or $\widetilde{\Theta}_e\gg \eta_e+\eta'_e$,
where $\tilde{f}'_1(\frac{\eta_e}{t},\frac{\eta'_e}{t}):=([\eta'_e/4t]+1) \exp( -t(\frac{\eta_e}{2t}-\frac{\eta'_e}{4t}-\frac{D+3}{2})^2)(\breve{\text{P}}(\frac{\eta'_e}{4t}-1))^{1/2} (\breve{\text{P}}(\frac{\eta'_e}{4t}))^{3/2}$. Also, for the case $\widetilde{\Theta}_e\simeq\eta_e+\eta'_e$ or $\widetilde{\Theta}_e\gg \eta_e+\eta'_e$, we can conclude that $\left|\frac{\text{FRHS\ of  Eq.}\eqref{h212121}\rangle }{||\breve{\Psi}_{\mathbb{H}'^o_e}||||\breve{\Psi}_{ \mathbb{H}^o_e}||}\right|$ is always suppressed exponentially by the factors $e^{-t(\frac{\eta'_e}{4t}-\frac{D+1}{2})^2}$ and $e^{-\frac{t}{2}(\frac{\eta'_e}{2t}-\frac{\eta_e}{2t})^2} e^{- [\frac{\eta'_e}{4t}]\widetilde{\Theta}_e}$ based on Eq.\eqref{H212121}.

The second term in the right-hand side (SRHS) of ``='' in Eq.\eqref{h212121} reads
\begin{eqnarray}\label{STFFh21}
&& \text{SRHS\ of  Eq.}\eqref{h212121}\\\nonumber
&=&
e^{\frac{(\eta_e)^2+(\eta'_e)^2+2t^2(D-1)^2}{4t}}e^{-\mathbf{i}\xi'^o_e}\sum_{N_e}(\dim(\pi_{N_e}))^{3/2} (\dim(\pi_{N_e-1}))^{1/2} \exp(-t(\frac{\eta_e}{2t} -d_{N_e-1})^2 -t(\frac{\eta'_e}{2t}-d_{N_e})^2)\\\nonumber
&&\cdot e^{\mathbf{i}d_{N_e-1}(\xi^o_e-\xi'^o_e)}(1-|\alpha_1(N_e)|^2)\langle N_e-1,V_{12}| u'^{-1}_eu_e|N_e-1,V_{12}\rangle\langle N_e-1,V_{12}|\tilde{u}^{-1}_e\tilde{u}'_e|N_e-1,V_{12}\rangle.
\end{eqnarray}
 It is easy to see
\begin{eqnarray}\label{STFFh2121}
&& \left|\text{SRHS\ of  Eq.}\eqref{h212121}\right|\\\nonumber
&\leq&
e^{\frac{(\eta_e)^2+(\eta'_e)^2+2t^2(D-1)^2}{4t}}\sum_{N_e}(\dim(\pi_{N_e}))^{3/2} (\dim(\pi_{N_e-1}))^{1/2} \\\nonumber
&&\cdot\exp(-t(\frac{\eta_e}{2t} -d_{N_e-1})^2 -t(\frac{\eta'_e}{2t}-d_{N_e})^2) (1-\frac{N_e(2N_e+D-3)}{(D+N_e-2)(2N_e+D-1)})\\\nonumber
&=&e^{\frac{(\eta_e)^2+(\eta'_e)^2+2t^2(D-1)^2}{4t}} e^{-\frac{t}{2}(\frac{\eta'_e}{2t}-\frac{\eta_e}{2t}-1)^2 } \sum_{[\tilde{k}_e]}(\widetilde{\text{P}}_1(\tilde{k}_e))^{1/4} \frac{\widetilde{\text{P}}_2(\tilde{k}_e)}{\widetilde{P}_3(\tilde{k}_e)}\left(\exp(-2t\tilde{k}_e^2)\right)
\end{eqnarray}
for large $\eta'_e$,
here $\tilde{k}_e$ is defined by $\tilde{k}_e:=d_{N_e}- \frac{\eta'_e}{4t}-\frac{\eta_e}{4t}-\frac{1}{2}=[\tilde{k}_e]+\text{mod}(\tilde{k}_e,1)$, $(\widetilde{\text{P}}_1(\tilde{k}_e))^{1/4}$ is defined by $(\widetilde{\text{P}}_1(\tilde{k}_e))^{1/4}=(\dim(\pi_{N_e}))^{3/2} (\dim(\pi_{N_e-1}))^{1/2}$, and  $\frac{\widetilde{\text{P}}_2(\tilde{k}_e)}{\widetilde{P}_3(\tilde{k}_e)}$ is defined by $\frac{\widetilde{\text{P}}_2(\tilde{k}_e)}{\widetilde{P}_3(\tilde{k}_e)}=1-\frac{N_e(2N_e+D-3)}{(D+N_e-2)(2N_e+D-1)}$.
Then by using the result of \textbf{Case III} discussed above,  we have
\begin{eqnarray}\label{STFFh21212121}
0&<&\left|\frac{\text{SRHS\ of  Eq.}\eqref{h212121} }{||\breve{\Psi}_{\mathbb{H}'^o_e}||||\breve{\Psi}_{ \mathbb{H}^o_e}||}\right|\\\nonumber
&\stackrel{\text{large}\ \eta'_e}{\lesssim}& \frac{e^{-\frac{t}{2}(\frac{\eta'_e}{2t}-\frac{\eta_e}{2t}-1)^2 } (\widetilde{\text{P}}_1(0))^{1/4}(\frac{\eta'_e+\eta_e}{\sqrt{t}}\mathcal{O}(e^{-\frac{{\eta_e'}^2}{8t}})+\mathcal{O}(\frac{t}{\eta'_e}))}{{ \breve{\text{P}}(\frac{\eta_e}{2t}) \breve{\text{P}}(\frac{\eta'_e}{2t})}}.
\end{eqnarray}
Then we can conclude that $\left|\frac{\text{SRHS\ of  Eq.}\eqref{h212121}\rangle }{||\breve{\Psi}_{\mathbb{H}'^o_e}||||\breve{\Psi}_{ \mathbb{H}^o_e}||}\right|$ is always suppressed by the factor $(\frac{\eta'_e+\eta_e}{\sqrt{t}}\mathcal{O}(e^{-\frac{{\eta_e'}^2}{8t}})+\mathcal{O}(\frac{t}{\eta'_e}))$ for $\eta_e\simeq\eta'_e$ and by the factor $e^{-\frac{t}{2}(\frac{\eta'_e}{2t}-\frac{\eta_e}{2t}-1)^2 }$ for $|\eta_e-\eta'_e|$ being large in the case of large $\eta'_e$.

The third term in the right-hand side (TRHS) of ``='' in Eq.\eqref{h212121} is given as
\begin{eqnarray}\label{FTFh21}
&&\text{TRHS\ of  Eq.}\eqref{h212121}\\\nonumber
&=& e^{\frac{(\eta_e)^2+(\eta'_e)^2+2t^2(D-1)^2}{4t}}e^{\mathbf{i}\xi'^o_e}\sum_{N_e} (\dim(\pi_{N_e}))^{3/2} (\dim(\pi_{N_e+1}))^{1/2} \exp(-t(\frac{\eta_e}{2t} -d_{N_e+1})^2 -t(\frac{\eta'_e}{2t}-d_{N_e})^2)\\\nonumber
&&\cdot |\alpha_2(N_e)|^2e^{\mathbf{i}d_{N_e+1}(\xi^o_e-\xi'^o_e)}\langle N_e+1,...| u'^{-1}_eu_e|N_e+1,V_{12}\rangle\langle N_e+1, V_{12}|\tilde{u}^{-1}_e\tilde{u}'_e|N_e+1,...\rangle,
\end{eqnarray}
wherein $0\leq |\alpha_2(N_e)|^2\leq1-\frac{N_e(2N_e+D-3)}{(D+N_e-2)(2N_e+D-1)}$. By using the result of the \textbf{Cases III} discussed above, Eq.\eqref{FTFh21} can be estimated following a similar procedure of Eq.\eqref{STFFh21}, which gives
\begin{eqnarray}\label{TTFFh21212121}
0&<&\left|\frac{\text{TRHS\ of  Eq.}\eqref{h212121} }{||\breve{\Psi}_{\mathbb{H}'^o_e}||||\breve{\Psi}_{ \mathbb{H}^o_e}||}\right|\\\nonumber
&\stackrel{\text{large}\ \eta'_e}{\lesssim}& \frac{e^{-\frac{t}{2}(\frac{\eta'_e}{2t}-\frac{\eta_e}{2t}+1)^2 } (\widetilde{\text{P}}(0))^{1/4}(\frac{\eta'_e+\eta_e}{\sqrt{t}}\mathcal{O}(e^{-\frac{{\eta_e'}^2}{8t}})+\mathcal{O}(\frac{t}{\eta'_e}))}{{ \breve{\text{P}}(\frac{\eta_e}{2t}) \breve{\text{P}}(\frac{\eta'_e}{2t})}},
\end{eqnarray}
here we defined $\tilde{k}_e:=d_{N_e}- \frac{\eta'_e}{4t}-\frac{\eta_e}{4t}+\frac{1}{2}$ and $(\widetilde{\text{P}}(\tilde{k}_e))^{1/4}:=(\dim(\pi_{N_e}))^{3/2} (\dim(\pi_{N_e+1}))^{1/2}$.  Then we can conclude that $\left|\frac{\text{TRHS\ of  Eq.}\eqref{h212121}\rangle }{||\breve{\Psi}_{\mathbb{H}'^o_e}||||\breve{\Psi}_{ \mathbb{H}^o_e}||}\right|$ is always suppressed by the factor $(\frac{\eta'_e+\eta_e}{\sqrt{t}}\mathcal{O}(e^{-\frac{{\eta_e'}^2}{8t}})+\mathcal{O}(\frac{t}{\eta'_e}))$ for $\eta_e\simeq\eta'_e$ and by the factor $e^{-\frac{t}{2}(\frac{\eta'_e}{2t}-\frac{\eta_e}{2t}+1)^2 }$ for $|\eta_e-\eta'_e|$ being large in the case of large $\eta'_e$.

Finally, by combining the results of Eqs.\eqref{H21212133}, \eqref{H212121},  \eqref{STFFh21212121} and \eqref{TTFFh21212121} we get Eq.\eqref{h2121fina2} immediately.
\\
\\
\textbf{Calculation of Eq.\eqref{h122100}:}
 The expression of Eq.\eqref{h122100} reads
\begin{eqnarray}\label{applica}
&& \left|\langle\breve{\Psi}_{\mathbb{H}^o_e}|(\widehat{u'^{-1}_e h_e\tilde{u}'_e})_{12,21}  |\breve{\Psi}_{\mathbb{H}'^o_e}\rangle\right|\\\nonumber
&\stackrel{\text{large}\ \eta'_e}{=}&e^{\frac{(\eta_e)^2+(\eta'_e)^2+2t^2(D-1)^2}{4t}}e^{\mathbf{i}\xi'^o_e}\sum_{N_e}(\dim(\pi_{N_e}))^{\frac{3}{2}} (\dim(\pi_{N_e+1}))^{\frac{1}{2}} \exp(-t(\frac{\eta_e}{2t} -d_{N_e+1})^2 -t(\frac{\eta'_e}{2t}-d_{N_e})^2)\\\nonumber
&&\cdot e^{\mathbf{i}d_{N_e+1}(\xi^o_e-\xi'^o_e)}\alpha_2(N_e)\langle N_e+1, V'_e|N_e+1, V_e\rangle\langle N_e+1, -\tilde{V}_e|N_e+1,...\rangle+\frac{1}{\sqrt{t}}\mathcal{O}(e^{-(\eta'_e)^2/(8t)})\\\nonumber
&\stackrel{\text{large}\ \eta'_e}{\lesssim}&e^{\frac{(\eta_e)^2+(\eta'_e)^2+2t^2(D-1)^2}{4t}}\\\nonumber
&&\cdot \sum_{N_e}(\dim(\pi_{N_e}))^{3/2} (\dim(\pi_{N_e+1}))^{1/2} \exp(-t(\frac{\eta_e}{2t} -d_{N_e+1})^2 -t(\frac{\eta'_e}{2t}-d_{N_e})^2) \sqrt{1-|\alpha_1(N_e)|^2}
\end{eqnarray}
Similar to Eq.\eqref{STFFh2121}, the result of the \textbf{Case III} discussed above is applicable for \eqref{applica} and we get
\begin{eqnarray}\label{TTFFh12211221}
0&<&\left|\frac{\langle\breve{\Psi}_{\mathbb{H}^o_e}|(\widehat{u'^{-1}_e h_e\tilde{u}'_e})_{12,21}  |\breve{\Psi}_{\mathbb{H}'^o_e}\rangle }{||\breve{\Psi}_{\mathbb{H}'^o_e}||||\breve{\Psi}_{ \mathbb{H}^o_e}||}\right|\\\nonumber
&\stackrel{\text{large}\ \eta'_e}{\lesssim}& \frac{e^{-\frac{t}{2}(\frac{\eta'_e}{2t}-\frac{\eta_e}{2t}+1)^2 } (\widetilde{\text{P}}(0))^{1/4}(\frac{\eta'_e+\eta_e}{\sqrt{t}}\mathcal{O}(e^{-\frac{{\eta_e'}^2}{8t}})+\mathcal{O}(\sqrt{\frac{t}{\eta'_e}}))}{{ \breve{\text{P}}(\frac{\eta_e}{2t}) \breve{\text{P}}(\frac{\eta'_e}{2t})}},
\end{eqnarray}
here $\tilde{k}_e$ is defined by $\tilde{k}_e:=d_{N_e}- \frac{\eta'_e}{4t}-\frac{\eta_e}{4t}+\frac{1}{2}$ and $(\widetilde{\text{P}}(\tilde{k}_e))^{1/4}:=(\dim(\pi_{N_e}))^{3/2} (\dim(\pi_{N_e+1}))^{1/2}$. Then, we can conclude that $\left|\frac{\langle\breve{\Psi}_{\mathbb{H}^o_e}|(\widehat{u'^{-1}_e h_e\tilde{u}'_e})_{12,21}  |\breve{\Psi}_{\mathbb{H}'^o_e}\rangle }{||\breve{\Psi}_{\mathbb{H}'^o_e}||||\breve{\Psi}_{ \mathbb{H}^o_e}||}\right|$ is always suppressed by the factor $(\frac{\eta'_e+\eta_e}{\sqrt{t}}\mathcal{O}(e^{-\frac{{\eta_e'}^2}{8t}})+\mathcal{O}(\sqrt{\frac{t}{\eta'_e}}))$ for $\eta_e\simeq\eta'_e$ and by the factor $e^{-\frac{t}{2}(\frac{\eta'_e}{2t}-\frac{\eta_e}{2t}+1)^2 }$ for $|\eta_e-\eta'_e|$ being large in the case of large $\eta'_e$.


\bibliographystyle{unsrt}

\bibliography{ref}

\begin{thebibliography}{10}

\bibitem{Long:2021lmd}
Gaoping Long, Xiangdong Zhang, and Cong Zhang.
\newblock {Twisted geometry coherent states in all dimensional loop quantum
  gravity: Construction and peakedness properties}.
\newblock {\em Phys. Rev. D}, 105(6):066021, 2022.

\bibitem{Rovelli_2006}
Carlo Rovelli.
\newblock Graviton propagator from background-independent quantum gravity.
\newblock {\em Physical Review Letters}, 97(15), Oct 2006.

\bibitem{Bianchi_2009}
Eugenio Bianchi, Elena Magliaro, and Claudio Perini.
\newblock Lqg propagator from the new spin foams.
\newblock {\em Nuclear Physics B}, 822(1-2):245–269, Nov 2009.

\bibitem{Bianchi_2010}
Eugenio Bianchi, Elena Magliaro, and Claudio Perini.
\newblock Coherent spin-networks.
\newblock {\em Physical Review D}, 82(2), Jul 2010.

\bibitem{Calcinari_2020}
Andrea Calcinari, Laurent Freidel, Etera Livine, and Simone Speziale.
\newblock Twisted geometries coherent states for loop quantum gravity.
\newblock {\em Classical and Quantum Gravity}, 38(2):025004, Dec 2020.

\bibitem{Freidel:2010aq}
Laurent Freidel and Simone Speziale.
\newblock {Twisted geometries: A geometric parametrisation of SU(2) phase
  space}.
\newblock {\em Phys. Rev. D}, 82:084040, 2010.

\bibitem{1994The}
B.~Hall.
\newblock The segal-bargmann "coherent state" transform for compact lie groups.
\newblock {\em Journal of Functional Analysis}, 122(1):103--151, 1994.

\bibitem{ThiemannComplexifierCoherentStates}
T~Thiemann.
\newblock {Complexifier coherent states for quantum general relativity}.
\newblock {\em Class. Quantum Gravity}, 23:2063--2117, mar 2006.

\bibitem{Thomas2001Gauge}
Thomas Thiemann.
\newblock Gauge field theory coherent states (gcs): I. general properties.
\newblock {\em Classical and Quantum Gravity}, 18(11), 2001.

\bibitem{2001Gauge}
T.~Thiemann and O.~Winkler.
\newblock Gauge field theory coherent states (gcs): Ii. peakedness properties.
\newblock {\em Classical and Quantum Gravity}, 18(14):2561--2636, 2001.

\bibitem{2000Gauge}
T.~Thiemann and O.~Winkler.
\newblock Gauge field theory coherent states (gcs) : Iii. ehrenfest theorems.
\newblock {\em classical and quantum gravity}, 18(21):2561--2636, 2000.

\bibitem{Han_2020}
Muxin Han and Hongguang Liu.
\newblock Effective dynamics from coherent state path integral of full loop
  quantum gravity.
\newblock {\em Physical Review D}, 101(4), Feb 2020.

\bibitem{Han_2020semiclassical}
Muxin Han and Hongguang Liu.
\newblock Semiclassical limit of new path integral formulation from reduced
  phase space loop quantum gravity.
\newblock {\em Physical Review D}, 102(2), Jul 2020.

\bibitem{Zhang:2021qul}
Cong Zhang, Shicong Song, and Muxin Han.
\newblock {First-Order Quantum Correction in Coherent State Expectation Value
  of Loop-Quantum-Gravity Hamiltonian}.
\newblock {\em Phys. Rev. D}, 105:064008, 2022.

\bibitem{Long:2021izw}
Gaoping Long and Yongge Ma.
\newblock {Effective dynamics of weak coupling loop quantum gravity}.
\newblock {\em Phys. Rev. D}, 105(4):044043, 2022.

\bibitem{Long:2021xjm}
Gaoping Long, Cong Zhang, and Xiangdong Zhang.
\newblock {Superposition type coherent states in all dimensional loop quantum
  gravity}.
\newblock {\em Phys. Rev. D}, 104(4):046014, 2021.

\bibitem{Long:2020euh}
Gaoping Long and Norbert Bodendorfer.
\newblock {Perelomov-type coherent states of SO($D+1$) in all-dimensional loop
  quantum gravity}.
\newblock {\em Phys. Rev. D}, 102(12):126004, 2020.

\bibitem{Livine:2007Nsfv}
Etera~R Livine and Simone Speziale.
\newblock New spinfoam vertex for quantum gravity.
\newblock {\em Physical Review D}, 76(8):084028, 2007.

\bibitem{Bodendorfer:Ha}
Norbert Bodendorfer, Thomas Thiemann, and Andreas Thurn.
\newblock New variables for classical and quantum gravity in all dimensions: I.
  \textsc{H}amiltonian analysis.
\newblock {\em Classical and Quantum Gravity}, 30(4):045001, 2013.

\bibitem{Bodendorfer:La}
Norbert Bodendorfer, Thomas Thiemann, and Andreas Thurn.
\newblock New variables for classical and quantum gravity in all dimensions:
  \textsc{II}. \textsc{L}agrangian analysis.
\newblock {\em Classical and Quantum Gravity}, 30(4):045002, 2013.

\bibitem{Bodendorfer:Qu}
Norbert Bodendorfer, Thomas Thiemann, and Andreas Thurn.
\newblock New variables for classical and quantum gravity in all dimensions:
  \textsc{III}. \textsc{Q}uantum theory.
\newblock {\em Classical and Quantum Gravity}, 30(4):045003, 2013.

\bibitem{Ashtekar2012Background}
Abhay Ashtekar and Jerzy Lewandowski.
\newblock Background independent quantum gravity: a status report.
\newblock {\em Classical and Quantum Gravity}, 21(15):R53--R152, 2012.

\bibitem{thiemann2007modern}
Thomas Thiemann.
\newblock {\em Modern canonical quantum general relativity}.
\newblock Cambridge University Press, 2007.

\bibitem{rovelli2007quantum}
Carlo Rovelli.
\newblock {\em Quantum gravity}.
\newblock Cambridge university press, 2007.

\bibitem{RovelliBook2}
Carlo Rovelli and Francesca Vidotto.
\newblock {\em {Covariant Loop Quantum Gravity: An Elementary Introduction to
  Quantum Gravity and Spinfoam Theory}}.
\newblock Cambridge University Press, 2014.

\bibitem{Han2005FUNDAMENTAL}
Muxin Han, M.~A. Yongge, and Weiming Huang.
\newblock Fundamental structure of loop quantum gravity.
\newblock {\em International Journal of Modern Physics D}, 16(09):1397--1474,
  2005.

\bibitem{Bodendorfer:SgI}
Norbert Bodendorfer, Thomas Thiemann, and Andreas Thurn.
\newblock Towards loop quantum supergravity (lqsg): I.
  \textsc{R}arita--\textsc{S}chwinger sector.
\newblock {\em Classical and Quantum Gravity}, 30(4):045006, 2013.

\bibitem{Bodendorfer:2011onthe}
Norbert Bodendorfer, Thomas Thiemann, and Andreas Thurn.
\newblock On the implementation of the canonical quantum simplicity constraint.
\newblock {\em Classical and Quantum Gravity}, 30(4):045005, 2013.

\bibitem{Dupuis:2010RSC}
Maite Dupuis and Etera~R Livine.
\newblock Revisiting the simplicity constraints and coherent intertwiners.
\newblock {\em Classical and Quantum Gravity}, 28(8):085001, 2011.

\bibitem{Kaminski:2009SFA}
Wojciech Kami{\'n}ski, Marcin Kisielowski, and Jerzy Lewandowski.
\newblock Spin-foams for all loop quantum gravity.
\newblock {\em Classical and Quantum Gravity}, 27(9):095006, 2010.

\bibitem{Engle:2007va}
Jonathan Engle, Roberto Pereira, and Carlo Rovelli.
\newblock Loop-quantum-gravity vertex amplitude.
\newblock {\em Physical review letters}, 99(16):161301, 2007.

\bibitem{Engle:2007LQGv}
Jonathan Engle, Etera Livine, Roberto Pereira, and Carlo Rovelli.
\newblock Lqg vertex with finite immirzi parameter.
\newblock {\em Nuclear Physics B}, 799(1-2):136--149, 2008.

\bibitem{Freidel:2007NSF}
Laurent Freidel and Kirill Krasnov.
\newblock A new spin foam model for 4d gravity.
\newblock {\em Classical and Quantum Gravity}, 25(12):125018, 2008.

\bibitem{FreidelBFDescriptionOf}
L.~Freidel, K.~Krasnov, and R.~Puzio.
\newblock {BF description of higher-dimensional gravity theories}.
\newblock {\em Adv. Theor. Math. Phys.}, 3:1289--1324, jan 1999.

\bibitem{Barrett:1997Rsn}
John~W Barrett and Louis Crane.
\newblock Relativistic spin networks and quantum gravity.
\newblock {\em Journal of Mathematical Physics}, 39(6):3296--3302, 1998.

\bibitem{PhysRevD.103.086016}
Gaoping Long and Chun-Yen Lin.
\newblock Geometric parametrization of $so\mathbf{(}d+1\mathbf{)}$ phase space
  of all dimensional loop quantum gravity.
\newblock {\em Phys. Rev. D}, 103:086016, Apr 2021.

\bibitem{Bianchi:2010Polyhedra}
Eugenio Bianchi, Pietro Dona, and Simone Speziale.
\newblock Polyhedra in loop quantum gravity.
\newblock {\em Physical Review D}, 83(4):044035, 2011.

\bibitem{long2020operators}
Gaoping Long and Yongge Ma.
\newblock {General geometric operators in all dimensional loop quantum
  gravity}.
\newblock {\em Phys. Rev. D}, 101(8):084032, 2020.

\bibitem{Long:2020agv}
Gaoping Long and Yongge Ma.
\newblock {Polytopes in all dimensional loop quantum gravity}.
\newblock {\em Eur. Phys. J. C}, 82(1):41, 2022.

\bibitem{Zhang:2015bxa}
Xiangdong Zhang.
\newblock {Higher dimensional Loop Quantum Cosmology}.
\newblock {\em Eur. Phys. J. C}, 76(7):395, 2016.

\bibitem{long2019coherent}
Gaoping Long, Chun-Yen Lin, and Yongge Ma.
\newblock Coherent intertwiner solution of simplicity constraint in all
  dimensional loop quantum gravity.
\newblock {\em Physical Review D}, 100(6):064065, 2019.

\bibitem{vilenkin2013representation}
NY~Vilenkin and Anatoli~Ulianovich Klimyk.
\newblock {\em Representation of Lie groups and special functions: Volume 2:
  Class I Representations, Special Functions, and Integral Transforms},
  volume~75.
\newblock Springer Science \& Business Media, 2013.

\bibitem{GeneralizedCoherentStates}
Askold Perelomov.
\newblock {\em Generalized coherent states and their applications}.
\newblock Springer Science \& Business Media, 2012.

\bibitem{1975Methods}
Michael Reed and Barry Simon.
\newblock {\em Methods of Modern Mathematical Physics Vol 2 - Fourier Analysis,
  Self Adjointness}.
\newblock New York: Academic, 1975.

\bibitem{Poissonsummation}
S.~Bochner.
\newblock {\em Vorlesungen über Fouriersche Integrale}.
\newblock Chelsea Publishing Company, 1948.

\bibitem{Giesel_2007}
K~Giesel and T~Thiemann.
\newblock Algebraic quantum gravity ({AQG}): {III}. semiclassical perturbation
  theory.
\newblock {\em Classical and Quantum Gravity}, 24(10):2565--2588, apr 2007.

\end{thebibliography}

\end{document}